\documentclass[fleqn,usenatbib]{mnras}

\usepackage{newtxtext,newtxmath}

\usepackage[T1]{fontenc}
\usepackage{ae,aecompl,amsmath}
\usepackage{xspace}
\usepackage[dvipsnames]{xcolor}

\usepackage{graphicx}	
\usepackage{amsmath}	
\usepackage{comment}    
\usepackage{float}
\usepackage{epstopdf}
\usepackage[colorinlistoftodos]{todonotes}  %
\usepackage{orcidlink}
\newcommand{\angstrom}{\textup{\AA}}

\title[Helium abundance in AGNs]{Cosmic evolution of the helium and oxygen abundances in type~2 Active Galactic Nuclei: Helium-loud AGNs}
\author[O.~L. Dors et al.]{O.~L. Dors$^{1}$\thanks{E-mail: olidors@univap.br}\orcidlink{0000-0003-4782-1570}, M.  Armah$^{1}$, G.~F. H\"agele$^{2,3}$\orcidlink{0000-0002-9011-8517}, M.~V. Cardaci$^{2,3}$\orcidlink{0000-0002-8856-602X}, P.~C. Santos$^{1}$, R.~A. Riffel$^{4}$,
G. Cresci$^{5}$,\newauthor{R. Riffel$^{6}$, L.~N. Marinho$^{6}$,
X. Ji$^{7,8}$, Y. Isobe$^{7,8}$, R. Maiolino$^{7,8,9}$, A. Marconi$^{5}$, A. Feltre$^{5}$, R.~L. Sanders$^{10}$,}
\newauthor{I.~N. Morais$^{1}$, G.~C. Almeida$^{1}$}
\\
$^{1}$Universidade do Vale do Para\'iba, Av. Shishima Hifumi, 2911, Cep
12244-000, S\~ao Jos\'e dos Campos, SP, Brazil \\
$^2$ Facultad de Ciencias Astron\'omicas y Geof\'{\i}sicas, Universidad Nacional de La Plata, Paseo del Bosque s/n, 1900 La Plata, Argentina.\\
$^3$ Instituto de Astrofísica de La Plata (CONICET-UNLP), La Plata, Avenida Centenario (Paseo del Bosque) S/N, B1900FWA, Argentina\\
$^{4}$ Departamento de F\'isica, Centro de Ci\^encias Naturais e Exatas, Universidade Federal de Santa Maria, 97105-900, Santa Maria, RS, Brazil\\
$^{5}$INAF – Osservatorio Astrofisico di Arcetri, largo E. Fermi 5, I-50127 Firenze, Italy\\
$^{6}$Departamento de Astronomia, Instituto de Física, Universidade Federal do Rio Grande do Sul, CP 15051, 91501-970, Porto Alegre, RS, Brazil\\
$^{7}$Kavli Institute for Cosmology, University of Cambridge, Madingley Road, Cambridge CB3 0HA, UK\\
$^{8}$Cavendish Laboratory – Astrophysics Group, University of Cambridge, 19
JJ Thomson Avenue, Cambridge CB3 0HE, UK\\
$^{9}$Department of Physics and Astronomy, University College London, Gower Street, London WC1E 6BT, UK\\
$^{10}$Department of Physics and Astronomy, University of Kentucky, 505 Rose Street, Lexington, KY 40506, USA}

\date{Accepted XXX. Received YYY; in original form ZZZ}

\pubyear{2026}

\begin{document}
\label{firstpage}
\pagerange{\pageref{firstpage}--\pageref{lastpage}}
\maketitle

\begin{abstract}

We derive the helium and oxygen abundances in the Narrow Line Regions (NLRs) of 84 Active Galactic Nuclei (AGNs) spanning two redshift ranges: 65 objects at $z \: < \: 0.2$ and 19 objects at  $2.8 \lesssim z \lesssim 6.8$. Spectroscopic data in the optical rest-frame range $ [3000 \: < \: \lambda (\angstrom) \: < \: 7000]$ from the JADES survey and from the literature were used to estimate $\text{He/H}$ and $\text{O/H}$ via the $T_{\rm e}$-method (7 objects) and
strong-emission line calibrations (19 objects). Our results indicate that the ionization degree in AGNs increases toward higher redshifts, exhibiting a trend similar to that observed in star-forming galaxies. We find that the majority of our sample of AGNs at $z \: > \: 2.8$ show oversolar helium abundances and subsolar oxygen abundances. We identified, via the $T_{\rm e}$-method, an object (goods-s-mediumhst-58850; $z=6.2615$) with the highest helium abundance estimated to date, i.e. $12+\log({\rm He/H})=11.64$, corresponding to $\rm (He/He_{\odot}) \sim 4.4$.
This result remains robust even when we adopt larger electron densities (varying $N_{\rm e}$ from $500$ to $10\:000$ $\rm cm^{-3}$) in abundance estimates using the $T_{\rm e}$-method and empirical strong-line methods. We found marginal evidence for a decline in the He/H abundance ($\sim 0.04$ dex per redshift unit) toward lower redshifts. In contrast, we found  evidence for an increase in $\text{O/H}$ toward the local universe at a rate of approximately 0.06 dex per redshift unit.

\end{abstract}




\section{Introduction}

Helium is the second most abundant element in the interstellar medium (ISM), and it exhibits emission lines  across distinct spectral ranges,  including ultraviolet (\ion{He}{ii}$\lambda1640$), optical (e.g. \ion{He}{i}$\lambda5876$, \ion{He}{ii}$\lambda4686$),
and infrared (e.g. \ion{He}{i}$\lambda10830$, $2.06\mu$m).
Estimates of the  total helium abundance relative to hydrogen,  defined as
\begin{equation}
y=\rm \left(\frac{He}{H}\right)_{total}=\frac{He^{0}+He^{+}+He^{2+}}{H^{0}+H^{+}},
\end{equation}
play an important
role in studies of  chemical evolution of galaxies, the  characterisation of
stellar nucleosynthesis, and the estimation of the primordial helium abundance \citep{1974ApJ...193..327P, 1992MNRAS.255..325P}.

The helium content in the gas phase of galaxies consists of two components: a constant abundance originating from primordial Big Bang nucleosynthesis \citep{1966ApJ...146..542P, 1967ApJ...148....3W, 1973ApJ...179..343W}, here defined as $y_{p}$, and another component ($\Delta y$) produced by stellar nucleosynthesis and subsequently ejected into the ISM, mainly through supernovae and the returned envelopes of intermediate-mass asymptotic giant branch (AGB) stars (e.g. \citealt{1974ApJ...193..327P}). Thus, the total helium abundance can be expressed as
\begin{equation}
y=y_{p}+\Delta y,
\end{equation}
where $\Delta y$ depends on the stellar Initial Mass Function (IMF), star formation rate (SFR), gas mass, evolutionary time, and other factors
(e.g. \citealt{2005MNRAS.358..521M, 2019A&A...630A.125V, 2025MNRAS.538.1517W}). 
For consistency with previous studies (e.g. \citealt{2003ApJ...591..801K, 2006ApJS..167..177D, 2008MNRAS.383..209H}), we define 
\begin{equation}
\Delta y= \rm \frac{He}{H}. 
\end{equation}

Galactic chemical evolution (GCE) models predict that helium is  distributed to
the ISM through the winds and core-collapse
supernovae (CCSN) of massive stars [$(M_{\star}/\rm M_{\odot})>$\,40] and  from ejected envelopes of
intermediate mass [$(M_{\star}/\rm M_{\odot})=$4-8] stars (see \citealt{2025MNRAS.538.1517W} and references therein). The helium launched by stars into the ISM  depends on how much
of the synthesized helium is ejected rather than converted into other chemical elements.
In this context, the He/H abundance   
can be used as a proxy to study the nucleosynthesis of
massive stars
in the early stages
($z \: > \: 5$) of galaxy formation (e.g. \citealt{2024ApJ...974..266Y}), where AGB stars  contribute little to the ISM helium enrichment. 

Estimates of  $y_{p}$ require precision of a few per cent ($\sim 1$ per cent, e.g. \citealt{2010IAUS..268..163F}) in its abundance value,  which can be obtained under certain circumstances
\citep{1992MNRAS.255..325P}, such as spectra with high signal-to-noise ratio (S/N), electron temperature ($T_{\rm e}$) and electron density
($N_{\rm e}$) determinations for the gas region occupied by both $\rm He^{+}$ and $\rm He^{2+}$ ions,
and  estimation of the fraction of $\rm He^{0}$ (e.g. \citealt{1986PASP...98.1061P, 1992RMxAA..24..155P, 2000MNRAS.311..329D, 2014MNRAS.445..778I, 2019ApJ...876...98V, 2020MNRAS.496.2726M, 2020ApJ...896...77H, 2021ApJ...922..170B}).
Otherwise, He/H estimates aimed at investigating the ISM enrichment by stars
can be carried out under less rigorous conditions as, for instance,  
simplified suppositions for $T_{\rm e}$ and $N_{\rm e}$ as well as helium
lines measured with  low S/N ($ \gtrsim \: 2$). In fact, \citet{2024ApJ...974..266Y}, using spectroscopic data of three galaxies at $z\sim 6$ observed with the James Webb Space Telescope (JWST), was unable to directly estimate $T_{\rm e}$ and $N_{\rm e}$ due to the non detection of emission-line ratios sensitive to these nebular parameters.
Thus, these authors estimated $T_{\rm e}$ and $N_{\rm e}$
indirectly adopting the modified version of the \textsc{ymcmc} code
\citep{2020ApJ...896...77H}, in which the Markov Chain
Monte Carlo (MCMC) algorithm is used to search for the best-fit parameters that reproduce the observed hydrogen and helium emission lines.  Using this methodology,  \citet{2024ApJ...974..266Y}
derived He/H with an
error  of $\sim 15$ per cent\footnote{For only one object (GLASS150008, $z=7.65$), \citet{2024ApJ...974..266Y} derived a He/H error of $\sim 70$ per cent.}, i.e. 
an uncertainty somewhat (a factor of $\sim 2$) higher than the one derived in precise abundance estimates of nearby galaxies (e.g. \citealt{2008MNRAS.383..209H}).

 Over the past decades, estimates of He/H as well as O/H have been obtained for large samples of local ($z \: < \: 0.4$)  star-forming regions (SFs, i.e. \ion{H}{ii} regions and star-forming galaxies; e.g. \citealt{2000MNRAS.311..329D, 2020ApJ...896...77H, 2021MNRAS.502.3045K}).  From these abundance estimates, a clear direct relation between He/H and the gas-phase metallicity (or O/H) has been established (e.g. \citealt{2006ApJS..167..177D}), with He/H ranging from $\sim0.5$ to $\sim1.0$ times the solar value 
for the metallicity range $0.03 \: \lesssim \: (Z/\rm Z_{\odot}) \: \lesssim \: 1.3$ (e.g. \citealt{2003ApJ...591..801K, 2022MNRAS.514.5506D}).

An opposite situation is found for the narrow-line regions (NLRs) of local Active Galactic Nuclei (AGNs), for which He/H has only recently been derived by \citet{2022MNRAS.514.5506D} for a sample of 65 Seyfert~2 galaxies with redshift $z \: < \: 0.2$\footnote{It appears that the first He/H estimates in AGNs [in the case of the broad-line region (BLR)] were carried out by \citet{1974ApJ...191..309S}, who derived (He/He$_{\odot}) \sim 1.0$.}.
These authors, using direct abundance estimates (via the $T_{\rm e}$-method), found He/H values ranging from $\sim0.6$ to $\sim2.5$ times the solar value for $0.3 \: \lesssim \: (Z/\rm Z_{\odot}) \: \lesssim \: 2.0$. 
 Despite the fact that \citet{2022MNRAS.514.5506D} showed that AGNs follow the  (He/H)–(O/H) relation in a similar manner to star-forming regions but for the high metallicity regime, a wider range of helium abundances was derived in comparison  with SFs, probably caused by different star formation rates and/or IMFs in these object classes (e.g. \citealt{2020ApJ...893...96B, 2022MNRAS.512.2867H}).

Thanks to the exceptional performance of the JWST, determinations of He/H abundances have become possible for high-redshift galaxies.
Recently, \citet{2024ApJ...974..266Y} presented the first high-$z$ He/H estimates for three  ($z \sim 6$) galaxies, likely SFs. These authors found helium abundances higher than the solar value [up to $\sim 1.7 \times \rm (He/He_{\odot}$)], significantly exceeding those derived for local galaxies. Subsequently, \citet{2025arXiv250717057B}, also using JWST observations, derived He/H for 20 SFs at redshifts between 1.6 and 3.3, identifying a subpopulation of four objects exhibiting high helium mass fractions inconsistent with enrichment from AGB stars, but instead  favoring early He enrichment from very massive stars ($M_{\star} \: \gtrsim \: 100 \: \rm M_{\odot}$). 

In contrast to SFs, He/H estimates are practically unknown for the NLRs of 
high-$z$ AGNs. These estimates, combined with those for local AGNs, are essential for understanding:
\begin{itemize}
    \item ISM helium enrichment in the high-metallicity regime, tracing the cosmic He/H in the central regions of galaxies,  
    \item the (He/H)–(O/H) relation across the Hubble time and
    \item the nucleosynthesis in stars of distinct masses.
\end{itemize}

In particular, AGNs serve as excellent tracers of the helium abundance owing to their high degree of ionization. This allows both the \ion{He}{i} and \ion{He}{ii} emission lines 
(see \citealt{1978ApJ...223...56K, 2015ApJS..217...12D}) to be measured more easily than in SFs, thereby yielding more accurate He/H estimates over a wide redshift range.

This study is motivated by three factors: ($i$) the availability of high-quality AGN spectroscopy from the JWST Advanced Deep Extragalactic Survey \cite[JADES;][]{2023arXiv230602465E, 2023arXiv231012340E, 2023ApJS..269...16R, 2024ApJ...964...71H, 2024A&A...690A.288B, 2025ApJS..277....4D}; ($ii$) recent advances in abundance diagnostics, including an updated implementation of the $T_{\rm e}$-method \citep{2020MNRAS.496.3209D}  and strong-line methods \citep{2021MNRAS.507..466D, 2024MNRAS.533L...1D}, developed for AGN narrow-line regions (NLRs); and ($iii$) the determination of direct helium and oxygen abundances for a relatively (65 objects) large sample of local ($z < 0.4$) AGNs \citep{2022MNRAS.514.5506D}, providing a baseline for comparison with high-$z$ AGNs.
Our aim is to estimate He/H and O/H abundances in type~2 AGNs over a wide redshift range and to assess their potential cosmic evolution.

The paper is organized as follows. In Section~\ref{meth} the methodology employed (observational data and abundance estimations) to derive  He/H and O/H abundances is presented. The results and the discussion are given in Sects.~\ref{res} and \ref{secdisc}, respectively.  
Our conclusions are summarised in Sect.~\ref{conc}.
Throughout this paper, we
adopt the cosmological parameters by \citet{2021A&A...652C...4P}:
$\rm H_{0}$ = 67.4 km s$^{-1}$ Mpc$^{-1}$ and $\Omega_{\rm m}= 0.315$. We
assume the solar values log(He/H)$_{\odot}=-1.0$ and log(O/H)$_{\odot}=-3.31$ \citep{2001ApJ...556L..63A},  as well as refer to metallicity ($Z$) and oxygen abundance (O/H) interchangeably \citep{1979MNRAS.189...95P}.

\section{Methodology}
\label{meth}

\begin{figure}
\includegraphics[angle=-90, width=1\columnwidth]{diag1.eps}
 \caption{Bottom panel: logarithm of [\ion{O}{iii}]$\lambda5007$/H$\beta$ versus [\ion{N}{ii}]$\lambda6584$/H$\alpha$.
 Red points represent 19 type~2 AGNs from the JADES survey (see Sect.~\ref{jadesec}) located in the
 redshift range $2.8 \: \leq \: z \: \leq \: 6.8$.
 Black points represent 65 local type~2  AGNs ($z \: < \: 0.2$) taken from \citet{2022MNRAS.514.5506D}.   Solid 
 and dashed curves represent the criteria
  proposed by \citet{2001ApJ...556..121K} and \citet{2003MNRAS.346.1055K}, respectively,  to separate AGNs from SFs (Eqs.~\ref{kewcr} and \ref{kaufc}). Solid line represents the criterion proposed by \citet{2010MNRAS.403.1036C} to separate AGNs from LINERs (Eq.~\ref{cidcr}).
 Dashed vertical line represents the BPT-valley defined by the Eq.~\ref{valleycr} and proposed 
 by \citet{2017ApJ...842...44K}. Upper panel:  logarithm of \ion{He}{ii}$\lambda4686$/H$\beta$ versus [\ion{N}{ii}]$\lambda6584$/H$\alpha$. Points are as in the bottom panel for the 7 objects for which the \ion{He}{ii}$\lambda4686$/H$\beta$ is available. The solid line indicates
 the criterion (Eq.~\ref{eqsep}) proposed by \citet{2012MNRAS.421.1043S}  to separate AGNs from SFs. Error bars indicate the uncertainty in the line-ratio intensities listed in Table~\ref{tabap1}.}
\label{fig1}
\end{figure}

The principal aim of this study is to estimate the helium and oxygen abundances in NLRs of AGNs over a wide range of redshifts.
To this end, we consider optical emission-line data from the literature of type~2 AGNs and apply the $T_{\rm e}$-method and strong-line methods to estimate the He/H and O/H abundances. In what follows, we present a description of the observational data and the methodology adopted for the abundance estimates. 

\subsection{Observational data}
\label{obsec} 
We consider spectroscopic observational data in the optical  [$3000 \: < \: \lambda(\angstrom) \: < \: 7000$] for type~2 AGNs  spanning two distinct redshift regimes:
$z \: < \: 0.2$
and $ 2.8 \: \leq \: z \: \leq \: 6.8$. Based on these intervals, the main sample is divided into two sub-samples, as described below.

\subsubsection{Local objects}
\label{locals}

These objects are located in the local universe ($z \: < \: 0.2$) and their emission-line intensities are the ones compiled from the literature by \citet{2022MNRAS.514.5506D}. 

 This sample consists of 65 type~2 AGNs, with data drawn from the Sloan Digital Sky Survey Data Release 15 \citep[SDSS DR15;][]{2000AJ....120.1579Y}  database\footnote{\url{https://dr15.sdss.org/optical/spectrum/search}}  and from the literature compilation by \citet{2022MNRAS.514.5506D}. To be included in this sample, the objects were required to meet the following selection criteria:
 \begin{itemize}
 \item Narrow emission lines within their spectra characterized by a Full Width at Half Maximum (FWHM) of less than $1000 \: \rm km \: s^{-1}$.
    \item Permitted and forbidden lines detected: [\ion{O}{ii}]$\lambda3727$, [\ion{O}{iii}]$\lambda4363$, H$\beta$, \ion{He}{ii}$\lambda4686$, [\ion{O}{iii}]$\lambda5007$, \ion{He}{i}$\lambda5876$, H$\alpha$, [\ion{N}{ii}]$\lambda6584$, and [\ion{S}{ii}]$\lambda\lambda6716, 6731$ doublet.
 \item  Detection threshold: All aforementioned lines were required to be measured with a $\rm (S/N)  >  2$.
  \end{itemize}
  
 For the SDSS data\footnote{An example SDSS spectrum is shown in Fig.~1 of \citet{2022MNRAS.514.5506D}.}, the line measurement, reddening correction, and stellar continuum subtraction were performed following the methodology described by \citet{2020MNRAS.492..468D}. The data compiled from the literature consist of narrow emission-line fluxes that were corrected for reddening either in the original studies or subsequently by \citet{2022MNRAS.514.5506D}.

The classification of the above objects  as AGNs  was performed by
\citet{2022MNRAS.514.5506D} through the classical    
[\ion{O}{iii}]$\lambda5007$/H$\beta$ versus [\ion{N}{ii}]$\lambda6584$/H$\alpha$ diagnostic diagram
\citep{1981PASP...93....5B} and adopting the criteria proposed by \citet{2001ApJ...556..121K} and \citet{2010MNRAS.403.1036C}.

\subsubsection{High redshift objects} 
\label{jadesec}

 This sample consists of  objects with  redshifts in the range of 
$ 2.8  \leq  z  \leq  6.8$ belonging to the JADES survey Data Release 4 (DR4)\footnote{\url{https://jades.herts.ac.uk/DR4/}} \citep{2025arXiv251001033C, 2025arXiv251001034S}.
The JADES DR4 consists of spectroscopic data from medium- and deep-depth NIRSpec/microshutter assembly observations of 5,190 targets, co\-ve\-ring the spectral range $0.6-5.5$ $\mu$m. These observations were obtained with low-dispersion ($R \approx 30-300$) CLEAR/PRISM and medium-resolution ($R = 500-1500$) gratings. Since our analysis isolates AGNs, which are inherently compact  at high redshifts (e.g. \citealt{2018PASJ...70S..15S}), we utilize the default 1D spectra from the JADES pipeline. Specifically, to maximize the S/N of these compact sources, we preferentially adopt the 3-pixel spectral extractions, which are recommended by the JADES collaboration as optimal for unresolved and compact targets.


To assess the reliability of our emission-line detections, we adopt the S/N ratios provided by the JADES DR4 catalog. These values are derived from a Bayesian Markov chain Monte Carlo (MCMC) spectral fitting framework, where the line significance is defined as the median of the posterior flux distribution divided by its $16^{\rm th}$ percentile lower bound.  We acknowledge that in the low-S/N limit, the posterior distributions of faint emission lines naturally exhibit asymmetric tails. Consequently, computing the ratio of two such low-S/N features can skew the resulting distribution to artificially high values, potentially scattering star-forming galaxies into the AGN locus. Because optical diagnostic diagrams are susceptible to this photometric scatter, we do not rely solely on these initial optical classifications to definitively confirm our sample. Instead, as detailed in Sect.~\ref{anadiag}, we require secondary confirmation through intrinsic emission-line kinematics and rest-UV diagnostics to robustly secure their AGN nature.

Targets from the initial parent sample were retained for our final analysis only if they met the following selection criteria:
\begin{itemize}
    \item A Bayesian $(\text{S/N}) \ge 2$, coupled with a strict classical detection baseline (i.e. a positive flux strictly greater than the nominal $1\sigma$ error), for all four primary diagnostic lines: H$\alpha$, H$\beta$, [\ion{O}{iii}]$\lambda$5007, and [\ion{N}{ii}]$\lambda$6584. 
 \item A Balmer decrement of (H$\alpha$/H$\beta) > 2.5$, ensuring reliable reddening corrections.  The emission lines were subsequently corrected for dust extinction, assuming a canonical \citet{1989ApJ...345..245C} Galactic extinction curve, adopting a total-to-selective extinction value of $R_V = A_V/E(B-V) = 3.1$ along with  Balmer decrement of $(\mathrm{H}\alpha/\mathrm{H}\beta) = 2.86$. Such an assumption is predicated on standard Case B recombination in nebular environments ($T_{\rm e} \approx 10^4$ K, $N_{\rm e} \sim 10^2$ cm$^{-3}$; \citealt{2006agna.book.....O}).

 \item Relative uncertainties of less than 50 per cent on the reddening-corrected intensities (normalized to H$\beta$) for all emission lines utilized in the analysis of a given object. The primary strong lines (H$\alpha$, H$\beta$, [\ion{O}{iii}] $\lambda5007$, [\ion{N}{ii}] $\lambda6584$) must meet this threshold to classify the sample. Furthermore, for an object to be included in specific diagnostic diagrams or abundance derivations requiring inherently fainter lines (e.g., [\ion{O}{ii}] $\lambda3727$, [\ion{O}{iii}] $\lambda4363$, \ion{He}{ii} $\lambda4686$, and \ion{He}{i} $\lambda5876$), those specific lines (normalized to H$\beta$) must independently satisfy this same $<50\%$ relative uncertainty constraint.
 
 \item To confirm that the position of each target on the diagnostic diagrams (Fig.~\ref{fig1}) is driven by genuine AGN activity rather than star formation, we compared our sample with comprehensive photoionization models with the ultraviolet emission-line ratios of confirmed AGNs (see below).
 \item Finally, measurements of the FWHM of H$\alpha$ indicate that our final sample consists predominantly of narrow-line type~2 AGNs.
 \end{itemize}

\subsection{Object classification}
\label{object_class}

 Before estimating the chemical abundances, it is necessary to  classify the targets as AGNs. To obtain abundance estimates through the $T_{\rm e}$-method, the object class must be known in order to correctly apply the relation between the electron temperatures of the low- ($T_{\rm low}$) and high-ionization ($T_{\rm high}$) regions. As showed by \citet{2020MNRAS.496.3209D}, this temperature relation differs significantly between AGNs and SFs. Furthermore, applying strong-line calibrations to derive He/H and O/H also  requires prior knowledge of the dominant ionizing source \citep{2024ApJ...977..187Z}.

\subsubsection{$N2$ diagram}

We classify our sample initially using the [\ion{O}{iii}]$\lambda5007$/H$\beta$ versus [\ion{N}{ii}]$\lambda6584$/H$\alpha$ diagnostic diagram \citep{1981PASP...93....5B}, hereafter the $N2$-diagram, shown in the bottom panel of Figure~\ref{fig1}. This panel presents the theoretical criterion (solid curve) proposed by \citet{2001ApJ...556..121K} to separate SFs from AGNs. These authors built photoionization model grids using spectral energy distributions (SEDs) from both the \textsc{pegase} v2.0 \citep{1997A&A...326..950F} and \textsc{starburst99} \citep{1999ApJS..123....3L} codes as ionizing sources. Based on these models, they established that objects satisfying
\begin{equation}
\label{kewcr}
\rm log([O\:III]\lambda5007/H\beta) \: > \: \frac{0.61}{log([N\:II]\lambda6584/H\alpha)-0.47}+1.19
\end{equation}
are classified as AGNs, and otherwise as SFs.

The dashed curve in Fig.~\ref{fig1} represents the empirical criterion proposed by \citet{2003MNRAS.346.1055K}. Using emission-line fluxes from the SDSS \citep{2000AJ....120.1579Y}, they concluded that emission-line galaxies form two well-separated sequences in the $N2$ diagram, effectively distinguishing the AGN sequence from that of SFs. Consequently, they empirically classified objects as AGNs if
\begin{equation}
\label{kaufc}
\rm log([O\:III]\lambda5007/H\beta) \: > \: \frac{0.61}{log([N\:II]\lambda6584/H\alpha)-0.05}+1.3
\end{equation}
and otherwise as SFs.

Furthermore, utilising SDSS observations, \citet{2010MNRAS.403.1036C} proposed the following criterion to separate AGNs from Low-Ionization Nuclear Emission-line Regions (LINERs), which is indicated by the solid straight line in Fig.~\ref{fig1}:
\begin{equation}
\label{cidcr}
\rm log([O\:III]\lambda5007/H\beta) \: > 1.10 \times log([N\:II]\lambda6584/H\alpha) + 0.46.
\end{equation}

Finally, the dashed vertical line in Fig.~\ref{fig1} marks the `BPT valley', a region defined by \citet{2017ApJ...842...44K} that lies above the maximum-starburst line and satisfies
\begin{equation}
\label{valleycr}
\rm log([\ion{N}{ii}]\lambda6584/H\alpha) \: < \: -0.5.
\end{equation}
By comparing photoionization models with SDSS data, these authors found that the majority ($\sim 60$ per cent) of the objects in the BPT valley are AGNs with lower metallicities than typical Seyfert nuclei.

\subsubsection{\ion{He}{ii} diagram}
As an additional selection criterion, in the upper panel of  Fig.~\ref{fig1},  we present a diagram of $\log$(\ion{He}{ii}$\lambda4686$/H$\beta$) versus $\log$([\ion{N}{ii}]$\lambda6584$/H$\alpha$) proposed by \citet{2012MNRAS.421.1043S}. 
These authors, by using spectroscopic data of galaxies taken from the SDSS, 
established that objects with 
\begin{equation}
\label{eqsep}
 \rm log(He\:II\lambda4686/H\beta) \: > \: \frac{1.0}{8.92\times \log([N\:II]\lambda6584/H\alpha)+1.32}-1.22
\end{equation}
are classified as AGNs and otherwise as SFs (see also \citealt{2022MNRAS.513.5134N}). 
The \ion{He}{ii}/H$\beta$ line ratio is strongly dependent on the ionization state of the gas. In particular, the presence of the 
\ion{He}{ii}$\lambda4686$ emission line indicates the existence of hard ionizing radiation, since the ionization potential of $\rm He^{+}$ is 54.4 eV, which is mainly produced by AGNs (e.g. \citealt{1994ApJ...435..171K, 2017ApJ...842...44K, 2024MNRAS.527.7217V}).

 Based on the selection criteria described in Sect.~\ref{jadesec} and the constraints imposed by equations~(\ref{kewcr})--(\ref{eqsep}), we selected 19 JADES AGNs spanning the redshift range $2.8 \leq z \leq 6.8$.  Within this JADES sample, we were able to measure the [\ion{O}{iii}]$\lambda4363$/H$\beta$ ratio for 11 objects and the \ion{He}{ii}$\lambda4686$/H$\beta$ ratio for 7 objects.  In Fig.~\ref{fig1}, these targets  are represented by red points, while the local AGNs are shown as black points. Table~\ref{tabap1} lists the observed H$\beta$ fluxes, the reddening-corrected narrow emission-line intensities (normalized to H$\beta = 1$), and the logarithm of the H$\beta$ luminosity for the 19 objects of the JADES AGN sample.

\subsubsection{Classification uncertainties}
\label{anadiag}

\begin{figure*}
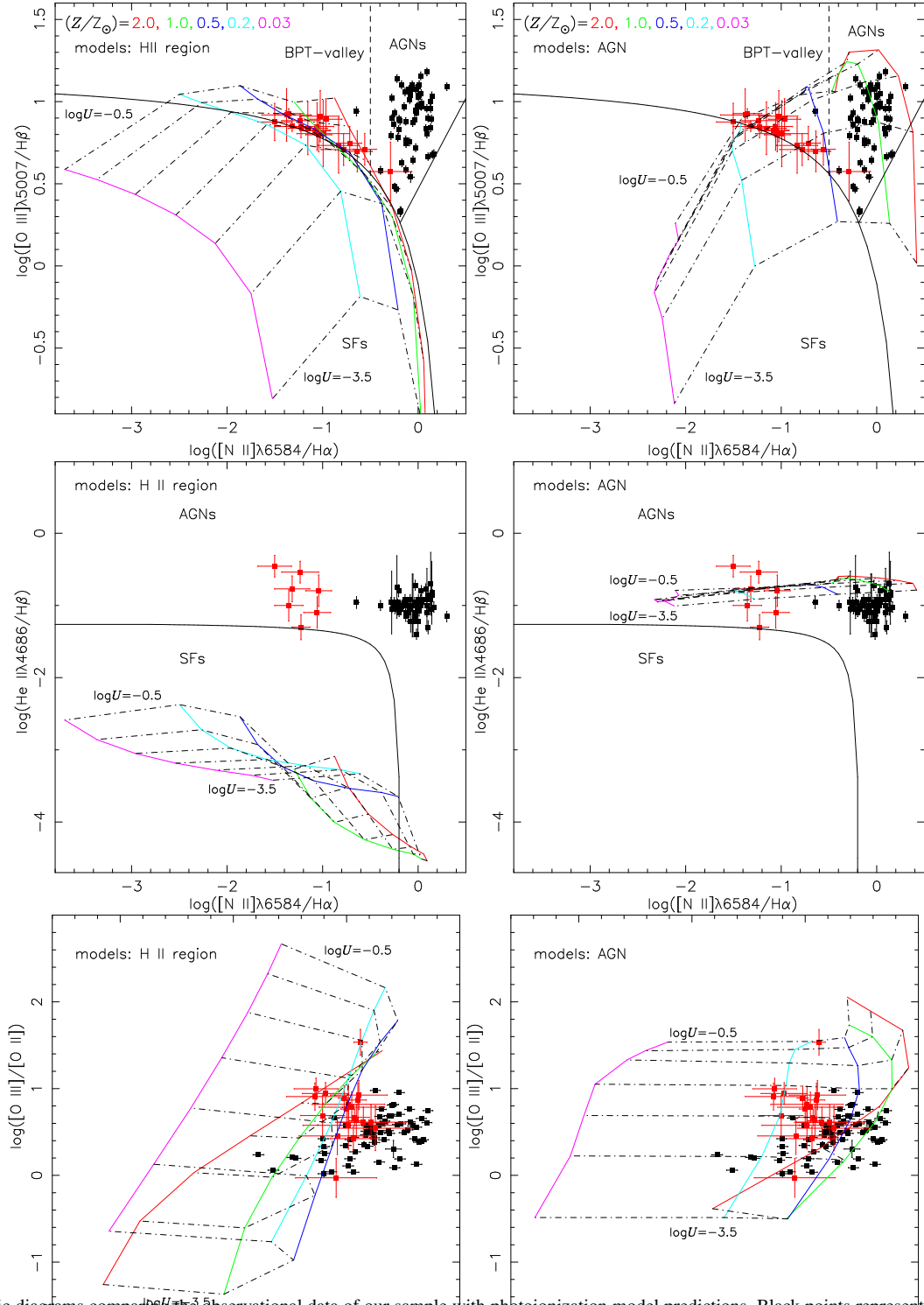

\includegraphics[angle=-90, width=0.8\columnwidth]{diag1_bp.eps}
\includegraphics[angle=-90, width=0.8\columnwidth]{diag1c.eps}
\includegraphics[angle=-90, width=0.8\columnwidth]{diag1d_bpass.eps}
\includegraphics[angle=-90, width=0.8\columnwidth]{diag1e.eps}
\includegraphics[angle=-90, width=0.8\columnwidth]{fig4_new2.eps}
\includegraphics[angle=-90, width=0.8\columnwidth]{fig4_new3.eps}
 \caption{ Diagnostic diagrams comparing the observational data of our sample with photoionization model predictions.
 Black points represent observational data of local AGNs and red points the JADES sample (see Sect.~\ref{obsec}).
 In the left and right panels results of photoionization models (see Sect.~\ref{anadiag}) simulating high-$z$ star-forming regions
 and AGNs, respectively, are shown. Solid colored lines connect models
 with same metallicity ($Z/\rm Z_{\odot}$), while point-dashed lines connect models with same logarithm of the ionization parameter ($\log U$), as indicated. In the top and middle panels the solid black lines  are as in Fig.~\ref{fig1}. In the bottom panels, $R_{23}$ = ([\ion{O}{ii}]$\lambda3727$ + [\ion{O}{iii}]$\lambda4959 + \lambda5007$)/H$\beta$. }
\label{fig0d}
\end{figure*}

\begin{figure*}
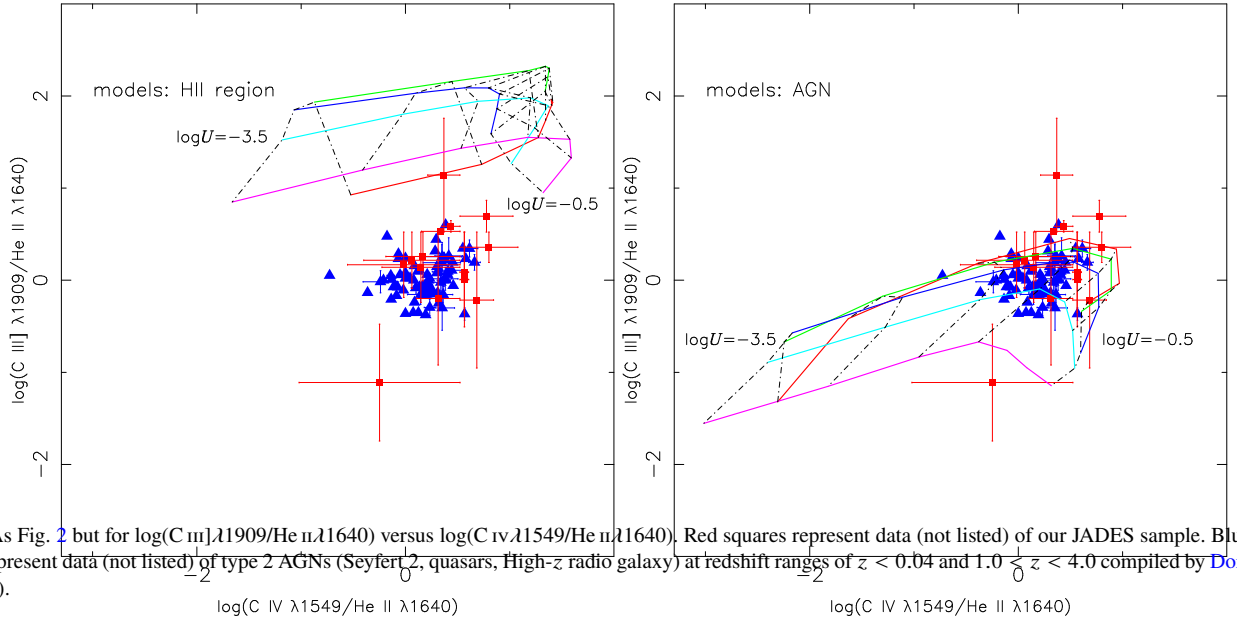

\includegraphics[angle=-90, width=0.45\textwidth]{fig2sfv.eps}
\includegraphics[angle=-90, width=0.45\textwidth]{fig2agnv.eps}
 \caption{As Fig.~\ref{fig0d} but for log(\ion{C}{iii}]$\lambda1909$/\ion{He}{ii}$\lambda1640$)
 versus log(\ion{C}{iv}$\lambda1549$/\ion{He}{ii}$\lambda1640$). Red squares represent data (not listed) of
 our JADES sample. Blue triangles represent data (not listed) of type~2 AGNs (Seyfert~2, quasars, High-$z$ radio galaxy)
 at redshift ranges of $z < 0.04$ and $1.0 < z < 4.0$ compiled by \citet{2014MNRAS.443.1291D}.}
\label{figuv}
\end{figure*}

\begin{figure}
\includegraphics[angle=-90, width=0.45\textwidth]{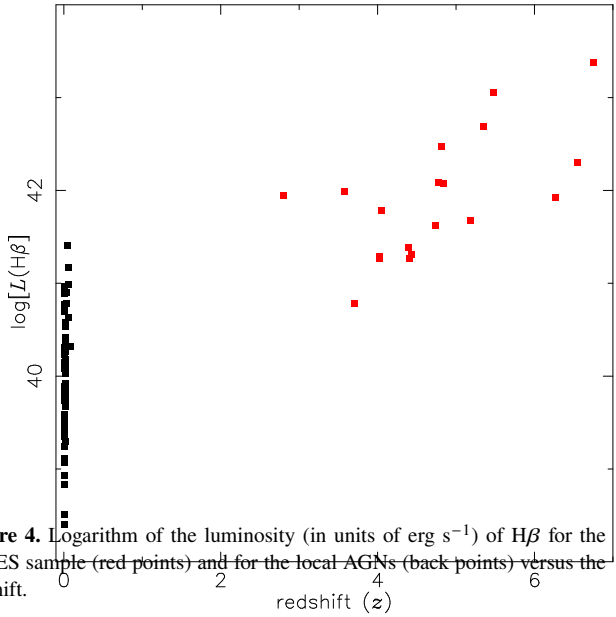}
 \caption{Logarithm of the luminosity (in units of erg s$^{-1}$) of H$\beta$ for the JADES sample (red points) and
 for the local AGNs (back points) versus the redshift.}
\label{figlum}
\end{figure}

\begin{table*}
\scriptsize
\addtolength{\tabcolsep}{-3pt}
\caption{ Reddening corrected emission line intensities (in relation to H$\beta$ = 1.0) of type~2 AGNs 
selected from the JADES sample described in Sect.~\ref{jadesec}. Flux of H$\beta$ [$F(\rm H\beta)$] is in units of
$\rm 10^{-20} \: erg\: cm^{2} \: s^{-1}$. Luminosity of H$\beta$ [$L(\rm H\beta)$] is in units of
$\rm erg \: s^{-1}$.  Full Width at Half Maximum (FWHM) of H$\alpha$ is in units of $\rm km \: s^{-1}$.}
\label{tabap1}
\begin{tabular}{@{}lcccccccccccc@{}}
\hline
 
          ID           &  Redshift & $F(\rm H\beta)$ &     FWHM              & [\ion{O}{ii}]       & [\ion{O}{iii}] & \ion{He}{ii}	& [\ion{O}{iii}] & \ion{He}{i}   & H$\alpha$	 & [\ion{N}{ii}]   & log[$L$(H$\beta$)] & E(B-V)   \\		      
	               &           &                 & $\lambda6563$         &  $\lambda3727$ & $\lambda4363$  & $\lambda4686$  &  $\lambda5007$ & $\lambda5876$  & $\lambda6563$ & $\lambda6584$	  &		       &	 \\
                         \hline
 n-mediumhst-607       &   5.1810  & $ 165 \pm 44 $  &   276                    &  $1.04\pm0.45$ &  $0.25\pm0.12$ &   ---	   &  $6.87\pm2.53$ &  $0.05\pm0.02$ & $ 2.86\pm0.93$ &   $0.20\pm0.07$ &	  41.68 	&  0.03   \\
 n-mediumhst-643       &   5.3472  & $1572 \pm301 $  &   272                    &  $1.84\pm0.56$ &  $0.26\pm0.10$ & $ 0.35\pm0.12$ &  $7.52\pm1.96$ &  $0.12\pm0.04$ & $ 2.86\pm0.66$ &   $0.09\pm0.03$ &	  42.69 	&  0.48   \\
 n-mediumhst-917       &   4.3986  & $ 124 \pm 35 $  &   367                    &  $1.89\pm0.86$ &	 ---	  &	---	   &  $5.41\pm2.10$ &  $0.10\pm0.05$ & $ 2.86\pm0.98$ &   $0.42\pm0.15$ &	  41.39 	&  0.02   \\
 n-mediumhst-954       &   6.7589  & $4435 \pm722 $  &   700                    &  $0.51\pm0.19$ &  $0.55\pm0.16$ &	---	   &  $5.12\pm1.14$ &  $0.05\pm0.02$ & $ 2.86\pm0.56$ &   $0.79\pm0.16$ &	  43.38 	&  0.99   \\
 s-mediumhst-9669      &   4.0239  & $ 121 \pm  8 $  &   288                    &  $1.65\pm0.19$ &  $0.24\pm0.07$ & $ 0.17\pm0.07$ &  $7.06\pm0.48$ &  $0.19\pm0.09$ & $ 2.50\pm0.19$ &   $0.12\pm0.05$ &	  41.29 	&  0.00   \\
 s-mediumhst-15357     &   4.0529  & $ 369 \pm 77 $  &   638                    &  $0.92\pm0.31$ &	 ---	  &	---	   &  $7.80\pm2.24$ &  $0.23\pm0.08$ & $ 2.86\pm0.73$ &   $0.31\pm0.09$ &	  41.78 	&  0.04   \\
 n-mediumjwst-29164    &   4.4059  & $  93 \pm 25 $  &   313                    &  $2.36\pm1.01$ &	 ---	  &	---	   &  $8.28\pm3.05$ &  $0.35\pm0.16$ & $ 2.86\pm0.93$ &   $0.12\pm0.06$ &	  41.27 	&  0.09   \\
 s-mediumjwst-38562    &   4.8203  & $1206 \pm207 $  &   583                    &  $0.63\pm0.19$ &  $0.18\pm0.06$ &	---	   &  $5.59\pm1.31$ &  $0.06\pm0.03$ & $ 2.86\pm0.59$ &   $0.55\pm0.11$ &	  42.47 	&  0.57   \\
 s-mediumjwst-41503    &   2.8049  & $1318 \pm230 $  &   324                    &  $2.37\pm0.66$ &  $0.16\pm0.07$ & $ 0.16\pm0.08$ &  $6.28\pm1.50$ &  $0.12\pm0.04$ & $ 2.86\pm0.60$ &   $0.26\pm0.06$ &	  41.95 	&  0.29   \\
 s-mediumjwst-53979    &   4.0290  & $ 114 \pm  9 $  &   310                    &  $1.03\pm0.16$ &  $0.40\pm0.12$ & $ 0.29\pm0.10$ &  $7.66\pm0.65$ &  $0.08\pm0.04$ & $ 2.75\pm0.25$ &   $0.16\pm0.07$ &	  41.27 	&  0.00   \\
 s-mediumhst-58656     &   4.7780  & $ 507 \pm124 $  &   308                    &  $1.15\pm0.46$ &	 ---	  &	---	   &  $7.05\pm2.35$ &  $0.10\pm0.05$ & $ 2.86\pm0.84$ &   $0.25\pm0.07$ &	  42.09 	&  0.28   \\
 s-mediumhst-58850     &   6.2615  & $ 184 \pm 12 $  &   271                    &  $0.24\pm0.06$ &  $0.19\pm0.05$ & $ 0.10\pm0.05$ &  $8.45\pm0.56$ &  $0.21\pm0.05$ & $ 2.71\pm0.19$ &   $0.12\pm0.04$ &	  41.92 	&  0.00   \\
 s-mediumjwst-61321    &   4.8424  & $ 466 \pm122 $  &   331                    &  $1.98\pm0.83$ &	 ---	  &	---	   &  $8.15\pm2.91$ &  $0.16\pm0.06$ & $ 2.86\pm0.90$ &   $0.27\pm0.09$ &	  42.07 	&  0.41   \\
 n-mediumjwst-78891    &   6.5479  & $ 393 \pm 43 $  &   271                    &  $1.50\pm0.27$ &  $0.09\pm0.02$ & $ 0.05\pm0.02$ &  $7.00\pm1.05$ &  $0.10\pm0.03$ & $ 2.86\pm0.38$ &   $0.17\pm0.03$ &	  42.30 	&  0.18   \\
 s-mediumjwst-172074   &   4.4303  & $ 101 \pm  5 $  &   315                    &  $1.14\pm0.11$ &  $0.15\pm0.05$ &	---	   &  $6.66\pm0.39$ &  $0.13\pm0.05$ & $ 2.656\pm0.1$ &   $0.22\pm0.08$ &    41.31		&  0.00   \\
 s-mediumjwst-181864   &   3.7022  & $  45 \pm  9 $  &   326                    &  $1.04\pm0.41$ &	 ---	  &	---	   &  $5.11\pm1.13$ &  $0.79\pm0.37$ & $ 2.71\pm0.65$ &   $0.45\pm0.22$ &	  40.78 	&  0.00   \\
 s-mediumjwst-184060   &   4.7360  & $ 178 \pm 10 $  &   256                    &  $0.85\pm0.08$ &  $0.26\pm0.06$ & $ 0.08\pm0.04$ &  $6.69\pm0.38$ &  $0.17\pm0.04$ & $ 2.85\pm0.17$ &   $0.25\pm0.08$ &	  41.62 	&  0.00   \\
 s-mediumjwst-185290   &   3.5849  & $ 788 \pm238 $  &   508                    &  $4.00\pm1.91$ &	  ---	  &	---	   &  $3.75\pm1.55$ &  $0.19\pm0.09$ & $ 2.86\pm1.04$ &   $1.47\pm0.54$ &	  41.99 	&  0.45   \\
 s-mediumjwst-204851   &   5.4797  & $3338 \pm704 $  &   713                    &  $0.61\pm0.25$ &	  ---	  &	---	   &  $4.98\pm1.43$ &  $0.08\pm0.04$ & $ 2.86\pm0.72$ &   $0.66\pm0.17$ &    43.05		&  0.87   \\
\hline
\end{tabular}	   								
\end{table*}

 \citet{2013ApJ...774L..10K} compared observational data of confirmed high-$z$ ($0.5 < z < 2.6$) star-forming galaxies with theoretical predictions to elucidate how optical diagnostic line ratios in galaxy ensembles evolve as a function of cosmic time in a $N2$-diagram. These authors found that high-$z$ SFs can be located slightly ($\sim 0.1$~dex) above or on the maximum starburst line. This behaviour may occur because high-$z$ SFs exhibit a larger ionization parameter, a higher electron density, and/or an ionising radiation field with a larger fraction of O$^{+}$-ionising to H-ionising photons than local SFs \citep{2013ApJ...774L..10K}. This result implies that a fraction of our selected objects in the $N2$-diagram (bottom panel of Fig.~\ref{fig1}) may be SFs. Regarding  AGNs, as pointed out by \citet{2024MNRAS.527.8193D}, a further limitation to classifications based on the $N2$-diagram is that low-metallicity ($Z \lesssim 0.2 \, {\rm Z}_{\odot}$) active nuclei may be misidentified as SFs since they occupy the SF-like region (see also \citealt{2006MNRAS.371.1559G, 2016MNRAS.456.3354F, 2023A&A...675A..74F, 2017ApJ...842...44K, 2022MNRAS.513.5134N, 2023MNRAS.522.5788O, 2023MNRAS.526.3610H, 2024MNRAS.527.1962B}). Therefore, although the $N2$-diagram provides useful constraints, its interpretation can be ambiguous, as SFs may be misclassified as AGNs and genuine low-metallicity active nuclei may be excluded.

To provide additional support for our classifications based on the $N2$ and \ion{He}{ii} diagrams (see Fig.~\ref{fig1}), we compare the observed line ratios of our samples (JADES objects and local AGNs) with predictions from photoionization models built using the \textsc{cloudy} code \citep{2017RMxAA..53..385F} to simulate both AGNs and SFs. For the AGN models, we adopt nebular parameters similar to those used by \citet{2025MNRAS.542.3181D}, summarised as follows: metallicities of $(Z/{\rm Z}_{\odot}) = 2.0$, 1.0, 0.5, 0.2, and 0.03; the logarithm of the ionization parameter ($U$) ranging from $-3.5$ to $-0.5$ (in steps of 0.5~dex); the SED optical to X-rays spectral index $\alpha_{ox} = -1.1$; an electron density $N_{\rm e} = 500 \, {\rm cm}^{-3}$; and abundance relations between oxygen and other elements (i.e. He, N, C) derived via the $T_{\rm e}$-method.

For the SF models, we employ parameters similar to those in \citet{2018MNRAS.479.2294D}, but the ionising SED is generated using \textsc{Bpass} code\footnote{\url{https://www.bpass.auckland.ac.nz/}} \citep{2017PASA...34...58E}, in which  incorporates binary mass transfer and its effect on stellar evolution pathways. We assumed a stellar cluster with an age of 1~Myr formed in an instantaneous burst, an Initial Mass Function  (IMF) slope of $-1.30$ for stars with mass from 0.1 to 0.5 $\rm M_{\odot}$, and a slope of $-2.35$ from 0.5 to 300 $\rm M_{\odot}$. The metallicity of the SED of each stellar cluster was matched with the nearest  gas metallicity\footnote{For a discussion of the impact of matching nebular and stellar metallicities on photoionization model predictions, see \citet{2011MNRAS.415.3616D} and references therein.}. 
\citet{2019ApJ...878....2D} compared different examples of stellar evolutionary synthesis as input parameters photoionization model grids. These authors found that
photoionization models assuming distinct SEDs predict  strong emission-line ratios (as those used in diagnostic diagrams) differing on average of  $\sim0.1$ dex, a value in 
order of the observational uncertainties of emission line ratios (e.g. \citealt{2008MNRAS.383..209H}).
Oxygen abundance scales and metallicity range  are identical to those in the AGN models. However, to reproduce SFs lying above the maximum starburst line, we follow the methodology of \citet{2022ApJ...926...80G}; i.e., we adopt an elevated\footnote{Typical star-forming galaxies exhibit much lower densities of $N_{\rm e} \sim 30 \, {\rm cm}^{-3}$ (e.g. \citealt{2017MNRAS.465.3220K}).} electron density of $1000 \, {\rm cm}^{-3}$ and a nitrogen abundance enhanced by $0.4$~dex relative to the (N/O)--(O/H) relation assumed in the AGN models.

In Fig.~\ref{fig0d}, the results of our simulations are compared with the observational data of our samples in the $N2$ and \ion{He}{ii} diagnostic diagrams. Furthermore, we consider the $O_{32}$ = [\ion{O}{iii}]$\lambda5007$/[\ion{O}{ii}]$\lambda3727$ versus $R_{23}$ = ([\ion{O}{ii}]$\lambda3727$ + [\ion{O}{iii}]$\lambda4959 + \lambda5007$)/H$\beta$ diagram, as proposed by \citet{1994ApJ...426..135M}. The $O_{32}$ ratio serves as a diagnostic of the ionization state of the gas \citep[e.g.][]{1991ApJ...380..140M}, whereas $R_{23}$ depends primarily on metallicity (O/H; \citealt{1979MNRAS.189...95P}) and, to a lesser extent, on the ionization state of the gas \citep{2001A&A...369..594P}. This diagnostic is particularly useful for galaxy classification because SFs and AGNs tend to exhibit distinct behaviors in the relation between oxygen abundance and $R_{23}$. Indeed, it is well established that in SFs, the $R_{23}$ intensity increases with O/H up to approximately 30 per cent of the solar metallicity, corresponding to $12+\log({\rm O/H}) \sim 8.0$ (e.g. \citealt{1984MNRAS.211..507E,1991ApJ...380..140M}). At higher metallicities, metal-line cooling becomes increasingly efficient, lowering the electron temperature and reducing the strength of the oxygen emission lines. As a result, $R_{23}$ decreases with increasing O/H, producing the well-known turnover point in the (O/H)--$R_{23}$ relation. Conversely, \citet{2015MNRAS.453.4102D} found that for AGNs, the turnover point in the (O/H)--$R_{23}$ relation occurs at a metallicity approximately 30 per cent above the solar, i.e. at $12+\log({\rm O/H}) \sim 8.8$. For lower metallicities, the relation between O/H and $R_{23}$ remains  monotonic. Because of this distinct (O/H)--$R_{23}$ relation (see also Fig.~7 of \citealt{2021MNRAS.507..466D}), diagnostic diagrams incorporating $R_{23}$ can help to distinguish these classes, as SFs do not reach the high $R_{23}$ values observed in AGNs.

In Fig.~\ref{fig0d}, our observational data are compared with the SF and AGN photoionization model results across the $N2$, \ion{He}{ii}, and $O_{32}$-$R_{23}$ diagrams. From this comparison, we note the following key points:
\begin{itemize}
\item $N2$ diagram: ($i$) SF photoionization models with $(Z/{\rm Z}_{\odot}) \gtrsim 0.2$, a high $N_{\rm e}$, and enhanced nitrogen abundances successfully reproduce the JADES-selected objects, but fail to reproduce local AGNs. Notably, it is not necessary to invoke high ionization parameters ($\log U > -1.5$) in these SF models. ($ii$) AGN photoionization models with $(Z/{\rm Z}_{\odot}) \gtrsim 0.2$ assuming standard nebular parameters also reproduce almost all of the JADES objects and the local AGNs.
\item \ion{He}{ii} diagram: ($i$) SF models predict $\log(\ion{He}{ii}/{\rm H}\beta)$ values that are lower than the observational data by approximately $3.5$~dex. \citet{2003A&A...397...71S} previously highlighted the necessity of including a secondary ionization source (e.g. X-ray radiation) in SF models to reproduce the \ion{He}{ii}/H$\beta$ ratios of local star-forming galaxies. However, even when incorporating extra heating and ionization sources, SF models can not predict $\log(\ion{He}{ii}/{\rm H}\beta) \gtrsim -1.2$ (e.g. \citealt{2003A&A...397...71S}), which are the values derived for our JADES and local AGN samples. ($ii$) AGN models reproduce both the JADES objects and local AGNs. A further refined match between the models and observational data can be achieved by varying the He/H abundance relative to the fixed helium–oxygen scaling assumed in the baseline photoionization models (not shown).
\item $O_{32}$-$R_{23}$ diagram: Approximately half of the JADES data can be reproduced by the SF models, while the AGN models span a wide range of parameter space consistent with the observed line ratios of both samples.
\end{itemize}

 In addition to the optical diagnostic diagrams, we analysed diagnostic diagrams involving high-ionization rest-frame UV emission lines. We utilized the $\log(\text{C\,\textsc{iii}]}\lambda1908 / \text{He\,\textsc{ii}}\lambda1640) $ versus $\log(\text{C\,\textsc{iv}}\lambda1549 / \text{He\,\textsc{ii}}\lambda1640)$ diagnostic diagram, which is effective at separating AGNs from star-forming galaxies due to its extreme sensitivity to the hardness of the ionizing continuum \citep[e.g.][]{2016MNRAS.456.3354F,2022MNRAS.513.5134N}.
In Fig.~\ref{figuv}, we present this UV diagnostic diagram for the high-$z$ JADES sample. The emission-line fluxes used in this analysis were also obtained from the JADES DR4 data release; while they are not presented here, they will be discussed in detail in a future study (Dors et al., in preparation).
This figure also includes the results of our photoionization models and observational data for 77 confirmed type~2 AGNs (Seyfert~2 galaxies, type~2 quasars, and high-$z$ radio galaxies) compiled by \citet{2014MNRAS.443.1291D}. As shown in Fig.~\ref{figuv}, the JADES targets lie within the established AGN locus defined by the photoionization model predictions and the observational data for type~2 AGNs. Furthermore, they are clearly separated from the region of parameter space occupied by the SF photo\-ioni\-za\-tion models.  This boundary clearly separates AGNs from SFs. As demonstrated by the extensive CLASSY sample analysis \citep{2024ApJ...962...95M}, even the most extreme starbursts lack the high-energy radiation field required to simultaneously produce the intense \ion{He}{ii} and \ion{C}{iv} emission needed to scatter into the AGN locus. 

Full Width at Half Maximum (FWHM) of permitted lines of our targets provide an additional kinematic consistency check. Accordingly, Table~\ref{tabap1} lists the H$\alpha$ FWHM values corrected for the wavelength-dependent instrumental line spread function (LSF), where we found values ranging between $\sim 200$ and $\rm \sim 800~\rm km \: s^{-1}$.

Based on the results presented above, although SF photoionization models incorporating enhanced electron densities and N/O abundance ratios can reproduce the JADES data in the $N2$ diagram, they can not account for the full extent of the observed $R_{23}$ range and fail to reproduce the observed \ion{He}{ii}/H$\beta$, \ion{C}{iii}]/\ion{He}{ii}, and \ion{C}{iv}/\ion{He}{ii} line ratios.
In contrast, AGN models consistently reproduce the bulk of the observational data across all considered diagnostic diagrams. Thus, taking these results into account and the fact that the H$\alpha$ line of the JADES objects exhibit  FWHM values in the range of 200--800 ${\rm km \: s^{-1}}$ (see Table~\ref{tabap1}), which are significantly broader than those observed in star-forming galaxies (e.g. \citealt{2008MNRAS.383..209H, 2011MNRAS.414.3288F, 2012MNRAS.422.3475H, 2014MNRAS.442.3565C, 2019ApJ...886L..28M}) and similar to the ones of type~2 AGNs (e.g. \citealt{1990ApJS...74..731W, 2012MNRAS.427.1266V, 2020MNRAS.496.3209D}), we infer that the JADES targets are predominantly driven by AGN activity rather than star formation.

Fig.~\ref{figlum} shows the logarithm of the H$\beta$ luminosity for the JADES (red points) and local (black points) AGNs as a function of redshift, illustrating the expected Malmquist bias. It is worth emphasising that the [\ion{O}{iii}]$\lambda4363$ emission line detected in the JADES spectra is not blended with H$\gamma$ \citep{2025arXiv251001034S}. This isolated feature enables a reliable and direct determination of the electron temperature and, consequently, the accurate estimation of ionic abundances via the $T_{\rm e}$-method. The unprecedented sensitivity and spectral resolution of the JADES survey thereby provide a unique opportunity to probe the chemical properties of galaxies across a wide redshift domain. Exploiting this advantage, \citet{2026arXiv260115964C} derived nitrogen abundances for star-forming galaxies in the redshift interval $1.5 < z < 7.0$, while \citet{2026MNRAS.547ag123I} measured the abundances of C, Fe, and $\alpha$-elements (O, Ne, Si, and Ar) in star-forming systems at $z = 4$--$7$. These studies highlight the power of JADES for advancing our understanding of chemical enrichment in the early universe. In the present study, we extend the application of JADES data to determine He and O abundances within the NLRs of AGNs over a broad redshift range, providing new constraints on the chemical evolution of ionized gas in AGN environments.

\begin{table*}
\addtolength{\tabcolsep}{-3pt}
\caption{Electron temperatures (in K), ionization correction factors (ICFs), and abundances [in units of 12+log(X/H)] for our JADES samples (see Sect.~\ref{jadesec}). Estimates
were derived by using the $T_{\rm e}$-method (see Sect.~\ref{abusubs}).}
\label{tabap2}
\begin{tabular}{@{}lcccccccccc@{}}
\hline
       ID               &    $T$(high)       &   $T$(low)     & $\rm O^{+}/H^{+}$ &  $\rm O^{+2}/H^{+}$ &    ICF(O)    &    O/H       &  $\rm He^{+}/H^{+}$  &  $\rm He^{2+}/H^{+}$  &  ICF(He)     &  He/H        \\
\hline
n-mediumhst-643  & $16215 \pm 3501$ & $12402 \pm 4916$ & $8.02 \pm 0.75$ & $8.01 \pm 0.33$ & $1.36 \pm 0.03$ & $8.58 \pm 0.51$ & $10.94 \pm 0.02$ & $10.49 \pm 0.02$ & $1.65 \pm 0.03$ & $11.29 \pm 0.02$ \\
s-mediumhst-9669 & $17071 \pm 2612$ & $12680 \pm 5104$ & $7.97 \pm 0.75$ & $7.89 \pm 0.17$ & $1.12 \pm 0.01$ & $8.40 \pm 0.50$ & $11.12 \pm 0.03$ & $10.18 \pm 0.02$ & $1.71 \pm 0.02$ & $11.40 \pm 0.03$ \\
s-mediumjwst-41503 & $14291 \pm 3308$ & $11810 \pm 4785$ & $8.22 \pm 0.76$ & $8.07 \pm 0.36$ & $1.17 \pm 0.01$ & $8.65 \pm 0.56$ & $10.94 \pm 0.02$ & $10.16 \pm 0.02$ & $1.68 \pm 0.03$ & $11.23 \pm 0.02$ \\
s-mediumjwst-53979 & $19880 \pm 3270$ & $13560 \pm 5711$ & $7.70 \pm 0.79$ & $7.78 \pm 0.18$ & $1.46 \pm 0.04$ & $8.35 \pm 0.49$ & $10.77 \pm 0.03$ & $10.42 \pm 0.02$ & $1.65 \pm 0.02$ & $11.15 \pm 0.02$ \\
s-mediumhst-58850 & $14141 \pm 1887$ & $11907 \pm 4584$ & $7.22 \pm 0.74$ & $8.16 \pm 0.17$ & $1.06 \pm 0.00$ & $8.31 \pm 0.26$ & $11.17 \pm 0.02$ & $9.94 \pm 0.01$  & $2.80 \pm 0.27$ & $11.64 \pm 0.04$ \\
n-mediumjwst-78891 & $11636 \pm 1395$ & $10891 \pm 3805$ & $8.12 \pm 0.68$ & $8.31 \pm 0.20$ & $1.06 \pm 0.00$ & $8.66 \pm 0.38$ & $10.84 \pm 0.02$ & $9.62 \pm 0.01$  & $1.90 \pm 0.03$ & $11.15 \pm 0.02$ \\
s-mediumjwst-184060 & $18239 \pm 2291$ & $12969 \pm 5188$ & $7.66 \pm 0.75$ & $7.79 \pm 0.13$ & $1.06 \pm 0.00$ & $8.18 \pm 0.45$ & $11.07 \pm 0.02$ & $9.87 \pm 0.02$  & $1.97 \pm 0.04$ & $11.39 \pm 0.03$ \\
\hline
\end{tabular}
\end{table*}

\begin{table}
\addtolength{\tabcolsep}{-3pt}
\caption{Helium and oxygen abundances for AGNs of our sample (see Sect.\ref{jadesec})
estimated through the empirical calibrations by   
\citet{2024MNRAS.533L...1D} and \citet{2021MNRAS.507..466D} (see Eqs.\ref{eqhec} and \ref{eqheo}),
respectively.}
\label{tabac2}
\centering
\begin{tabular}{@{}lcc@{}}
\hline
 
          ID           & 12+log(He/H)            & 12+log(O/H)            \\		    
	               &                         &   	                    \\
 \hline                                                                           
 n-mediumhst-607       &  $10.91 \pm  0.14$	  &     $8.36  \pm  0.36$	  \\	
 n-mediumhst-643       &  $11.18 \pm  0.12$	  &     $8.48  \pm  0.25$	  \\	
 n-mediumhst-917       &  $11.12 \pm  0.16$	  &     $8.43  \pm  0.32$	  \\	
 n-mediumhst-954       &  $10.91 \pm  0.14$	  &     $8.23  \pm  0.24$	  \\	
 s-mediumhst-9669      &  $11.32 \pm  0.15$	  &     $8.44  \pm  0.12$	  \\	
 s-mediumhst-15357     &  $11.38 \pm  0.12$	  &     $8.38  \pm  0.30$	  \\	
 n-mediumjwst-29164    &  $11.50 \pm  0.15$	  &     $8.55  \pm  0.34$	  \\	
 s-mediumjwst-38562    &  $10.97 \pm  0.17$	  &     $8.26  \pm  0.25$	  \\	
 s-mediumjwst-41503    &  $11.18 \pm  0.12$	  &     $8.50  \pm  0.21$	  \\	
 s-mediumjwst-53979    &  $11.05 \pm  0.17$	  &     $8.39  \pm  0.13$	  \\	
 s-mediumhst-58656     &  $11.12 \pm  0.16$	  &     $8.38  \pm  0.33$	  \\	
 s-mediumhst-58850     &  $11.35 \pm  0.09$	  &     $8.33  \pm  0.13$	  \\	
 s-mediumjwst-61321    &  $11.27 \pm  0.13$	  &     $8.51  \pm  0.33$	  \\	
 n-mediumjwst-78891    &  $11.12 \pm  0.11$	  &     $8.42  \pm  0.17$	  \\	
 s-mediumjwst-172074   &  $11.20 \pm  0.13$	  &     $8.37  \pm  0.11$	  \\	
 s-mediumjwst-181864   &  $11.75 \pm  0.15$	  &     $8.31  \pm  0.22$	  \\	
 s-mediumjwst-184060   &  $11.28 \pm  0.09$	  &     $8.33  \pm  0.11$	  \\	
 s-mediumjwst-185290   &  $11.32 \pm  0.15$	  &     $8.66  \pm  0.26$	  \\	
 s-mediumjwst-204851   &  $11.05 \pm  0.17$	  &     $8.24  \pm  0.29$	  \\	
\hline
\end{tabular}	   								
\end{table}

\subsection{Abundance estimates}
\label{abusubs}

\subsubsection{$T_{\rm e}$-method}
\label{Te_meth}
For the 7 objects for which the [\ion{O}{iii}]$\lambda4363$/H$\beta$ and \ion{He}{ii}$\lambda4686$/H$\beta$ line ratios are available (see Table~\ref{tabap1}), the He/H and O/H abundances were derived
following the same metho\-do\-lo\-gy as described by \citet{2022MNRAS.514.5506D}, i.e. the use of the direct or $T_{\rm e}$-method, based on estimating the electron temperature from the $R_{\rm O3}=$[\ion{O}{iii}]$(1.33 \times \lambda5007)/\lambda4363$ line ratio, for our type~2 AGNs belonging to the JADES sample. This is the same approach used to derive the abundances for the local sample (see Sect.~\ref{locals}) considered in this work as comparison.

Initially, using the AGN observational data listed in  Table~\ref{tabap1}, we calculated the electron temperature in the high-ionization zone, $T_{\rm high}$, from $R_{\rm O3}$. For this purpose, we used version 1.1.13 of the \textsc{pyneb} code \citep{2015A&A...573A..42L}. It is not possible to estimate the electron density due to the lack of the [\ion{S}{ii}]$\lambda6716,\lambda6731$ emission lines in our data, therefore, we derive $T_{\rm high}$ assuming the typical value $N_{\rm e} = 500 \: \rm cm^{-3}$ 
for NLRs in local AGNs (e.g. \citealt{2018A&A...618A...6K, 2020MNRAS.492..468D, 2021MNRAS.507...74R, 2024ApJ...960..108Z}).
We discuss the effect of the assumed density in Sect.~\ref{helsec}.

Since it is not possible to directly estimate $T_{\rm low}$ due to the lack of measurements of the [\ion{N}{ii}]$\lambda5755$ auroral line, we derive it using the theoretical relation proposed by \citet{2020MNRAS.496.3209D}, given by
  \begin{equation}
\label{t2t3agn}
t_{\mathrm{low}}=({\rm a} \times t_{\mathrm{high}}^{3})+({\rm b} \times t_{\mathrm{high}}^{2})+({\rm c} \times t_{\mathrm{high}})+{\rm d},
\end{equation}
where $\rm a=0.17$, $\rm b=-1.07$, $\rm c=2.07$ and  $\rm d=-0.33$,
 $t_{\mathrm{low}}$ and $t_{\mathrm{high}}$ represent $T_{\rm low}$ and $T_{\rm high}$, respectively, in units of $10^{4}$ K. 
The values of $T_{\rm high}$ were assumed in the calculations of the abundances of He$^{2+}$ and O$^{2+}$, while $T_{\rm low}$ in the 
He$^{+}$ and O$^{+}$ estimates. Thereafter, we assume the expressions

\begin{equation}
    \frac{\mathrm{He}}{\mathrm{H}} = \mathrm{ICF}( \mathrm{He^{0}} ) \times 
    \left( \frac{ \mathrm{He^{+}} + \mathrm{He^{2+}} }{ \mathrm{H^+} } \right)
\end{equation}
and
\begin{equation}
    \frac{\mathrm{O}}{\mathrm{H}} = \mathrm{ICF}( \mathrm{O} ) \times 
    \left( \frac{ \mathrm{O^{+}} + \mathrm{O^{2+}} }{ \mathrm{H^+} } \right)
\end{equation}
\noindent to estimate the total abundances. The uncertainties in the abundances were derived by propagating the errors of the emission-line ratios. The $\rm He^{+}$ and $\rm He^{2+}$ abundances were calculated using the \ion{He}{i}$\lambda5876$ and \ion{He}{ii}$\lambda4686$ lines, respectively. The $\rm O^{+}$ and $\rm O^{2+}$ abundances were derived from the [\ion{O}{ii}]$\lambda3727$ and [\ion{O}{iii}]$\lambda5007$ lines, respectively.
The ICF($\rm He^{0}$) and ICF(O) values were obtained according to the methodology developed by  \citet{2022MNRAS.514.5506D}. Table~\ref{tabap2}   presents the electron temperatures, ICFs, ionic and total abundances of the 7 JADES AGNs.

\subsubsection{Strong-line methods}

As the $T_{\rm e}$-method could be applied to only 7 of the 19 JADES AGNs, we estimated He/H and O/H abundances for the full high-$z$ sample using the two empirical calibrations described below:
\begin{itemize}
\item Helium calibration:
\citet{2024MNRAS.533L...1D} used $T_{\rm e}$-method and derived He/H estimates in a local sample of 65 AGNs (the same sample described in Sect.~\ref{locals}). These authors  obtained the empirical calibration
\begin{equation}
\label{eqhec}
\rm 12+\log(He/H)=(0.703\pm0.05) \times \log\left(\frac{\ion{He}{i}\lambda5876}{H\beta}\right)+11.83\pm0.05.
\end{equation}

\item Oxygen calibration:
\citet{2021MNRAS.507..466D} used the $T_{\rm e}$-method to derive O/H estimates in a local sample of 91 AGNs, whose observational data were taken from the SDSS DR7 \citep{2009ApJS..182..543A}, and derived the empirical calibration
\begin{equation}
\label{eqheo}
12+\log({\rm O/H}) = (-1.00\pm0.09 \times P) + (0.036\pm0.003 \times R_{23}) + 8.80\pm0.06,
\end{equation}
where $P$=([\ion{O}{iii}]$\lambda4959+\lambda5007$)/$R_{23}$.  
\end{itemize}
The error in the abundance estimates were derived considering the emission-line uncertainties and the error of the coefficients of the expressions above. Table~\ref{tabac2} lists the abundance values derived for our JADES sample.

\section{Results}
\label{res}

\begin{figure*}
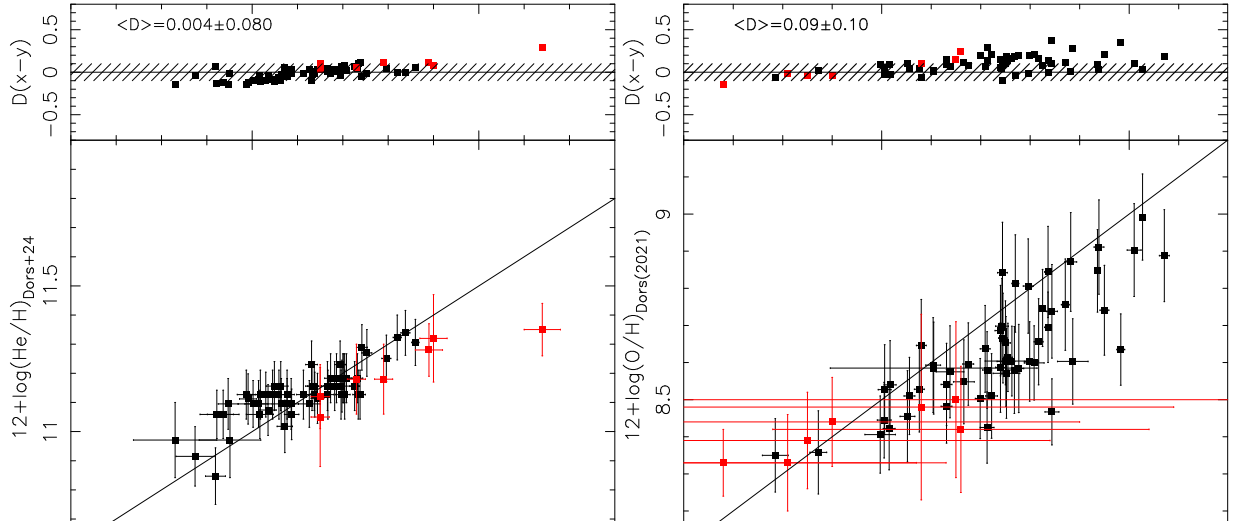

\includegraphics[angle=-90, width=0.45\textwidth]{comp2.eps}
\includegraphics[angle=-90, width=0.45\textwidth]{comp1.eps}
 \caption{Left bottom panel: Comparison between He/H abundances estimated via the empirical
 calibration proposed by  \citet{2024MNRAS.533L...1D} versus those via $T_{\rm e}$ method
 (see Sect.~\ref{Te_meth}). Red and black points represent the JADES and local objects, respectively. Solid line represent the equality between the estimates. Left top panel: Difference between the estimates (D=x-y) versus
 those via the $T_{\rm e}$-method. Line represents the null difference between the estimates, while hatched area represents the typical uncertainty ($\rm0.1$ dex)  in NLR abundance estimates
 (e.g. \citealt{2022MNRAS.514.5506D}). Right panels. As the left panels but for O/H estimates
 (y-axis) obtained via the calibration proposed by \citet{2021MNRAS.507..466D}.
 The average difference (<D>) between the estimates is indicated in each panel.}
\label{figcheo}
\end{figure*}

Before analysing our abundance results, it is necessary to verify whether the strong-line methods proposed by \citet{2024MNRAS.533L...1D} and \citet{2021MNRAS.507..466D} for estimating the He/H and O/H abundances, respectively, yield values consistent  with those derived using the 
$T_{\rm e}$-method for our samples. This verification is important (e.g. \citealt{2026ApJ..1003..228S}) because these indirect methods were calibrated using observational data of local objects ($z < 0.4$), whose ISM conditions may differ from those of high-$z$ objects (e.g. \citealt{2013ApJ...774L..10K}). In the left-hand panel of Fig.~\ref{figcheo}, the He/H abundances derived from the calibration of \citet{2024MNRAS.533L...1D} (Eq.~\ref{eqhec}) are compared with the corresponding $T_{\rm e}$-based estimates. Similarly, the right-hand panel shows a comparison between the O/H abundances obtained from the calibration of \citet{2021MNRAS.507..466D} (Eq.~\ref{eqheo}) and those derived using the $T_{\rm e}$ method. In both panels, the differences between the two sets of estimates are of the order of 0.004 dex and 0.1 dex for helium and oxygen, respectively. Therefore, although only seven $T_{\rm e}$-based abundance determinations are available for the JADES sample, the empirical calibrations given by equations~\ref{eqhec} and \ref{eqheo} appear to provide abundance estimates for high-$z$ AGNs consistent with those obtained from the $T_{\rm e}$-method.

\begin{figure}
\includegraphics[angle=0, width=1\columnwidth]{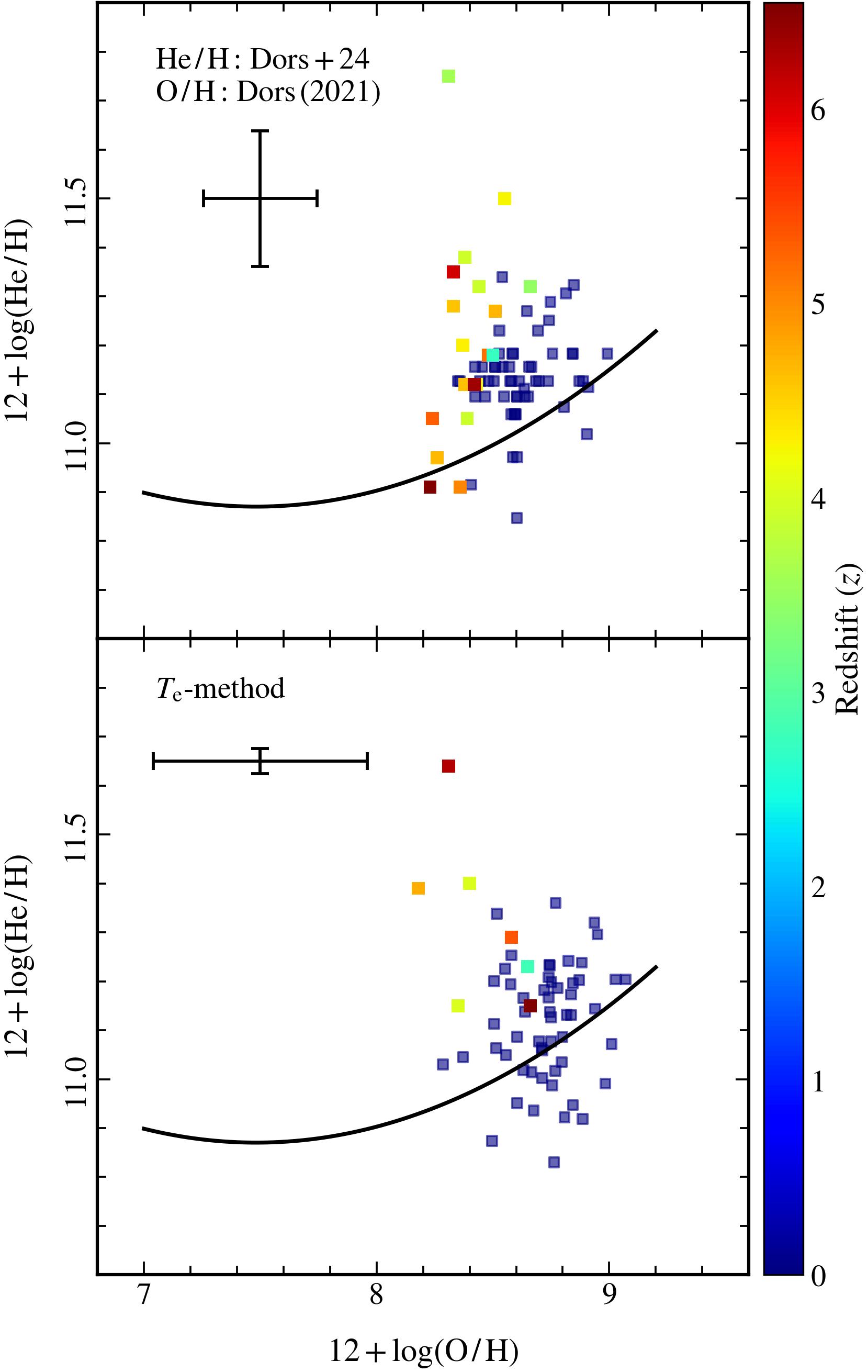}
\caption{Helium [in units of 12+log(He/H)] versus oxygen [in units of 12+log(O/H)] abundances for our JADES (see Sect.~\ref{jadesec}) and local (see Sect.~\ref{locals}) samples. Points represent type~2 AGNs classified using diagnostic diagrams (see Fig.~\ref{fig1}).
Bottom panel: Abundance estimates derived via the $T_{\rm e}$-method (see Sect.~\ref{abusubs}).
Upper panel: Abundance estimates derived via the empirical calibrations (Eqs.~\ref{eqhec} and \ref{eqheo}) proposed by \citet{2024MNRAS.533L...1D} and \citet{2021MNRAS.507..466D}. Points are color-coded according to the redshift indicated in the colored bar. Error bars in the bottom panel represent the mean errors in the JADES estimates of He/H and O/H via the $T_{\rm e}$-method, which are 0.03 and 0.46 dex, respectively. The mean errors in He/H and O/H in the upper panel for the JADES estimates are 0.14  and 0.24 dex, respectively. In both panels, the curve represents the abundance relation derived by \citet{2022MNRAS.514.5506D} by applying the $T_{\rm e}$-method to local SFs and AGNs, as represented by Eq.~\ref{fit1}.}
\label{fig3}
\end{figure}

In the bottom panel of Fig.~\ref{fig3}, the 12+log(He/H) versus 12+log(O/H) values derived through the $T_{\rm e}$-method are presented for the JADES sample, along with local AGNs. The color bar indicates the redshift of each object. Also in this plot, the (He/H)-(O/H) relation derived by \citet{2022MNRAS.514.5506D} by applying  $T_{\rm e}$-method for local SFs and AGNs represented by 
\begin{equation}
\label{fit1}
\begin{split}
{\rm w} =&(0.1215\pm0.0422) \times {\rm x^{2} } -
(1.8183\pm0.6977) \times {\rm x}  \\
&+ \mathrm{ (17.6732\pm2.8798)},
\end{split}
\end{equation}
where w=12+log(He/H) and x=12+log(O/H) is shown. The following results are obtained:
\begin{itemize}
\item High-$z$ AGNs do not follow the (He/H)–(O/H) relation (Eq.~\ref{fit1}) established for local objects (AGNs+SFs).
\item The helium abundances in high-$z$ AGNs tend to be higher than those of local AGNs.
\item The maximum helium abundance value is $ 12+\log({\rm He/H})_{T_{\rm e}}=11.64\pm0.04$ [or $\rm (He/He_{\odot})\sim4.4$] derived for the object s-mediumhst-58850 ($z=6.2615$), a value somewhat ($\sim 0.3$ dex) higher than the one estimated from the empirical calibration,
i.e. $\rm 12+\log(He/H)_{Dors+24}=11.36\pm0.09$ [or $\rm (He/He_{\odot})\sim2.3$] (see Tables~\ref{tabap2} and \ref{tabac2}).
\end{itemize}
In the upper panel of Fig.~\ref{fig3}, we present the He/H versus O/H abundances derived from the empirical relations described above. The following results are obtained:
\begin{itemize}
\item The results are consistent with those obtained from the
$T_{\rm e}$-method: high-$z$ AGNs exhibit higher He/H values than local AGNs and 
do not follow the local (He/H)–(O/H) relation.
\item The maximum helium abundance is
$\rm 12+\log(He/H)_{Dors+24}=11.75\pm0.15$ [or $\rm (He/He_{\odot})\sim5.6$] estimated for the object s-mediumjwst-181864 $z=3.7022$ (see Table~\ref{tabac2}). It was
not possible apply the $T_{\rm e}$-method for this object due to the not detection
of the [\ion{O}{iii}]$\lambda4363$ line.
\end{itemize}
Discrepancies between abundances derived with different techniques can reach $\sim 0.7$ dex (see \citealt{2008ApJ...681.1183K, 2010arXiv1004.5251L, 2020MNRAS.496.3209D}). 
Thus, as the He/H value derived for s-mediumjwst-181864 can be somewhat uncertain
due to the indirect method used, we assume that the maximumm value obtained
for the objects in our sample is the one for s-mediumhst-58850 ($z=6.2615$), i.e. $ 12+\log({\rm He/H})_{T_{\rm e}}=11.64\pm0.04$ [or $\rm (He/He_{\odot})\sim4.4$].
In any case, the abundances derived by
the distinct methods shown in Fig.~\ref{fig3} indicate a distinct cosmic enrichment history of helium and oxygen in NLRs of AGNs.

It is worth noting that the uncertainties on the He/H abundances are generally smaller than those for O/H, despite the helium emission lines typically possessing lower signal-to-noise ratios. This behaviour is a direct consequence of the atomic physics governing the respective emission features. As outlined in Sect.~\ref{Te_meth}, estimating the total abundances for both elements requires utilizing the $T_{\rm high}-T_{\rm low}$ relation, where the inferred $T_{\rm low}$ is applied to both the O$^+$ and He$^+$ ionic zones. Because $T_{\rm low}$ must be theoretically derived from $T_{\rm high}$ (Eq.~\ref{t2t3agn}), its propagated uncertainty is substantial. However, oxygen abundances are derived from collisionally excited lines, which depend exponentially on the electron temperature. This exponential dependence heavily magnifies the propagated $T_{\rm low}$ errors, driving up the total O/H uncertainty. Conversely, helium abundances are derived from optical recombination lines, which possess a very weak power-law dependence on $T_{\rm e}$. Due to this temperature insensitivity, the helium derivations are largely shielded from the severe error propagation introduced by the $T_{\rm high}-T_{\rm low}$ relation. Furthermore, for the subset of objects where both methods could be applied, the abundances derived from the strong-line calibrations are generally consistent with the $T_{\rm e}$-method estimates within $\sim 0.2$ dex, indicating that the observed evolutionary trends are not strictly driven by the systematic uncertainties of a single calibration.

\begin{figure*}
\includegraphics[angle=-90, width=0.45\textwidth]{evol_he.eps}
\includegraphics[angle=-90, width=0.45\textwidth]{evol_ox.eps}
 \caption{Helium and oxygen abundances versus redshift for the JADES (red points) and local (black points) AGNs. In the left panels, the blue lines represent fits to the data points (Eqs.~\ref{evolhe} and \ref{evolhe2}). The triangle represents the helium estimate 
 [12+log(He/H)=$10.92^{+0.08}_{-0.06}$] for the quasar HS,1700+6416 ($z=1.724$) by \citet{2018NatAs...2..957C}, derived via the \ion{He}{i} [$500 \: < \: \lambda(\angstrom) \: < \: 540$] absorption lines (see Sect.~\ref{helsec}). Dashed lines represent the solar helium abundance, 12+log(He/H)$_{\odot}=11.0$ \citep{2001ApJ...556L..63A}.
Right panels: same as the left panels, but for the oxygen abundance. 
Dashed lines represent the solar oxygen abundance, 12+log(O/H)$_{\odot}=8.69$ \citep{2001ApJ...556L..63A}.
Solid lines represent fits to the data points (Eqs.~\ref{evolox} and \ref{evolox2}). In each panel, the method used to estimate the abundances, the Spearman  coefficient ($\rho$) and $p$-values   are indicated.}
\label{fig4}
\end{figure*}

In Fig.~\ref{fig4}, the He/H (left panels) and O/H (right panels) abundances derived by the $T_{\rm e}$-method and via the empirical calibrations
are plotted as a function of redshift for the JADES and local AGN samples.
 It can be seen that the helium abundance for $ 2.8 \: < \: z\: < \: 6.8$ covers a very similar range for each $z$ value.  The Spearman Coefficient ($\rho$) and the $p$-values, shown in Fig.~\ref{fig4}, indicate a weak correlation ($p=0.008$, $\sim 2.6\sigma$) and a statistically marginal, non-significant correlation ($p=0.131$, $\sim 1\sigma$) between the He/H and $z$, for estimates from $T_{\rm e}$ and strong-line methods, respectively.

We perform a  linear fitting to the points (represented by blue lines) yielding
\begin{equation}
\label{evolhe}
12+\log\left({\rm \frac{He}{H}}\right)_{T_{\rm e}}=0.043(\pm 0.009) \times z + 11.11(\pm0.01) 
\end{equation}
and
\begin{equation}
\label{evolhe2}
12+\log\left({\rm \frac{He}{H}}\right)_{{\rm Dors+24}}=0.011(\pm 0.006) \times z + 11.13(\pm0.01). 
\end{equation}

In the right panels of Fig.~\ref{fig4},
the (O/H)-$z$ relations derived from the $T_{\rm e}$-method and from the empirical calibration are shown.  In contrast to helium, we note a decrease in O/H with increasing redshift. 
The Spearman coefficient ($\rho$)  and the $p$-values indicate weak and moderate 
correlations between the O/H and $z$ for estimates from $T_{\rm e}$ and strong-line methods, respectively.
A linear fit to the points (blue line) yields
\begin{equation}
\label{evolox}
12+\log\left({\rm \frac{O}{H}}\right)_{T_{\rm e}}=-0.057(\pm 0.013) \times z + 8.74(\pm0.02)
\end{equation}
and
\begin{equation}
\label{evolox2}
12+\log\left({\rm \frac{O}{H}}\right)_{{\rm Dors(2021)}}=-0.048(\pm 0.007) \times z + 8.62(\pm0.01).
\end{equation}
We note that objects at $z \: > \: 3.5$, those with the highest helium abundance, present O/H abundances similar to those of local AGNs. This result, as well as the slight decrease of O/H with $z$, may be somewhat biased because the $N2$-diagram excludes AGNs with  $\rm 12+\log(O/H) \: \lesssim \:8.0$ or $(Z/\rm Z_{\odot}) \: \lesssim \: 0.2$ (see Figure~12 of \citealt{2024MNRAS.527.8193D}). 

\begin{figure}
\includegraphics[angle=-90, width=0.49\textwidth]{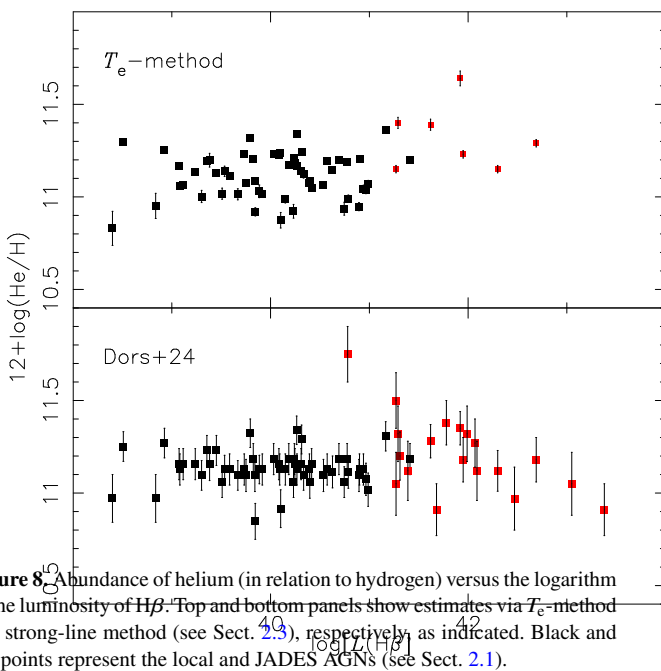}
 \caption{Abundance of helium (in relation to hydrogen) versus the logarithm of the luminosity of H$\beta$. 
Top and bottom panels show estimates via $T_{\rm e}$-method and strong-line method (see Sect.~\ref{abusubs}), respectively, as indicated.
 Black and red points
 represent the local and JADES AGNs (see Sect.~\ref{obsec}).}
\label{figluhe}
\end{figure}

The occurrence of helium-enriched AGN candidates may be associated with the elevated luminosities and, consequently, larger masses of the hosting galaxy of these objects, potentially introducing a selection bias. To investigate this possibility, Fig.~\ref{figluhe} shows the helium abundances derived using the strong-line method (top panel) and the $T_{\rm e}$-method (bottom panel) as a function of the logarithm of the H$\beta$ luminosity. It can be seen that the highest He/H values, irrespective of the abundance determination method adopted, are not associated with the most luminous objects. Therefore, these results suggest a genuine enhancement of the ISM helium abundance rather than an effect arising from mass or luminosity selection.

\begin{figure}
\includegraphics[angle=-90, width=0.45\textwidth]{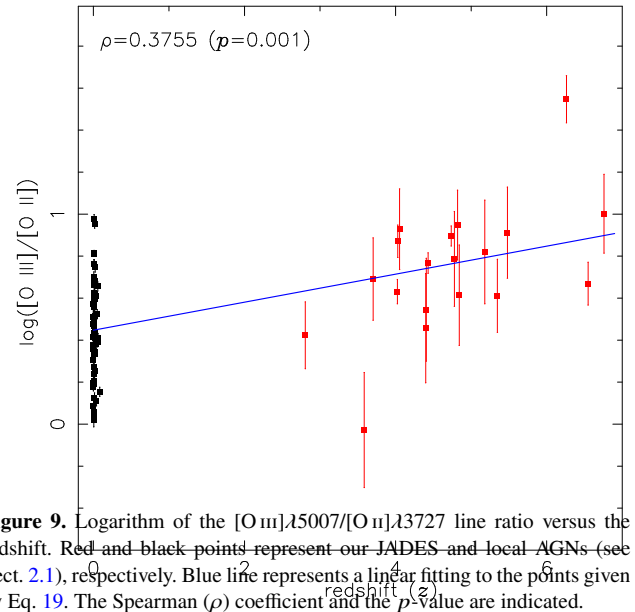}
 \caption{Logarithm of the [\ion{O}{iii}]$\lambda5007$/[\ion{O}{ii}]$\lambda3727$ line ratio versus the
    redshift. Red and black points represent our JADES and local AGNs (see Sect.~\ref{obsec}), respectively. Blue line represents a linear
    fitting to the points given by Eq.~\ref{fitpedro}. The Spearman ($\rho$) coefficient
 and the $p$-value are indicated.}
\label{figionz}
\end{figure}


\section{Discussion}
\label{secdisc}

\subsection{Ionization degree at high-\MakeLowercase{\textit{z}}}

Before discussing our abundance results, it is useful to compare the ionization degree between
local and high-$z$ AGNs. It is known that SFs at $z \: > \: 1.0$ present
a higher ionization degree than local ones, although the source of this difference is still under debate in the literature (e.g. \citealt{2013ApJ...774L..10K, 2022ApJ...926...80G, 2023ApJ...955...54S}). The JADES and the local AGN observational data (see Sect.~\ref{obsec}) provide an excellent opportunity to test whether AGNs, like SFs, present distinct ionization degrees along the Hubble time. Unfortunately, it is not possible to derive the ionization parameter $U$
for most objects belonging to the JADES sample because these present, in general,
values of $\log([\ion{O}{iii}] / [\ion{O}{ii}])$ outside the validity range of 
calibrations, such as the one proposed by \citet{2020MNRAS.492.5675C},
 valid for $-1.5 \: < \: \log([\ion{O}{iii}]/[\ion{O}{ii}]) \: < \: 0.5$ (see also \citealt{2024ApJ...977..187Z}).

As an alternative, in Fig.~\ref{figionz} we analyze the behavior of the ionization degree as a function of redshift, assuming the [\ion{O}{iii}]$\lambda5007$/[\ion{O}{ii}]$\lambda3727$ line ratio as a proxy for this quantity \citep{1994ApJ...426..135M}.  It can be seen that, despite the scatter the Spearman Coefficient ($\rho$) 
value indicating a weak correlation, a clear trend is present, with the logarithm of [\ion{O}{iii}]/[\ion{O}{ii}] increasing with $z$; that is, high-$z$ AGNs tend to have a higher ionization degree than local ones.
A fitting to the points in Fig.~\ref{figionz} results in
\begin{equation}
\label{fitpedro}
    \log(O_{32}) = 0.066(\pm 0.012 ) \times z + 0.447 (\pm 0.029),
\end{equation}
where $O32=([\ion{O}{iii}]\lambda5007 / [\ion{O}{ii}]\lambda3727)$, which is represented in this figure by a blue line.

\subsection{Cosmic metallicity evolution}

Along the past decades, several studies have attempted to determine
the metallicity and elemental abundances of high-redshift galaxies,
yielding conflicting results. In particular, some works report an in-
crease in O/H toward lower redshifts, while others find little to no
evident evolution. A summary of some of these studies is presented
below.

\paragraph*{Metallicity evolution:}
It appears that the most compelling evidence for a clear evolution of $Z$ with decreasing redshift is found for Damped Lyman-$\alpha$ systems (DLAs). For instance, \citet{2025ApJ...991..228H} recently conducted a survey targeting DLAs along the line of sight to high-$z$ quasars in order to measure $Z$ and elemental abundances (see also \citealt{2003ApJ...595L...9P, 2006MNRAS.370...43M, 2012ApJ...755...89R, 2016ApJ...830..158M, 2026arXiv260416751W}).
The metallicity measurements by \citet{2025ApJ...991..228H}
are determined from undepleted elements (Zn, S, or O) for 
the range $0.0 \: < \: z \: < \: 6.0$ and, translation to oxygen abundance, results in  
\begin{equation}
\label{eqhuy}
12+\log({\rm O/H})_{{\rm DLA}}=-0.19 \times z + 8.03.
\end{equation}
It is worth mentioning that the above results are not derived for individual active or star-forming galaxies, but rather for intervening objects along the line of sight.
However, these results are based on a different method (from absorption lines) than the one (from emission lines) considered in the present study, and they seem to be in consonance with some cosmic chemical evolution models that predict a metallicity increment with time (see \citealt{2013ApJ...772...93K, 2014MNRAS.443.1291D}).

Additional evidence for chemical evolution at high redshift comes from nebular-line studies.
\citet{2019A&A...626A...9M}, who estimated oxygen abundances by comparing observations with photoionization model predictions of UV emission lines, found a moderate evolution of O/H in  NLRs over the interval $1.5 \: < \: z \: < \: 3.0$ (for $z \: \lesssim \: 0.4$ see  \citealt{2023ApJ...955..141C, 2023MNRAS.520.1687A}). Although \citet{2019A&A...626A...9M} did not provide an explicit analytical expression for the $Z$--$z$ relation, we use their results (see their Fig.~6) to derive
\begin{equation}
\label{eqmig}
12+\log({\rm O/H})_{{\rm NLR}}= -0.285 \times z + 8.62.
\end{equation}

\paragraph*{No metallicity evolution:}
\citet{2020ApJ...898..105O} found no evolution in the \ion{Fe}{ii}(UV)/\ion{Mg}{ii}$\lambda2798$\footnote{The \ion{Fe}{ii} flux is integrated over the emission lines of this ion emitted between 2200\angstrom\ and 3090\angstrom. The \ion{Mg}{ii}$\lambda2798$ flux represents the sum of the doublet at 2796\angstrom\ and 2803\angstrom.} emission-line ratio — commonly used as a cosmic clock tracing iron enrichment \citep{1993ApJ...418...11H, 1998ApJ...507L.113Y} — in the Broad Line Regions (BLRs) of quasars over the redshift range between $\sim0.0$ and $\sim8.0$ (see also \citealt{2002ApJ...564..592H, 2003ApJ...596L.155M, 2011ApJ...739...56D, 2018MNRAS.480..345X, 2019ApJ...874...22S, 2022ApJ...925..121W, 2024ApJ...975..214J}).
Interestingly, most $Z$ estimates for NLRs based on comparisons between ultraviolet emission lines predicted by photoionization models and those observed in type~2 AGNs similarly indicate no detectable metallicity evolution with cosmic time (e.g. \citealt{2006A&A...447..863N, 2009A&A...503..721M, 2014MNRAS.443.1291D, 2025MNRAS.542.3181D}).

Although no convincing evidence for metallicity evolution has been obtained for neither BLRs nor NLRs, the mean $Z$ inferred for BLRs (see \citealt{1993ApJ...418...11H, 2003ApJ...589..722D, 2006A&A...447..157N, 2009A&A...494L..25J, 2014MNRAS.439..771B, 2017ApJ...834..203S, 2021ApJ...910..115S, 2022MNRAS.513.1801L}) is significantly higher by a factor of $\sim5$ than that derived for NLRs (see \citealt{2006A&A...447..863N, 2009A&A...503..721M, 2018A&A...616L...4M, 2014MNRAS.443.1291D, 2016MNRAS.456.3354F, 2018ApJ...856...46R, 2019MNRAS.486.5853D, 2020MNRAS.496.2191F, 2023ApJ...955..141C, 2024MNRAS.535..881J, 2024PASA...41...99O}).

For consistency with the present study, we adopt as representative of the ``no evolution'' scenario in AGNs the results obtained by \citet{2025MNRAS.542.3181D}. These authors applied a semi-empirical calibration between narrow UV emission lines and $Z$, finding, for a sample of 91 type~2 AGNs spanning two redshift ranges, the following values: for $z \: < \: 0.1$, 12+log(O/H)$=8.22\pm0.18$, and for $1.2 \: < \: z \: < \: 3.8$, 12+log(O/H)$=8.44\pm0.41$.

\begin{figure}
\includegraphics[angle=0, width=0.47\textwidth]{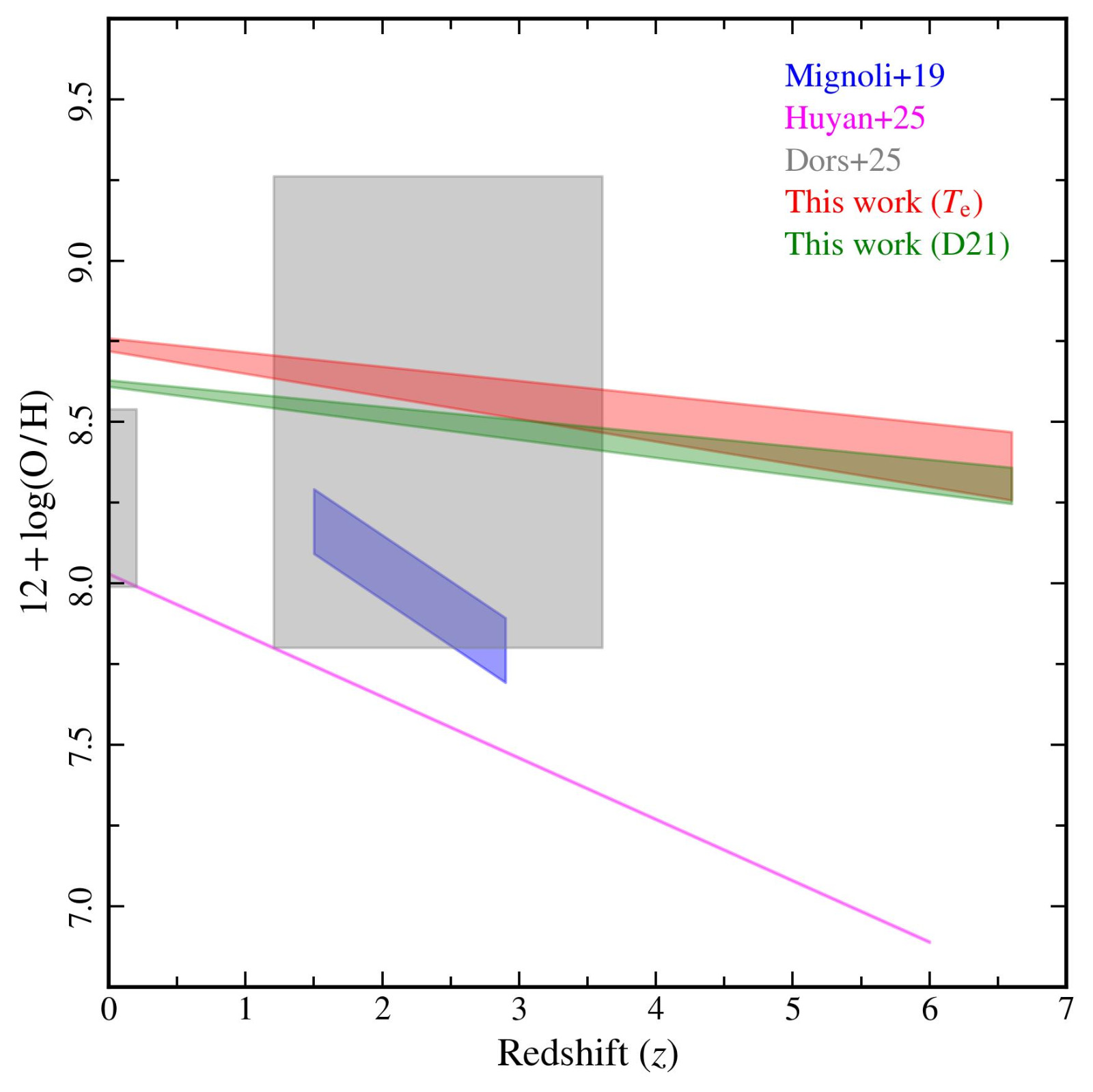}
 \caption{Relation between the oxygen abundance [in units of 12+log(O/H)] versus 
 the redshift. Red and green areas represent results obtained in the present work (Eqs.~\ref{evolox} and \ref{evolox2}). Gray area represents estimates for type~2 AGNs from the semi-empirical
 calibration between UV emission lines and $Z$ proposed by \citet{2025MNRAS.542.3181D}.
 Blue area represents estimates for NLRs (Eq.~\ref{eqmig}) based on comparisons between observations and 
 photoionization model predictions by  \citet{2019A&A...626A...9M}.   Magenta line  represents estimates for Damped Lyman-$\alpha$ systems (DLAs) by  \citet{2025ApJ...991..228H} and based on  absorption lines (Eq.~\ref{eqhuy}).}
\label{fig6n}
\end{figure}

Our abundance estimates obtained using the $T_{\rm e}$-method and strong-line methods for the NLRs of AGNs over the wide redshift range
$0.0 \: \lesssim \: z \: \lesssim \: 7.0$ provide unprecedented constraints on the cosmic chemical evolution of this class of objects. In Fig.~\ref{fig6n},  we present 12+log(O/H) as a function of redshift ($z$), comparing our results (Eq.~\ref{evolox}  and \ref{evolox2}) with those derived for DLAs (Eq.~\ref{eqhuy}) by \citet{2025ApJ...991..228H}, for NLRs (Eq.~\ref{eqmig}) by \citet{2019A&A...626A...9M} and the mean values by \citet{2025MNRAS.542.3181D}. The shaded regions represent the uncertainties in the (O/H)--$z$ relations reported by each  study considered.
For \citet{2019A&A...626A...9M}, we adopt an O/H uncertainty of $\pm 0.1$ dex, i.e. a typical value from direct abundance determinations for NLRs (e.g. \citealt{2024MNRAS.534.3040D}) and star-forming systems (e.g. \citealt{2008MNRAS.383..209H}).

We note the following:
\begin{itemize}
\item Our (O/H)--$z$ relations overlap with the region covered by the UV-based metallicity estimates for type~2 quasars ($1.2 \: < \: z \: < \: 3.8$) derived by \citet{2025MNRAS.542.3181D}, despite the fact that these authors did not report any metallicity evolution. In contrast, lower O/H abundances by $\sim 1.0$ dex are derived when UV estimates by \citet{2025MNRAS.542.3181D} at $z \: < \: 0.1$ are compared with our relations.

\item The results of \citet{2019A&A...626A...9M} exhibit a steeper decline of O/H with increasing redshift and abundances lower by $\sim 1.5$ dex relative to our measurements. However, their extrapolated value at $z=0$, i.e. 12+log(O/H)$\sim 8.7$ (corresponding to $Z\sim Z_{\odot}$), is consistent with our findings.

\item DLA metallicities remain systematically lower by $\sim 1.0$ dex, in average, than the values derived for NLRs across the entire redshift range (see also \citealt{2010IAUS..265..147E}).
\end{itemize}

The discrepancies shown in Fig.~\ref{fig6n} can be attributed to several factors, the most important being the method adopted to estimate the metallicity and the characteristics of the samples considered. For instance, \citet{2025MNRAS.542.3181D} showed that their UV semi-empirical calibration yields $Z$ estimates consistent (within a mean difference of $\sim 0.1$ dex) with those obtained from the $T_{\rm e}$-method based on optical emission lines, although this comparison was possible for only six AGNs at $z \sim 0$. Thus, the distinct metallicity trends inferred for the JADES sources  and the type~2 AGNs analysed by \citet{2025MNRAS.542.3181D} may arise from differences in the underlying samples. Indeed, while the JADES sample spans redshift bins with relatively uniform source counts (see Fig.~\ref{figlum}), the observational compilation of \citet{2025MNRAS.542.3181D} is concentrated primarily in the interval $1.2 \: < \: z \: < \: 3.8$ (98 objects) and only 8 objects at $z \: < \: 0.1$.

The discrepancy between the metallicity estimates of \citet{2019A&A...626A...9M} and those obtained for the JADES sample  likely results from a combination of methodology and sample selection effects. \citet{2019A&A...626A...9M} inferred $Z$ by comparing observations with photoionization model predictions in diagnostic diagrams involving carbon lines (i.e. \ion{C}{iv}$\lambda1550$ and \ion{C}{iii}]$\lambda1909$). The models employed by these authors adopted input parameters similar to those of \citet{2016MNRAS.456.3354F}, in which the carbon abundance scales linearly with oxygen. As shown by \citet{2025MNRAS.540.1608D}, a constant C/O ratio is not representative for NLRs, and photoionization models imposing a fixed C/O value yield metallicity estimates that differ systematically from those generated using a more realistic (C/O)–(O/H) relation, such as the one derived by \citet{2025MNRAS.540.1608D}. Likewise, the observational sample of \citet{2019A&A...626A...9M} lies within the restricted redshift range $1.5 \: < \: z \: < \: 3.0$, whereas the our JADES sample covers a significantly broader interval ($2.8 \: \lesssim \: z \lesssim \: 6.8$).

Finally, metallicity estimates in DLAs by \citet{2025ApJ...991..228H} are based on absorption lines, which are less sensitive to variations in electron temperature \citep{2021MNRAS.506L..11R} and electron density (e.g. \citealt{2024A&A...684A..53B}) present in NLRs, and are unaffected by uncertainties inherent to photoionization modelling (e.g. \citealt{2024ApJ...977..187Z}). However, abundances derived from absorption lines are subject to dust depletion, whose magnitude depends on the considered element. For instance, for the elements (e.g. O, S) used as metallicity tracers for DLAs by Huyan and collaborators, the depletion onto dust grains can reach $\sim 0.3$ dex in the most metal-rich or dusty systems (e.g. \citealt{2002A&A...391...21P, 2011A&A...530A..33V, 2016A&A...596A..97D}),
a lower value than the discrepancy ($\sim 1.0$ dex) derived when compared to our estimates. Thus, the different metallicity evolution inferred for DLAs and NLRs may, in part,  reflect a genuinely distinct chemical enrichment history of their respective ISM, rather that systematic differences due to the applied methodologies or sample selection biases.
 Furthermore, it must be noted that the depletion of elements like oxygen onto dust grains also systematically affects abundances derived from emission-line methods, adding a layer of universal uncertainty to both techniques.

\subsection{Cosmic helium abundance evolution}
\label{helsec}

As previously reported, the helium abundance for a large AGN sample was estimated (via the $T_{\rm e}$-method) only by \citet{2022MNRAS.514.5506D} for 65 local ($z < 0.2$) Seyfert 2 nuclei. These authors found that AGNs follow the (He/H)–(O/H) abundance relation derived for SFs, but restricted to the high-metallicity regime [$\rm 12+\log(O/H) \: \gtrsim \: 8.7$], with He/H ranging from 0.60 to 2.50 times the solar value, while $\sim$85 per cent of their sample presented oversolar abundance values. At high redshift ($z \: > \: 1.0$), the He/H abundances in AGNs are poorly known. For instance, \citet{1971ApJ...167L..27W}, using observational data from \citet{1971ApJ...163..235B} and direct estimates, argued that high-$z$ quasars tend to have solar helium abundances (see also \citealt{1978ApJ...223...56K, 1991PASP..103..888C}). More recently, \citet{2024ApJ...974..266Y}, using optical spectroscopic data for three systems (GS-NDG9422, GLASS150008, RXCJ2248-ID) at $z \: \sim \: 6$ obtained with the JWST, derived abundances in the range $\rm 11.00 \: \lesssim \: 12+\log(He/H) \: \lesssim \: 11.30$, or $\rm 1.0 \: \lesssim \: (He/He_{\odot}) \: \lesssim \: 2.0$. According to these authors, these values are comparable to, or significantly larger than, those in local galaxies.

\begin{figure}
\includegraphics[angle=-90, width=0.45\textwidth]{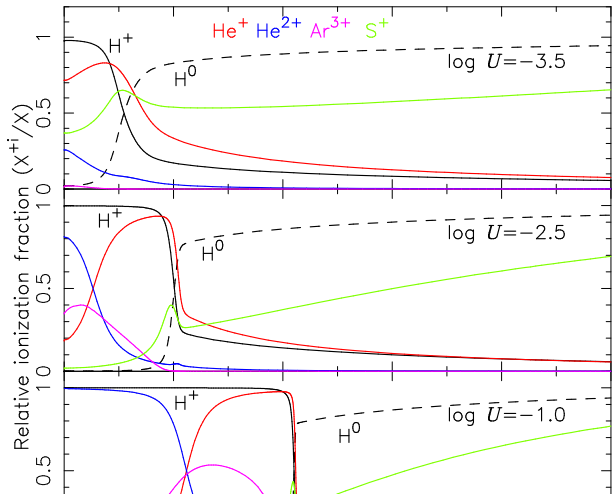}
 \caption{Relative ionic abundance fraction for distinct ions ($\rm X^{i} /X$) versus the nebular radius (normalized by the outermost radius) predicted by photoionization
models simulating NLRs of AGNs. The nebular parameters assumed in the
models are spherical geometry, solar metallicity, $\alpha_{ox}=-1.1$, electron density profile defined by Eq.~\ref{eqmich}, and three values
for $\log U= -1.0$, $-2.5$ and $-3.5$.}
\label{fig6m}
\end{figure}

Our direct and strong-line method He/H estimates in AGNs over a wide redshift range (see Fig.~\ref{fig4}) show that solar or oversolar helium abundances were obtained  for most of the high-$z$ objects, with the highest abundance reaching $12+\log({\rm He/H})_{T_{\rm e}}=11.64$ or $\rm (He/He_{\odot}) \sim 4.4$ and $12+\log({\rm He/H})_{\rm Dors+24}=11.75$ or $\rm (He/He_{\odot})\sim 5.6$, revealing a population of `helium-loud' AGNs at $z > 2.8$. These values are higher than the results of \citet{2024ApJ...974..266Y} for high-$z$ objects at a similar redshift range.

\begin{figure}
\includegraphics[angle=0, width=0.49\textwidth]{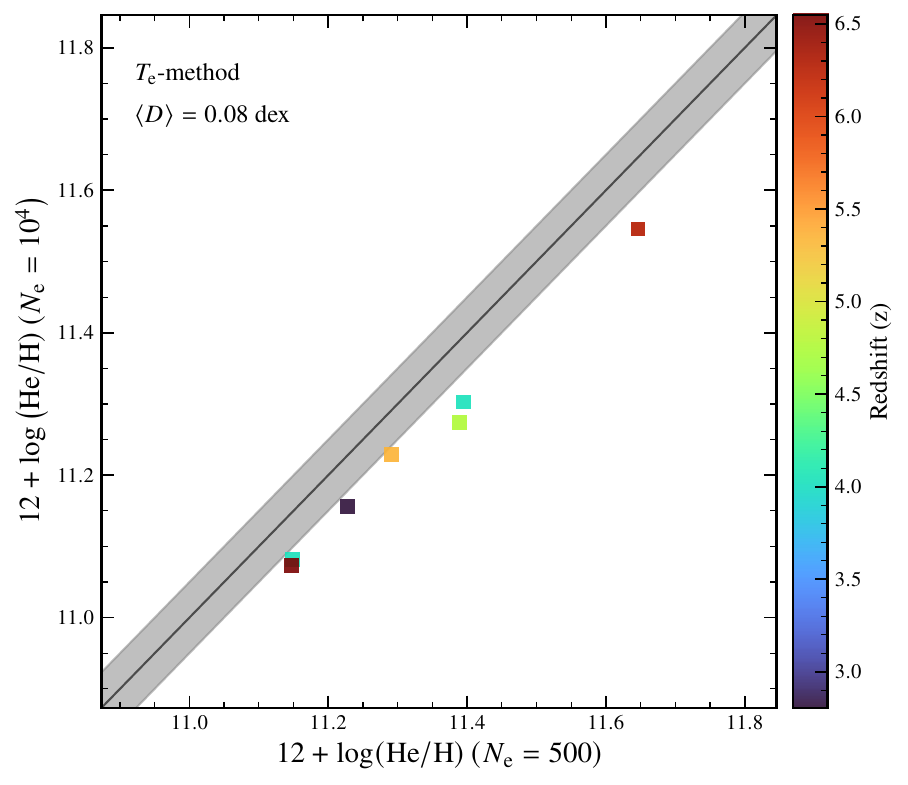}
 \caption{Comparison between helium abundance [in units of 12+log(He/H)] for our JADES AGN sample (see Sect.~\ref{jadesec}) 
 derived through
 the $T_{\rm e}$-method (see Sect.~\ref{abusubs}) assuming $N_{\rm e} =10\,000 \:\rm cm^{-3}$
 versus those for $N_{\rm e} =500 \:\rm cm^{-3}$. Line represents the equality between the estimates and the cyan area the typical error ($\pm0.1$ dex) in the estimates.
 Points
are color-coded according the redshift indicated in the colored bar. The mean difference $<D>$ between the estimates is shown.}
\label{fig6n2}
\end{figure}

The high He/H values found in the present work may be overestimated because we do not account for the presence of high-density clouds in the NLRs. For example, $N_{\rm e}$ estimates in NLRs based on the [\ion{Ar}{iv}]$\lambda4711/\lambda4730$ ratio, which traces the electron density in the inner and high-excitation zones (e.g. \citealt{2019ApJ...880...16K}), yield higher density values than those derived from the [\ion{S}{ii}] ratio. In fact, \citet{2012MNRAS.427.1266V}, using a large sample of NLRs whose observational data were taken from the SDSS, found that $N_{\rm e}$ estimates from [\ion{Ar}{iv}] fall within $200 \: \lesssim \: N_{\rm e} \: ({\rm cm^{-3}}) \: \lesssim \: 10\,000$, while those from [\ion{S}{ii}] lie in the range $30 \: \lesssim \: N_{\rm e} \: ({\rm cm^{-3}}) \: \lesssim \: 1800$ (see also \citealt{2017MNRAS.471..562C, 2021MNRAS.500.2666C, 2024A&A...684A..53B}). Another concern arises from the results of \citet{2025MNRAS.541.1707T}, who, using deep JWST/NIRSpec spectra of star-forming galaxies, reported an increase in $N_{\rm e}$ from $\sim 270 \: \rm cm^{-3}$ at $z=2.3$ to $\sim 500 \: \rm cm^{-3}$ at $z=5.3$ (see also \citealt{2025ApJ...993..204H}). The difference in electron density between local and high-$z$ galaxies may be related to the compact morphology of high-$z$ systems (e.g. \citealt{2023ApJ...956..139I, 2024MNRAS.527.6110O}).
Furthermore, \citet{2024ApJ...974..266Y} noted that for the three JWST-observed objects at $z \: \sim \: 6$, two scenarios are possible: ($i$) oversolar He/H abundances if the low-density regime ($\lesssim 1000 \: \rm cm^{-3}$) is assumed; and ($ii$) He/H values similar to those of local galaxies if $N_{\rm e} \sim 10^{5} \:\rm cm^{-3}$ is adopted in the abundance estimates.

To analyse the effect of high $N_{\rm e}$ values in our helium estimates, we carried out two tests. First, we built photoionization models to investigate the ionization structure of the helium ($\rm He^{+}$, $\rm He^{2+}$) and argon ($\rm Ar^{3+}$) ions in NLRs. For this purpose, we adopted in the models the same nebular parameters as in \citet{2025MNRAS.540.1608D}: a spherical geometry with an inner radius of 3 pc, solar metallicity, a SED characterized by a slope of $\alpha_{ox} = -1.1$, and three values for the ionization parameter: $\log U = -1.0$, $-2.5$ and $-3.5$ dex. These $\log U$ values encompass those found for local (e.g. \citealt{2020MNRAS.492.5675C, 2025A&A...696A.229P}) and high-$z$ (e.g. \citealt{2025MNRAS.542.3181D}) NLRs of AGNs, while  
$\alpha_{ox}=-1.1$ is about the mean value derived from detailed photoionization modeling of local AGNs \citep{2017MNRAS.468L.113D}.
We assumed that the NLRs follow a density profile as the one derived by \citet{2018ApJ...856...46R}, who obtained spatially resolved spectroscopic data with the \textit{Hubble Space Telescope} for the NLR of the Seyfert 2 galaxy Markarian~573. Adopting this electron density profile and considering the maximum $N_{\rm e}$ value of $10\,000 \:\rm cm^{-3}$ inferred from the [\ion{Ar}{iv}]$\lambda4711/\lambda4730$ ratio by \citet{2012MNRAS.427.1266V} for $z \sim 0$, we propose
\begin{equation}
\label{eqmich}
    N_{\rm e} ({\rm cm^{-3}})_{{\rm NLR}}= 10\:000 \times  r^{-0.5},
\end{equation}
where $r$ is the AGN radius. The predicted ionization structure from our simulation is shown in Fig.~\ref{fig6m}, where, for $\log U = -1.0$ and $-2.5$ dex (lower and middle panels), we note the following:
\begin{itemize}
\item Due to the low Ionization Potential (IP) of $\rm S^{+}$ (10.36 eV), this ion is located mainly in the partially ionized regions, i.e. in an outer zone relative to those occupied by $\rm He^{+}$ and $\rm He^{2+}$. In this scenario, it is not appropriate to adopt the $N_{\rm e}$ values derived from the [\ion{S}{ii}]$\lambda6716/\lambda6731$ line ratio in the helium abundance estimates, otherwise, the $\rm He^{+}$ and $\rm He^{2+}$ abundances can be overestimated.
\item Electron density values obtained from the [\ion{Ar}{iv}]$\lambda4711/\lambda4730$ ratio appear to be more representative for the $\rm He^{+}$ and $\rm He^{2+}$ zones.
\end{itemize}
For $\log U = -3.5$ (upper panel of Fig.~\ref{fig6m}) we can appreciate that this effect is not so remarkable due to the presence the [\ion{Ar}{iv}] lines is almost negligible and the [\ion{S}{ii}] lines are also originated in the inner part of the nebula. 
We repeat (not shown) the simulation adopting a softer SED ($\alpha_{ox}=-1.5$), yielding similar results than the ones previously derived with $\alpha_{ox}=-1.1$.

Our second test  consists of recalculating the He/H abundances assuming the maximum $N_{\rm e}$ value derived by \citet{2012MNRAS.427.1266V}, i.e.
$N_{\rm e} = 10\,000 \: \rm cm^{-3}$, and compare them with those obtained (see Fig.~\ref{fig3}) under the low-density regime ($N_{\rm e} = 500 \: \rm cm^{-3}$). This comparison is shown in Fig.~\ref{fig6n2}. The mean He/H abundance difference ($D$) is of the order of $\sim 0.1$ dex, with the largest values reaching $\sim 0.2$ dex.
From this comparison, it can be seen that even if the $\rm He^{+}$ and $\rm He^{2+}$ ions are located in layers with $N_{\rm e} = 10\,000 \: \rm cm^{-3}$, high He/H abundances are still obtained, with a maximum value adopting the $T_{\rm e}$-method of $12+\log({\rm He/H}) \sim 11.54$, or $\rm (He/He_{\odot}) \sim 3.5$. Thus, we confirm the existence of AGNs with oversolar He/H abundances (helium-loud AGNs) at $z  > 2.8$, independently of the electron density assumed in the helium abundance determinations. Similarly enhanced abundances of other elements have also been reported in high-$z$ systems, such as those found by \citet{2024MNRAS.535..881J}, who derived $\log(\mathrm{N/O}) = 0.42$, or $(\mathrm{N/O})/(\mathrm{N/O})_{\odot} \sim 20$, for the system GS-3073 at $z \sim 5.5$ (see also \citealt{2024A&A...687L..11S, 2025ApJ...995L..44Y, 2025arXiv251204043Z, 2025ApJ...994L..29Z, 2025ApJ...989...75N, 2025arXiv250801372M, 2025MNRAS.540.2991A, 2025A&A...697A..89C, 2025arXiv250204817Z, 2025ApJ...994...65N, 2026MNRAS.tmp...20M, 2026arXiv260115964C}).

Helium abundance measurements from emission or absorption lines in AGNs at high redshifts $(z \: > \: 1.0)$ are rare in the literature, which makes a comparison with our results difficult. In particular, it appears that the only helium abundance based on absorption lines in an AGN was obtained by \citet{2018NatAs...2..957C}, who used the \ion{He}{i} absorption features in the range $500 \: < \: \lambda(\angstrom) \: < \: 540$ and derived, for the quasar HS\,1700+6416 ($z = 1.724$), $12+\log({\rm He/H}) = 10.92^{+0.08}_{-0.06}$ or $\rm (He/He_{\odot})=0.83^{+0.17}_{-0.10}$.  This estimate is shown in the left panels of  Fig.~\ref{fig4} as a triangle. Although it is nearby consistent with our results, additional helium abundance measurements based on absorption lines in AGNs are required to confirm this agreement.

\section{Conclusion}
\label{conc}

We estimate helium and oxygen abundances (relative to hydrogen) in the Narrow Line Regions (NLRs) of 84 Active Galactic Nuclei (AGNs) spanning two redshift ranges: $z\: < \:0.2$ and $2.8 \: \lesssim \: z \: \lesssim \: 6.8$.    For this purpose, optical $ [3000 \: < \: \lambda (\angstrom) \: < \: 7000]$ narrow (Full Width at Half Maximum lower than $1000 \: \rm km \: s^{-1}$) emission-line fluxes available in the JADES DR4 survey and in the literature were used to determine He/H and O/H via direct estimations of the electron temperature  ($T_{\rm e}$-method) and through strong-line methods. The main results of our study are summarised below: 

\begin{itemize}
\item[--] We find that the ionization degree in AGNs increases toward higher redshifts in a  manner similar to that observed in star-forming galaxies (e.g. \citealt{2013ApJ...774L..10K}).

\item[--] We have identified a population of AGNs at $z > 2.8$ with super-solar helium abundances, which we refer to as helium-loud AGNs. Assuming a low electron-density regime ($N_{\rm e} = 500 \: {\rm cm}^{-3}$), the highest helium abundance to date derived via the $T_{\rm e}$-method was estimated in the JADES object goods-s-mediumhst-58850 ($z=6.2615$). For this target, we derive $12+\log({\rm He/H}) = 11.64$, corresponding to $\rm (He/He_{\odot}) \sim 4.4$.

\item[--]  If clouds with $N_{\rm e}=10\:000 \: \rm cm^{-3}$ are present in the NLRs, the abundances still reach $12+\log({\rm He/H}) \sim 11.54$ [$\rm (He/He_{\odot}) \sim 3.5$].

\item[--] We found evidence for a decline in the He/H abundance 
($\sim0.04$ dex per redshift unit) toward lower redshifts. 

\item[--] We found evidence for an increase in O/H toward lower redshifts, at a rate of $\sim 0.06$ dex per unit redshift.

\item[--]  The use of empirical calibrations based on strong-emission lines provides additional support for the cosmic evolution of helium and oxygen abundances in type~2 AGNs derived via the $T_{\rm e}$-method.

\end{itemize}

\section*{Acknowledgements}
OLD is grateful to Funda\c cão de Amparo à
Pesquisa do Estado de São Paulo (FAPESP) and Conselho Nacional
de Desenvolvimento Científico e Tecnológico (CNPq). MA gratefully acknowledges support from Fundação de Amparo à Pesquisa do Estado de São Paulo (FAPESP, Processo:	
2024/03727-3). RAR  acknowledges the support from the Conselho Nacional de Desenvolvimento Científico e Tecnológico (CNPq; Projects 303450/2022-3,  and 403398/2023-1),  the Coordenação de Aperfeiçoamento de Pessoal de Nível Superior (CAPES; Project 88887.894973/2023-00), and Fundação de Amparo à Pesquisa do Estado do Rio Grande do Sul (FAPERGS).
LNM received support from the CAPES scholarship.
\section*{Data Availability}

The data underlying this article will be shared on reasonable request
to the corresponding author.

\bibliographystyle{mnras}
\bibliography{refs} 

@ARTICLE{2021A&A...652C...4P,
       author = {{Planck Collaboration} and {Aghanim}, N. and {Akrami}, Y. and {Ashdown}, M. and {Aumont}, J. and {Baccigalupi}, C. and {Ballardini}, M. and {Banday}, A.~J. and {Barreiro}, R.~B. and {Bartolo}, N. and {Basak}, S. and {Battye}, R. and {Benabed}, K. and {Bernard}, J. -P. and {Bersanelli}, M. and {Bielewicz}, P. and {Bock}, J.~J. and {Bond}, J.~R. and {Borrill}, J. and {Bouchet}, F.~R. and {Boulanger}, F. and {Bucher}, M. and {Burigana}, C. and {Butler}, R.~C. and {Calabrese}, E. and {Cardoso}, J. -F. and {Carron}, J. and {Challinor}, A. and {Chiang}, H.~C. and {Chluba}, J. and {Colombo}, L.~P.~L. and {Combet}, C. and {Contreras}, D. and {Crill}, B.~P. and {Cuttaia}, F. and {de Bernardis}, P. and {de Zotti}, G. and {Delabrouille}, J. and {Delouis}, J. -M. and {Di Valentino}, E. and {Diego}, J.~M. and {Dor{\'e}}, O. and {Douspis}, M. and {Ducout}, A. and {Dupac}, X. and {Dusini}, S. and {Efstathiou}, G. and {Elsner}, F. and {En{\ss}lin}, T.~A. and {Eriksen}, H.~K. and {Fantaye}, Y. and {Farhang}, M. and {Fergusson}, J. and {Fernandez-Cobos}, R. and {Finelli}, F. and {Forastieri}, F. and {Frailis}, M. and {Fraisse}, A.~A. and {Franceschi}, E. and {Frolov}, A. and {Galeotta}, S. and {Galli}, S. and {Ganga}, K. and {G{\'e}nova-Santos}, R.~T. and {Gerbino}, M. and {Ghosh}, T. and {Gonz{\'a}lez-Nuevo}, J. and {G{\'o}rski}, K.~M. and {Gratton}, S. and {Gruppuso}, A. and {Gudmundsson}, J.~E. and {Hamann}, J. and {Handley}, W. and {Hansen}, F.~K. and {Herranz}, D. and {Hildebrandt}, S.~R. and {Hivon}, E. and {Huang}, Z. and {Jaffe}, A.~H. and {Jones}, W.~C. and {Karakci}, A. and {Keih{\"a}nen}, E. and {Keskitalo}, R. and {Kiiveri}, K. and {Kim}, J. and {Kisner}, T.~S. and {Knox}, L. and {Krachmalnicoff}, N. and {Kunz}, M. and {Kurki-Suonio}, H. and {Lagache}, G. and {Lamarre}, J. -M. and {Lasenby}, A. and {Lattanzi}, M. and {Lawrence}, C.~R. and {Le Jeune}, M. and {Lemos}, P. and {Lesgourgues}, J. and {Levrier}, F. and {Lewis}, A. and {Liguori}, M. and {Lilje}, P.~B. and {Lilley}, M. and {Lindholm}, V. and {L{\'o}pez-Caniego}, M. and {Lubin}, P.~M. and {Ma}, Y. -Z. and {Mac{\'\i}as-P{\'e}rez}, J.~F. and {Maggio}, G. and {Maino}, D. and {Mandolesi}, N. and {Mangilli}, A. and {Marcos-Caballero}, A. and {Maris}, M. and {Martin}, P.~G. and {Martinelli}, M. and {Mart{\'\i}nez-Gonz{\'a}lez}, E. and {Matarrese}, S. and {Mauri}, N. and {McEwen}, J.~D. and {Meinhold}, P.~R. and {Melchiorri}, A. and {Mennella}, A. and {Migliaccio}, M. and {Millea}, M. and {Mitra}, S. and {Miville-Desch{\^e}nes}, M. -A. and {Molinari}, D. and {Montier}, L. and {Morgante}, G. and {Moss}, A. and {Natoli}, P. and {N{\o}rgaard-Nielsen}, H.~U. and {Pagano}, L. and {Paoletti}, D. and {Partridge}, B. and {Patanchon}, G. and {Peiris}, H.~V. and {Perrotta}, F. and {Pettorino}, V. and {Piacentini}, F. and {Polastri}, L. and {Polenta}, G. and {Puget}, J. -L. and {Rachen}, J.~P. and {Reinecke}, M. and {Remazeilles}, M. and {Renzi}, A. and {Rocha}, G. and {Rosset}, C. and {Roudier}, G. and {Rubi{\~n}o-Mart{\'\i}n}, J.~A. and {Ruiz-Granados}, B. and {Salvati}, L. and {Sandri}, M. and {Savelainen}, M. and {Scott}, D. and {Shellard}, E.~P.~S. and {Sirignano}, C. and {Sirri}, G. and {Spencer}, L.~D. and {Sunyaev}, R. and {Suur-Uski}, A. -S. and {Tauber}, J.~A. and {Tavagnacco}, D. and {Tenti}, M. and {Toffolatti}, L. and {Tomasi}, M. and {Trombetti}, T. and {Valenziano}, L. and {Valiviita}, J. and {Van Tent}, B. and {Vibert}, L. and {Vielva}, P. and {Villa}, F. and {Vittorio}, N. and {Wandelt}, B.~D. and {Wehus}, I.~K. and {White}, M. and {White}, S.~D.~M. and {Zacchei}, A. and {Zonca}, A.},
        title = "{Planck 2018 results. VI. Cosmological parameters (Corrigendum)}",
      journal = {\aap},
         year = 2021,
        month = aug,
       volume = {652},
          eid = {C4},
        pages = {C4},
          doi = {10.1051/0004-6361/201833910e},
       adsurl = {https://ui.adsabs.harvard.edu/abs/2021A&A...652C...4P}
}

@ARTICLE{2017PASA...34...58E,
       author = {{Eldridge}, J.~J. and {Stanway}, E.~R. and {Xiao}, L. and {McClelland}, L.~A.~S. and {Taylor}, G. and {Ng}, M. and {Greis}, S.~M.~L. and {Bray}, J.~C.},
        title = "{Binary Population and Spectral Synthesis Version 2.1: Construction, Observational Verification, and New Results}",
      journal = {\pasa},
         year = 2017,
        month = nov,
       volume = {34},
          eid = {e058},
        pages = {e058},
          doi = {10.1017/pasa.2017.51},
archivePrefix = {arXiv},
       eprint = {1710.02154},
 primaryClass = {astro-ph.SR},
       adsurl = {https://ui.adsabs.harvard.edu/abs/2017PASA...34...58E}
}

@ARTICLE{2019ApJ...878....2D,
       author = {{D'Agostino}, Joshua J. and {Kewley}, Lisa J. and {Groves}, Brent and {Byler}, Nell and {Sutherland}, Ralph S. and {Nicholls}, David and {Leitherer}, Claus and {Stanway}, Elizabeth R.},
        title = "{Comparison of Theoretical Starburst Photoionization Models for Optical Diagnostics}",
      journal = {\apj},
         year = 2019,
        month = jun,
       volume = {878},
       number = {1},
          eid = {2},
        pages = {2},
          doi = {10.3847/1538-4357/ab1d5e},
archivePrefix = {arXiv},
       eprint = {1905.09528},
 primaryClass = {astro-ph.GA},
       adsurl = {https://ui.adsabs.harvard.edu/abs/2019ApJ...878....2D}
}

@ARTICLE{2020MNRAS.492.5675C,
       author = {{Carvalho}, S.~P. and {Dors}, O.~L. and {Cardaci}, M.~V. and {H{\"a}gele}, G.~F. and {Krabbe}, A.~C. and {P{\'e}rez-Montero}, E. and {Monteiro}, A.~F. and {Armah}, M. and {Freitas-Lemes}, P.},
        title = "{Chemical abundances of Seyfert 2 AGNs - II. N2 metallicity calibration based on SDSS}",
      journal = {\mnras},
         year = 2020,
        month = mar,
       volume = {492},
       number = {4},
        pages = {5675-5683},
          doi = {10.1093/mnras/staa193},
archivePrefix = {arXiv},
       eprint = {2001.07126},
 primaryClass = {astro-ph.GA},
       adsurl = {https://ui.adsabs.harvard.edu/abs/2020MNRAS.492.5675C}
}

@ARTICLE{1973ApJ...179..343W,
       author = {{Wagoner}, Robert V.},
        title = "{Big-Bang Nucleosynthesis Revisited}",
      journal = {\apj},
         year = 1973,
        month = jan,
       volume = {179},
        pages = {343-360},
          doi = {10.1086/151873},
       adsurl = {https://ui.adsabs.harvard.edu/abs/1973ApJ...179..343W}
}

@ARTICLE{1966ApJ...146..542P,
       author = {{Peebles}, P.~J.~E.},
        title = "{Primordial Helium Abundance and the Primordial Fireball. II}",
      journal = {\apj},
         year = 1966,
        month = nov,
       volume = {146},
        pages = {542},
          doi = {10.1086/148918},
       adsurl = {https://ui.adsabs.harvard.edu/abs/1966ApJ...146..542P}
}

@ARTICLE{1967ApJ...148....3W,
       author = {{Wagoner}, Robert V. and {Fowler}, William A. and {Hoyle}, F.},
        title = "{On the Synthesis of Elements at Very High Temperatures}",
      journal = {\apj},
         year = 1967,
        month = apr,
       volume = {148},
        pages = {3},
          doi = {10.1086/149126},
       adsurl = {https://ui.adsabs.harvard.edu/abs/1967ApJ...148....3W}
}

@ARTICLE{1974ApJ...193..327P,
       author = {{Peimbert}, M. and {Torres-Peimbert}, S.},
        title = "{Chemical composition of H II regions in the Large Magellanic Cloud and its cosmological implications.}",
      journal = {\apj},
         year = 1974,
        month = oct,
       volume = {193},
        pages = {327-333},
          doi = {10.1086/153166},
       adsurl = {https://ui.adsabs.harvard.edu/abs/1974ApJ...193..327P}
}

@ARTICLE{2025MNRAS.538.1517W,
       author = {{Weller}, Miqaela K. and {Weinberg}, David H. and {Johnson}, James W.},
        title = "{Modelling the Galactic Chemical Evolution of Helium}",
      journal = {\mnras},
         year = 2025,
        month = apr,
       volume = {538},
       number = {3},
        pages = {1517-1534},
          doi = {10.1093/mnras/staf373},
archivePrefix = {arXiv},
       eprint = {2404.08765},
 primaryClass = {astro-ph.GA},
       adsurl = {https://ui.adsabs.harvard.edu/abs/2025MNRAS.538.1517W}
}

@ARTICLE{1992MNRAS.255..325P,
       author = {{Pagel}, B.~E.~J. and {Simonson}, E.~A. and {Terlevich}, R.~J. and {Edmunds}, M.~G.},
        title = "{The primordial helium abundance from observations of extragalactic HII regions.}",
      journal = {\mnras},
         year = 1992,
        month = mar,
       volume = {255},
        pages = {325-345},
          doi = {10.1093/mnras/255.2.325},
       adsurl = {https://ui.adsabs.harvard.edu/abs/1992MNRAS.255..325P}
}

@ARTICLE{2019ApJ...876...98V,
       author = {{Valerdi}, Mabel and {Peimbert}, Antonio and {Peimbert}, Manuel and {Sixtos}, Andr{\'e}s},
        title = "{Determination of the Primordial Helium Abundance Based on NGC 346, an H II Region of the Small Magellanic Cloud}",
      journal = {\apj},
         year = 2019,
        month = may,
       volume = {876},
       number = {2},
          eid = {98},
        pages = {98},
          doi = {10.3847/1538-4357/ab14e4},
archivePrefix = {arXiv},
       eprint = {1904.01594},
 primaryClass = {astro-ph.GA},
       adsurl = {https://ui.adsabs.harvard.edu/abs/2019ApJ...876...98V}
}

@ARTICLE{2021ApJ...922..170B,
       author = {{Berg}, Danielle A. and {Chisholm}, John and {Erb}, Dawn K. and {Skillman}, Evan D. and {Pogge}, Richard W. and {Olivier}, Grace M.},
        title = "{Characterizing Extreme Emission-line Galaxies. I. A Four-zone Ionization Model for Very High-ionization Emission}",
      journal = {\apj},
         year = 2021,
        month = dec,
       volume = {922},
       number = {2},
          eid = {170},
        pages = {170},
          doi = {10.3847/1538-4357/ac141b},
archivePrefix = {arXiv},
       eprint = {2105.12765},
 primaryClass = {astro-ph.GA},
       adsurl = {https://ui.adsabs.harvard.edu/abs/2021ApJ...922..170B}
}

@ARTICLE{2024ApJ...974..266Y,
       author = {{Yanagisawa}, Hiroto and {Ouchi}, Masami and {Watanabe}, Kuria and {Matsumoto}, Akinori and {Nakajima}, Kimihiko and {Yajima}, Hidenobu and {Nagamine}, Kentaro and {Takahashi}, Koh and {Nakane}, Minami and {Tominaga}, Nozomu and {Umeda}, Hiroya and {Fukushima}, Hajime and {Harikane}, Yuichi and {Isobe}, Yuki and {Ono}, Yoshiaki and {Xu}, Yi and {Zhang}, Yechi},
        title = "{Strong He I Emission Lines in High N/O Galaxies at z {\ensuremath{\sim}} 6 Identified in JWST Spectra: High He/H Abundance Ratios or High Electron Densities?}",
      journal = {\apj},
         year = 2024,
        month = oct,
       volume = {974},
       number = {2},
          eid = {266},
        pages = {266},
          doi = {10.3847/1538-4357/ad72ec},
archivePrefix = {arXiv},
       eprint = {2405.01823},
 primaryClass = {astro-ph.GA},
       adsurl = {https://ui.adsabs.harvard.edu/abs/2024ApJ...974..266Y}
}

@ARTICLE{2020ApJ...896...77H,
       author = {{Hsyu}, Tiffany and {Cooke}, Ryan J. and {Prochaska}, J. Xavier and {Bolte}, Michael},
        title = "{The PHLEK Survey: A New Determination of the Primordial Helium Abundance}",
      journal = {\apj},
         year = 2020,
        month = jun,
       volume = {896},
       number = {1},
          eid = {77},
        pages = {77},
          doi = {10.3847/1538-4357/ab91af},
archivePrefix = {arXiv},
       eprint = {2005.12290},
 primaryClass = {astro-ph.GA},
       adsurl = {https://ui.adsabs.harvard.edu/abs/2020ApJ...896...77H}
}

@INPROCEEDINGS{2010IAUS..268..163F,
       author = {{Ferland}, Gary J. and {Izotov}, Yuri and {Peimbert}, Antonio and {Peimbert}, Manuel and {Porter}, Ryan L. and {Skillman}, Evan and {Steigman}, Gary},
        title = "{What is $^{4}$He from H II regions? What needs to be done to better understand the systematic errors?}",
    booktitle = {Light Elements in the Universe},
         year = 2010,
       editor = {{Charbonnel}, Corinne and {Tosi}, Monica and {Primas}, Francesca and {Chiappini}, Cristina},
       series = {IAU Symposium},
       volume = {268},
        month = apr,
        pages = {163-167},
          doi = {10.1017/S1743921310004011},
       adsurl = {https://ui.adsabs.harvard.edu/abs/2010IAUS..268..163F}
}

@ARTICLE{2020MNRAS.496.2726M,
       author = {{M{\'e}ndez-Delgado}, J.~E. and {Esteban}, C. and {Garc{\'\i}a-Rojas}, J. and {Arellano-C{\'o}rdova}, K.~Z. and {Valerdi}, M.},
        title = "{Helium abundances and its radial gradient from the spectra of H II regions and ring nebulae of the Milky Way}",
      journal = {\mnras},
         year = 2020,
        month = aug,
       volume = {496},
       number = {3},
        pages = {2726-2742},
          doi = {10.1093/mnras/staa1705},
archivePrefix = {arXiv},
       eprint = {2006.06577},
 primaryClass = {astro-ph.SR},
       adsurl = {https://ui.adsabs.harvard.edu/abs/2020MNRAS.496.2726M}
}

@ARTICLE{2021MNRAS.507..466D,
       author = {{Dors}, Oli L.},
        title = "{Chemical abundances in Seyfert galaxies - VI. Empirical abundance calibration}",
      journal = {\mnras},
         year = 2021,
        month = oct,
       volume = {507},
       number = {1},
        pages = {466-474},
          doi = {10.1093/mnras/stab2166},
       adsurl = {https://ui.adsabs.harvard.edu/abs/2021MNRAS.507..466D}
}

@ARTICLE{2024MNRAS.533L...1D,
       author = {{Dors}, O.~L. and {Almeida}, G.~C. and {Oliveira}, C.~B. and {Flury}, S.~R. and {Riffel}, R. and {Riffel}, R.~A. and {Cardaci}, M.~V. and {H{\"a}gele}, G.~F. and {Ilha}, G.~S. and {Krabbe}, A.~C. and {Arellano-C{\'o}rdova}, K.~Z. and {Santos}, P.~C. and {Morais}, I.~N.},
        title = "{Emirical calibration for helium abundance determinations in active galactic nuclei}",
      journal = {\mnras},
         year = 2024,
        month = sep,
       volume = {533},
       number = {1},
        pages = {L1-L5},
          doi = {10.1093/mnrasl/slae052},
archivePrefix = {arXiv},
       eprint = {2406.03259},
 primaryClass = {astro-ph.GA},
       adsurl = {https://ui.adsabs.harvard.edu/abs/2024MNRAS.533L...1D}
}

@ARTICLE{1978ApJ...223...56K,
       author = {{Koski}, A.~T.},
        title = "{Spectrophotometry of Seyfert 2 galaxies and narrow-line radio galaxies.}",
      journal = {\apj},
         year = 1978,
        month = jul,
       volume = {223},
        pages = {56-73},
          doi = {10.1086/156235},
       adsurl = {https://ui.adsabs.harvard.edu/abs/1978ApJ...223...56K}
}

@ARTICLE{2015ApJS..217...12D,
       author = {{Dopita}, Michael A. and {Shastri}, Prajval and {Davies}, Rebecca and {Kewley}, Lisa and {Hampton}, Elise and {Scharw{\"a}chter}, Julia and {Sutherland}, Ralph and {Kharb}, Preeti and {Jose}, Jessy and {Bhatt}, Harish and {Ramya}, S. and {Jin}, Chichuan and {Banfield}, Julie and {Zaw}, Ingyin and {Juneau}, St{\'e}phanie and {James}, Bethan and {Srivastava}, Shweta},
        title = "{Probing the Physics of Narrow Line Regions in Active Galaxies. II. The Siding Spring Southern Seyfert Spectroscopic Snapshot Survey (S7)}",
      journal = {\apjs},
         year = 2015,
        month = mar,
       volume = {217},
       number = {1},
          eid = {12},
        pages = {12},
          doi = {10.1088/0067-0049/217/1/12},
archivePrefix = {arXiv},
       eprint = {1501.02022},
 primaryClass = {astro-ph.GA},
       adsurl = {https://ui.adsabs.harvard.edu/abs/2015ApJS..217...12D}
}

@ARTICLE{2000AJ....120.1579Y,
       author = {{York}, Donald G. and {Adelman}, J. and {Anderson}, Jr., John E. and {Anderson}, Scott F. and {Annis}, James and {Bahcall}, Neta A. and {Bakken}, J.~A. and {Barkhouser}, Robert and {Bastian}, Steven and {Berman}, Eileen and {Boroski}, William N. and {Bracker}, Steve and {Briegel}, Charlie and {Briggs}, John W. and {Brinkmann}, J. and {Brunner}, Robert and {Burles}, Scott and {Carey}, Larry and {Carr}, Michael A. and {Castander}, Francisco J. and {Chen}, Bing and {Colestock}, Patrick L. and {Connolly}, A.~J. and {Crocker}, J.~H. and {Csabai}, Istv{\'a}n and {Czarapata}, Paul C. and {Davis}, John Eric and {Doi}, Mamoru and {Dombeck}, Tom and {Eisenstein}, Daniel and {Ellman}, Nancy and {Elms}, Brian R. and {Evans}, Michael L. and {Fan}, Xiaohui and {Federwitz}, Glenn R. and {Fiscelli}, Larry and {Friedman}, Scott and {Frieman}, Joshua A. and {Fukugita}, Masataka and {Gillespie}, Bruce and {Gunn}, James E. and {Gurbani}, Vijay K. and {de Haas}, Ernst and {Haldeman}, Merle and {Harris}, Frederick H. and {Hayes}, J. and {Heckman}, Timothy M. and {Hennessy}, G.~S. and {Hindsley}, Robert B. and {Holm}, Scott and {Holmgren}, Donald J. and {Huang}, Chi-hao and {Hull}, Charles and {Husby}, Don and {Ichikawa}, Shin-Ichi and {Ichikawa}, Takashi and {Ivezi{\'c}}, {\v{Z}}eljko and {Kent}, Stephen and {Kim}, Rita S.~J. and {Kinney}, E. and {Klaene}, Mark and {Kleinman}, A.~N. and {Kleinman}, S. and {Knapp}, G.~R. and {Korienek}, John and {Kron}, Richard G. and {Kunszt}, Peter Z. and {Lamb}, D.~Q. and {Lee}, B. and {Leger}, R. French and {Limmongkol}, Siriluk and {Lindenmeyer}, Carl and {Long}, Daniel C. and {Loomis}, Craig and {Loveday}, Jon and {Lucinio}, Rich and {Lupton}, Robert H. and {MacKinnon}, Bryan and {Mannery}, Edward J. and {Mantsch}, P.~M. and {Margon}, Bruce and {McGehee}, Peregrine and {McKay}, Timothy A. and {Meiksin}, Avery and {Merelli}, Aronne and {Monet}, David G. and {Munn}, Jeffrey A. and {Narayanan}, Vijay K. and {Nash}, Thomas and {Neilsen}, Eric and {Neswold}, Rich and {Newberg}, Heidi Jo and {Nichol}, R.~C. and {Nicinski}, Tom and {Nonino}, Mario and {Okada}, Norio and {Okamura}, Sadanori and {Ostriker}, Jeremiah P. and {Owen}, Russell and {Pauls}, A. George and {Peoples}, John and {Peterson}, R.~L. and {Petravick}, Donald and {Pier}, Jeffrey R. and {Pope}, Adrian and {Pordes}, Ruth and {Prosapio}, Angela and {Rechenmacher}, Ron and {Quinn}, Thomas R. and {Richards}, Gordon T. and {Richmond}, Michael W. and {Rivetta}, Claudio H. and {Rockosi}, Constance M. and {Ruthmansdorfer}, Kurt and {Sandford}, Dale and {Schlegel}, David J. and {Schneider}, Donald P. and {Sekiguchi}, Maki and {Sergey}, Gary and {Shimasaku}, Kazuhiro and {Siegmund}, Walter A. and {Smee}, Stephen and {Smith}, J. Allyn and {Snedden}, S. and {Stone}, R. and {Stoughton}, Chris and {Strauss}, Michael A. and {Stubbs}, Christopher and {SubbaRao}, Mark and {Szalay}, Alexander S. and {Szapudi}, Istvan and {Szokoly}, Gyula P. and {Thakar}, Anirudda R. and {Tremonti}, Christy and {Tucker}, Douglas L. and {Uomoto}, Alan and {Vanden Berk}, Dan and {Vogeley}, Michael S. and {Waddell}, Patrick and {Wang}, Shu-i. and {Watanabe}, Masaru and {Weinberg}, David H. and {Yanny}, Brian and {Yasuda}, Naoki and {SDSS Collaboration}},
        title = "{The Sloan Digital Sky Survey: Technical Summary}",
      journal = {\aj},
         year = 2000,
        month = sep,
       volume = {120},
       number = {3},
        pages = {1579-1587},
          doi = {10.1086/301513},
archivePrefix = {arXiv},
       eprint = {astro-ph/0006396},
 primaryClass = {astro-ph},
       adsurl = {https://ui.adsabs.harvard.edu/abs/2000AJ....120.1579Y}
}

@ARTICLE{2024ApJ...977..187Z,
       author = {{Zhu}, Peixin and {Kewley}, Lisa J. and {Sutherland}, Ralph S.},
        title = "{Theoretical Diagnostics for Narrow-line Regions of Active Galactic Nuclei}",
      journal = {\apj},
         year = 2024,
        month = dec,
       volume = {977},
       number = {2},
          eid = {187},
        pages = {187},
          doi = {10.3847/1538-4357/ad8f37},
archivePrefix = {arXiv},
       eprint = {2411.04103},
 primaryClass = {astro-ph.GA},
       adsurl = {https://ui.adsabs.harvard.edu/abs/2024ApJ...977..187Z}
}

@ARTICLE{2000MNRAS.311..329D,
       author = {{Deharveng}, L. and {Pe{\~n}a}, M. and {Caplan}, J. and {Costero}, R.},
        title = "{Oxygen and helium abundances in Galactic Hii regions - II. Abundance gradients}",
      journal = {\mnras},
         year = 2000,
        month = jan,
       volume = {311},
       number = {2},
        pages = {329-345},
          doi = {10.1046/j.1365-8711.2000.03030.x},
       adsurl = {https://ui.adsabs.harvard.edu/abs/2000MNRAS.311..329D}
}

@ARTICLE{1992RMxAA..24..155P,
       author = {{Peimbert}, M. and {Torres-Peimbert}, S. and {Ruiz}, M.~T.},
        title = "{The chemical composition of the galactic H II region M17.}",
      journal = {\rmxaa},
         year = 1992,
        month = oct,
       volume = {24},
        pages = {155-177},
       adsurl = {https://ui.adsabs.harvard.edu/abs/1992RMxAA..24..155P}
}

@ARTICLE{1986PASP...98.1061P,
       author = {{Pena}, Miriam},
        title = "{The ionization structure of helium in H II region complexes.}",
      journal = {\pasp},
         year = 1986,
        month = oct,
       volume = {98},
        pages = {1061-1065},
          doi = {10.1086/131873},
       adsurl = {https://ui.adsabs.harvard.edu/abs/1986PASP...98.1061P}
}

@INPROCEEDINGS{2010IAUS..265..147E,
       author = {{Erb}, Dawn K.},
        title = "{Chemical Abundances in Star-Forming Galaxies at High Redshift}",
    booktitle = {Chemical Abundances in the Universe: Connecting First Stars to Planets},
         year = 2010,
       editor = {{Cunha}, Katia and {Spite}, Monique and {Barbuy}, Beatriz},
       series = {IAU Symposium},
       volume = {265},
        month = mar,
        pages = {147-154},
          doi = {10.1017/S1743921310000438},
archivePrefix = {arXiv},
       eprint = {0912.0313},
 primaryClass = {astro-ph.CO},
       adsurl = {https://ui.adsabs.harvard.edu/abs/2010IAUS..265..147E}
}

@ARTICLE{2021MNRAS.507...74R,
       author = {{Ruschel-Dutra}, D. and {Storchi-Bergmann}, T. and {Schnorr-M{\"u}ller}, A. and {Riffel}, R.~A. and {Dall'Agnol de Oliveira}, B. and {Lena}, D. and {Robinson}, A. and {Nagar}, N. and {Elvis}, M.},
        title = "{AGNIFS survey of local AGN: GMOS-IFU data and outflows in 30 sources}",
      journal = {\mnras},
         year = 2021,
        month = oct,
       volume = {507},
       number = {1},
        pages = {74-89},
          doi = {10.1093/mnras/stab2058},
archivePrefix = {arXiv},
       eprint = {2107.07635},
 primaryClass = {astro-ph.GA},
       adsurl = {https://ui.adsabs.harvard.edu/abs/2021MNRAS.507...74R}
}

@ARTICLE{2023MNRAS.520.1687A,
       author = {{Armah}, Mark and {Riffel}, Rog{\'e}rio and {Dors}, O.~L. and {Oh}, Kyuseok and {Koss}, Michael J. and {Ricci}, Claudio and {Trakhtenbrot}, Benny and {Valerdi}, Mabel and {Riffel}, Rogemar A. and {Krabbe}, Angela C.},
        title = "{Oxygen abundances in the narrow line regions of Seyfert galaxies and the metallicity-luminosity relation}",
      journal = {\mnras},
         year = 2023,
        month = apr,
       volume = {520},
       number = {2},
        pages = {1687-1703},
          doi = {10.1093/mnras/stad217},
archivePrefix = {arXiv},
       eprint = {2301.07596},
 primaryClass = {astro-ph.GA},
       adsurl = {https://ui.adsabs.harvard.edu/abs/2023MNRAS.520.1687A}
}

@ARTICLE{2009A&A...503..721M,
       author = {{Matsuoka}, K. and {Nagao}, T. and {Maiolino}, R. and {Marconi}, A. and {Taniguchi}, Y.},
        title = "{Chemical evolution of high-redshift radio galaxies}",
      journal = {\aap},
         year = 2009,
        month = sep,
       volume = {503},
       number = {3},
        pages = {721-730},
          doi = {10.1051/0004-6361/200811478},
archivePrefix = {arXiv},
       eprint = {0905.1581},
 primaryClass = {astro-ph.CO},
       adsurl = {https://ui.adsabs.harvard.edu/abs/2009A&A...503..721M}
}

@ARTICLE{2012MNRAS.421.1043S,
       author = {{Shirazi}, Maryam and {Brinchmann}, Jarle},
        title = "{Strongly star forming galaxies in the local Universe with nebular He II{\ensuremath{\lambda}}4686 emission}",
      journal = {\mnras},
         year = 2012,
        month = apr,
       volume = {421},
       number = {2},
        pages = {1043-1063},
          doi = {10.1111/j.1365-2966.2012.20439.x},
archivePrefix = {arXiv},
       eprint = {1201.1290},
 primaryClass = {astro-ph.CO},
       adsurl = {https://ui.adsabs.harvard.edu/abs/2012MNRAS.421.1043S}
}

@ARTICLE{2024MNRAS.535..881J,
       author = {{Ji}, Xihan and {{\"U}bler}, Hannah and {Maiolino}, Roberto and {D'Eugenio}, Francesco and {Arribas}, Santiago and {Bunker}, Andrew J. and {Charlot}, St{\'e}phane and {Perna}, Michele and {Rodr{\'\i}guez Del Pino}, Bruno and {B{\"o}ker}, Torsten and {Cresci}, Giovanni and {Curti}, Mirko and {Kumari}, Nimisha and {Lamperti}, Isabella},
        title = "{GA-NIFS: an extremely nitrogen-loud and chemically stratified galaxy at z   5.55}",
      journal = {\mnras},
         year = 2024,
        month = nov,
       volume = {535},
       number = {1},
        pages = {881-908},
          doi = {10.1093/mnras/stae2375},
archivePrefix = {arXiv},
       eprint = {2404.04148},
 primaryClass = {astro-ph.GA},
       adsurl = {https://ui.adsabs.harvard.edu/abs/2024MNRAS.535..881J}
}

@ARTICLE{2009ApJS..182..543A,
       author = {{Abazajian}, Kevork N. and {Adelman-McCarthy}, Jennifer K. and {Ag{\"u}eros}, Marcel A. and {Allam}, Sahar S. and {Allende Prieto}, Carlos and {An}, Deokkeun and {Anderson}, Kurt S.~J. and {Anderson}, Scott F. and {Annis}, James and {Bahcall}, Neta A. and {Bailer-Jones}, C.~A.~L. and {Barentine}, J.~C. and {Bassett}, Bruce A. and {Becker}, Andrew C. and {Beers}, Timothy C. and {Bell}, Eric F. and {Belokurov}, Vasily and {Berlind}, Andreas A. and {Berman}, Eileen F. and {Bernardi}, Mariangela and {Bickerton}, Steven J. and {Bizyaev}, Dmitry and {Blakeslee}, John P. and {Blanton}, Michael R. and {Bochanski}, John J. and {Boroski}, William N. and {Brewington}, Howard J. and {Brinchmann}, Jarle and {Brinkmann}, J. and {Brunner}, Robert J. and {Budav{\'a}ri}, Tam{\'a}s and {Carey}, Larry N. and {Carliles}, Samuel and {Carr}, Michael A. and {Castander}, Francisco J. and {Cinabro}, David and {Connolly}, A.~J. and {Csabai}, Istv{\'a}n and {Cunha}, Carlos E. and {Czarapata}, Paul C. and {Davenport}, James R.~A. and {de Haas}, Ernst and {Dilday}, Ben and {Doi}, Mamoru and {Eisenstein}, Daniel J. and {Evans}, Michael L. and {Evans}, N.~W. and {Fan}, Xiaohui and {Friedman}, Scott D. and {Frieman}, Joshua A. and {Fukugita}, Masataka and {G{\"a}nsicke}, Boris T. and {Gates}, Evalyn and {Gillespie}, Bruce and {Gilmore}, G. and {Gonzalez}, Belinda and {Gonzalez}, Carlos F. and {Grebel}, Eva K. and {Gunn}, James E. and {Gy{\"o}ry}, Zsuzsanna and {Hall}, Patrick B. and {Harding}, Paul and {Harris}, Frederick H. and {Harvanek}, Michael and {Hawley}, Suzanne L. and {Hayes}, Jeffrey J.~E. and {Heckman}, Timothy M. and {Hendry}, John S. and {Hennessy}, Gregory S. and {Hindsley}, Robert B. and {Hoblitt}, J. and {Hogan}, Craig J. and {Hogg}, David W. and {Holtzman}, Jon A. and {Hyde}, Joseph B. and {Ichikawa}, Shin-ichi and {Ichikawa}, Takashi and {Im}, Myungshin and {Ivezi{\'c}}, {\v{Z}}eljko and {Jester}, Sebastian and {Jiang}, Linhua and {Johnson}, Jennifer A. and {Jorgensen}, Anders M. and {Juri{\'c}}, Mario and {Kent}, Stephen M. and {Kessler}, R. and {Kleinman}, S.~J. and {Knapp}, G.~R. and {Konishi}, Kohki and {Kron}, Richard G. and {Krzesinski}, Jurek and {Kuropatkin}, Nikolay and {Lampeitl}, Hubert and {Lebedeva}, Svetlana and {Lee}, Myung Gyoon and {Lee}, Young Sun and {French Leger}, R. and {L{\'e}pine}, S{\'e}bastien and {Li}, Nolan and {Lima}, Marcos and {Lin}, Huan and {Long}, Daniel C. and {Loomis}, Craig P. and {Loveday}, Jon and {Lupton}, Robert H. and {Magnier}, Eugene and {Malanushenko}, Olena and {Malanushenko}, Viktor and {Mandelbaum}, Rachel and {Margon}, Bruce and {Marriner}, John P. and {Mart{\'\i}nez-Delgado}, David and {Matsubara}, Takahiko and {McGehee}, Peregrine M. and {McKay}, Timothy A. and {Meiksin}, Avery and {Morrison}, Heather L. and {Mullally}, Fergal and {Munn}, Jeffrey A. and {Murphy}, Tara and {Nash}, Thomas and {Nebot}, Ada and {Neilsen}, Jr., Eric H. and {Newberg}, Heidi Jo and {Newman}, Peter R. and {Nichol}, Robert C. and {Nicinski}, Tom and {Nieto-Santisteban}, Maria and {Nitta}, Atsuko and {Okamura}, Sadanori and {Oravetz}, Daniel J. and {Ostriker}, Jeremiah P. and {Owen}, Russell and {Padmanabhan}, Nikhil and {Pan}, Kaike and {Park}, Changbom and {Pauls}, George and {Peoples}, Jr., John and {Percival}, Will J. and {Pier}, Jeffrey R. and {Pope}, Adrian C. and {Pourbaix}, Dimitri and {Price}, Paul A. and {Purger}, Norbert and {Quinn}, Thomas and {Raddick}, M. Jordan and {Re Fiorentin}, Paola and {Richards}, Gordon T. and {Richmond}, Michael W. and {Riess}, Adam G. and {Rix}, Hans-Walter and {Rockosi}, Constance M. and {Sako}, Masao and {Schlegel}, David J. and {Schneider}, Donald P. and {Scholz}, Ralf-Dieter and {Schreiber}, Matthias R. and {Schwope}, Axel D. and {Seljak}, Uro{\v{s}} and {Sesar}, Branimir and {Sheldon}, Erin and {Shimasaku}, Kazu and {Sibley}, Valena C. and {Simmons}, A.~E. and {Sivarani}, Thirupathi and {Allyn Smith}, J. and {Smith}, Martin C. and {Smol{\v{c}}i{\'c}}, Vernesa and {Snedden}, Stephanie A. and {Stebbins}, Albert and {Steinmetz}, Matthias and {Stoughton}, Chris and {Strauss}, Michael A. and {SubbaRao}, Mark and {Suto}, Yasushi and {Szalay}, Alexander S. and {Szapudi}, Istv{\'a}n and {Szkody}, Paula and {Tanaka}, Masayuki and {Tegmark}, Max and {Teodoro}, Luis F.~A. and {Thakar}, Aniruddha R. and {Tremonti}, Christy A. and {Tucker}, Douglas L. and {Uomoto}, Alan and {Vanden Berk}, Daniel E. and {Vandenberg}, Jan and {Vidrih}, S. and {Vogeley}, Michael S. and {Voges}, Wolfgang and {Vogt}, Nicole P. and {Wadadekar}, Yogesh and {Watters}, Shannon and {Weinberg}, David H. and {West}, Andrew A. and {White}, Simon D.~M. and {Wilhite}, Brian C. and {Wonders}, Alainna C. and {Yanny}, Brian and {Yocum}, D.~R.},
        title = "{The Seventh Data Release of the Sloan Digital Sky Survey}",
      journal = {\apjs},
         year = 2009,
        month = jun,
       volume = {182},
       number = {2},
        pages = {543-558},
          doi = {10.1088/0067-0049/182/2/543},
archivePrefix = {arXiv},
       eprint = {0812.0649},
 primaryClass = {astro-ph},
       adsurl = {https://ui.adsabs.harvard.edu/abs/2009ApJS..182..543A}
}

@ARTICLE{2025A&A...697A..89C,
       author = {{Curti}, Mirko and {Witstok}, Joris and {Jakobsen}, Peter and {Kobayashi}, Chiaki and {Curtis-Lake}, Emma and {Hainline}, Kevin and {Ji}, Xihan and {D'Eugenio}, Francesco and {Chevallard}, Jacopo and {Maiolino}, Roberto and {Scholtz}, Jan and {Carniani}, Stefano and {Arribas}, Santiago and {Baker}, William M. and {Bhatawdekar}, Rachana and {Boyett}, Kristan and {Bunker}, Andrew J. and {Cameron}, Alex and {Cargile}, Phillip A. and {Charlot}, St{\'e}phane and {Eisenstein}, Daniel J. and {Ji}, Zhiyuan and {Johnson}, Benjamin D. and {Kumari}, Nimisha and {Maseda}, Michael V. and {Robertson}, Brant and {Silcock}, Maddie S. and {Tacchella}, Sandro and {{\"U}bler}, Hannah and {Venturi}, Giacomo and {Williams}, Christina C. and {Willmer}, Christopher N.~A. and {Willott}, Chris},
        title = "{JADES: The star formation and chemical enrichment history of a luminous galaxy at z {\ensuremath{\sim}} 9.43 probed by ultra-deep JWST/NIRSpec spectroscopy}",
      journal = {\aap},
         year = 2025,
        month = may,
       volume = {697},
          eid = {A89},
        pages = {A89},
          doi = {10.1051/0004-6361/202451410},
archivePrefix = {arXiv},
       eprint = {2407.02575},
 primaryClass = {astro-ph.GA},
       adsurl = {https://ui.adsabs.harvard.edu/abs/2025A&A...697A..89C}
}

@ARTICLE{2025ApJ...994L..29Z,
       author = {{Zhu}, Peixin and {Kewley}, Lisa J. and {Hsiao}, Tiger Yu-Yang and {Trussler}, James},
        title = "{Only Nitrogen-enhanced Galaxies Have Detectable Ultraviolet Nitrogen Emission Lines at High Redshift}",
      journal = {\apjl},
         year = 2025,
        month = nov,
       volume = {994},
       number = {1},
          eid = {L29},
        pages = {L29},
          doi = {10.3847/2041-8213/ae1c43},
archivePrefix = {arXiv},
       eprint = {2511.03681},
 primaryClass = {astro-ph.GA},
       adsurl = {https://ui.adsabs.harvard.edu/abs/2025ApJ...994L..29Z}
}

@ARTICLE{2024A&A...687L..11S,
       author = {{Schaerer}, D. and {Marques-Chaves}, R. and {Xiao}, M. and {Korber}, D.},
        title = "{Discovery of a new N-emitter in the epoch of reionization}",
      journal = {\aap},
         year = 2024,
        month = jul,
       volume = {687},
          eid = {L11},
        pages = {L11},
          doi = {10.1051/0004-6361/202450721},
archivePrefix = {arXiv},
       eprint = {2406.08408},
 primaryClass = {astro-ph.GA},
       adsurl = {https://ui.adsabs.harvard.edu/abs/2024A&A...687L..11S}
}

@ARTICLE{2025arXiv250204817Z,
       author = {{Zhang}, Yechi and {Morishita}, Takahiro and {Stiavelli}, Massimo},
        title = "{Potential Nitrogen Enrichment via Direct-Collapse Wolf-Rayet Stars in a $z=4.7$ Star-Forming Galaxy}",
      journal = {arXiv e-prints},
         year = 2025,
        month = feb,
          eid = {arXiv:2502.04817},
        pages = {arXiv:2502.04817},
          doi = {10.48550/arXiv.2502.04817},
archivePrefix = {arXiv},
       eprint = {2502.04817},
 primaryClass = {astro-ph.GA},
       adsurl = {https://ui.adsabs.harvard.edu/abs/2025arXiv250204817Z}
}

@ARTICLE{2025MNRAS.540.2991A,
       author = {{Arellano-C{\'o}rdova}, K.~Z. and {Cullen}, F. and {Carnall}, A.~C. and {Scholte}, D. and {Stanton}, T.~M. and {Kobayashi}, C. and {Martinez}, Z. and {Berg}, D.~A. and {Barrufet}, L. and {Begley}, R. and {Donnan}, C.~T. and {Dunlop}, J.~S. and {Hamadouche}, M.~L. and {McLeod}, D.~J. and {McLure}, R.~J. and {Rowlands}, K. and {Shapley}, A.~E.},
        title = "{The JWST EXCELS survey: direct estimates of C, N, and O abundances in two relatively metal-rich galaxies at z ≃ 5}",
      journal = {\mnras},
         year = 2025,
        month = jul,
       volume = {540},
       number = {4},
        pages = {2991-3007},
          doi = {10.1093/mnras/staf855},
archivePrefix = {arXiv},
       eprint = {2412.10557},
 primaryClass = {astro-ph.GA},
       adsurl = {https://ui.adsabs.harvard.edu/abs/2025MNRAS.540.2991A}
}

@ARTICLE{2025arXiv250801372M,
       author = {{Morishita}, Takahiro and {Stiavelli}, Massimo and {Mason}, Charlotte A. and {Tripodi}, Roberta and {Chiaberge}, Marco and {Schuldt}, Stefan and {Willott}, Chris J. and {Zhang}, Yechi},
        title = "{A Nitrogen-rich AGN Powering a Large Ionizing Bubble at z=8.63}",
      journal = {arXiv e-prints},
         year = 2025,
        month = aug,
          eid = {arXiv:2508.01372},
        pages = {arXiv:2508.01372},
          doi = {10.48550/arXiv.2508.01372},
archivePrefix = {arXiv},
       eprint = {2508.01372},
 primaryClass = {astro-ph.GA},
       adsurl = {https://ui.adsabs.harvard.edu/abs/2025arXiv250801372M}
}

@ARTICLE{2025ApJ...989...75N,
       author = {{Napolitano}, Lorenzo and {Castellano}, Marco and {Pentericci}, Laura and {Vignali}, Cristian and {Gilli}, Roberto and {Fontana}, Adriano and {Santini}, Paola and {Treu}, Tommaso and {Calabr{\`o}}, Antonello and {Llerena}, Mario and {Piconcelli}, Enrico and {Zappacosta}, Luca and {Mascia}, Sara and {Tripodi}, Roberta and {Arrabal Haro}, Pablo and {Bergamini}, Pietro and {Bakx}, Tom J.~L.~C. and {Dickinson}, Mark and {Glazebrook}, Karl and {Henry}, Alaina and {Leethochawalit}, Nicha and {Mazzolari}, Giovanni and {Merlin}, Emiliano and {Morishita}, Takahiro and {Nanayakkara}, Themiya and {Paris}, Diego and {Puccetti}, Simonetta and {Roberts-Borsani}, Guido and {Rojas Ruiz}, Sofia and {Rosati}, Piero and {Vanzella}, Eros and {Vito}, Fabio and {Vulcani}, Benedetta and {Wang}, Xin and {Yoon}, Ilsang and {Zavala}, Jorge A.},
        title = "{The Dual Nature of GHZ9: Coexisting Active Galactic Nuclei and Star Formation Activity in a Remote X-Ray Source at z = 10.145}",
      journal = {\apj},
         year = 2025,
        month = aug,
       volume = {989},
       number = {1},
          eid = {75},
        pages = {75},
          doi = {10.3847/1538-4357/ade706},
archivePrefix = {arXiv},
       eprint = {2410.18763},
 primaryClass = {astro-ph.GA},
       adsurl = {https://ui.adsabs.harvard.edu/abs/2025ApJ...989...75N}
}

@ARTICLE{2025ApJ...994...65N,
       author = {{Nakane}, Minami and {Ouchi}, Masami and {Nakajima}, Kimihiko and {Ono}, Yoshiaki and {Harikane}, Yuichi and {Isobe}, Yuki and {Nomoto}, Ken'ichi and {Ishigaki}, Miho N. and {Yanagisawa}, Hiroto and {Kashino}, Daichi and {Tominaga}, Nozomu and {Takahashi}, Koh and {Nishigaki}, Moka and {Takeda}, Yui and {Watanabe}, Kuria},
        title = "{Fe Abundances of Early Galaxies at z = 9─12 Derived with Deep JWST Spectra}",
      journal = {\apj},
         year = 2025,
        month = nov,
       volume = {994},
       number = {1},
          eid = {65},
        pages = {65},
          doi = {10.3847/1538-4357/ae04e6},
archivePrefix = {arXiv},
       eprint = {2503.11457},
 primaryClass = {astro-ph.GA},
       adsurl = {https://ui.adsabs.harvard.edu/abs/2025ApJ...994...65N}
}

@ARTICLE{2025arXiv251204043Z,
       author = {{Zhu}, Peixin and {Trussler}, James and {Kewley}, Lisa J.},
        title = "{The Nature of Nitrogen Enhanced High Redshift Galaxies}",
      journal = {arXiv e-prints},
         year = 2025,
        month = dec,
          eid = {arXiv:2512.04043},
        pages = {arXiv:2512.04043},
          doi = {10.48550/arXiv.2512.04043},
archivePrefix = {arXiv},
       eprint = {2512.04043},
 primaryClass = {astro-ph.GA},
       adsurl = {https://ui.adsabs.harvard.edu/abs/2025arXiv251204043Z}
}

@ARTICLE{2025ApJ...995L..44Y,
       author = {{Yang}, Chenwei and {Liu}, Bo and {Jiang}, Peng and {Shi}, Xiheng and {Zhou}, Yipeng and {Zhao}, Yaqi and {Zhou}, Xingyu and {Pan}, Xiang and {Zhou}, Hongyan},
        title = "{A Close Look at the Closest Nitrogen-loud Active Galaxy Akn 564}",
      journal = {\apjl},
         year = 2025,
        month = dec,
       volume = {995},
       number = {2},
          eid = {L44},
        pages = {L44},
          doi = {10.3847/2041-8213/ae25f1},
       adsurl = {https://ui.adsabs.harvard.edu/abs/2025ApJ...995L..44Y}
}

@ARTICLE{2026arXiv260115964C,
       author = {{Cameron}, Alex J. and {Carreira}, Courtney and {Simmonds}, Charlotte and {Bunker}, Andrew J. and {Saxena}, Aayush and {Carniani}, Stefano and {Charlot}, St{\'e}phane and {Chevallard}, Jacopo and {Curtis-Lake}, Emma and {Hainline}, Kevin and {Hausen}, Ryan and {Ji}, Xihan and {Ji}, Zhiyuan and {Johnson}, Benjamin D. and {Rinaldi}, Pierluigi and {Robertson}, Brant and {Scholtz}, Jan and {Silcock}, Maddie S. and {Tacchella}, Sandro and {Trussler}, James A.~A. and {{\"U}bler}, Hannah and {Williams}, Christina C. and {Willmer}, Christopher N.~A. and {Willott}, Chris and {Witstok}, Joris},
        title = "{JADES: Evolution of nitrogen abundances in star-forming galaxies from z \raisebox{-0.5ex}\textasciitilde 1.5-7}",
      journal = {arXiv e-prints},
         year = 2026,
        month = jan,
          eid = {arXiv:2601.15964},
        pages = {arXiv:2601.15964},
          doi = {10.48550/arXiv.2601.15964},
archivePrefix = {arXiv},
       eprint = {2601.15964},
 primaryClass = {astro-ph.GA},
       adsurl = {https://ui.adsabs.harvard.edu/abs/2026arXiv260115964C}
}

@ARTICLE{2026MNRAS.tmp...20M,
       author = {{McClymont}, William and {Tacchella}, Sandro and {Smith}, Aaron and {Kannan}, Rahul and {Garaldi}, Enrico and {Puchwein}, Ewald and {Isobe}, Yuki and {Ji}, Xihan and {Shen}, Xuejian and {Wang}, Zihao and {Belokurov}, Vasily and {Borrow}, Josh and {D'Eugenio}, Francesco and {Keating}, Laura and {Maiolino}, Roberto and {Monty}, Stephanie and {Vogelsberger}, Mark and {Zier}, Oliver},
        title = "{The THESAN-ZOOM project: Mystery N/O more - uncovering the origin of peculiar chemical abundances and a not-so-fundamental metallicity relation at 3 < z < 12}",
      journal = {\mnras},
         year = 2026,
        month = jan,
          doi = {10.1093/mnras/stag016},
archivePrefix = {arXiv},
       eprint = {2507.08787},
 primaryClass = {astro-ph.GA},
       adsurl = {https://ui.adsabs.harvard.edu/abs/2026MNRAS.tmp...20M}
}

@ARTICLE{2024MNRAS.534.3040D,
       author = {{Dors}, O.~L. and {Cardaci}, M.~V. and {H{\"a}gele}, G.~F. and {Valerdi}, M. and {Ilha}, G.~S. and {Oliveira}, C.~B. and {Riffel}, R.~A. and {Flury}, S.~R. and {Arellano-C{\'o}rdova}, K.~Z. and {Storchi-Bergmann}, T. and {Riffel}, R. and {Almeida}, G.~C. and {Morais}, I.~N.},
        title = "{Direct estimates of nitrogen abundance for Seyfert 2 nuclei}",
      journal = {\mnras},
         year = 2024,
        month = nov,
       volume = {534},
       number = {4},
        pages = {3040-3054},
          doi = {10.1093/mnras/stae2253},
archivePrefix = {arXiv},
       eprint = {2405.13906},
 primaryClass = {astro-ph.GA},
       adsurl = {https://ui.adsabs.harvard.edu/abs/2024MNRAS.534.3040D}
}

@ARTICLE{1971ApJ...167L..27W,
       author = {{Williams}, Robert E.},
        title = "{The Helium Abundance in Quasi-Stellar Objects}",
      journal = {\apjl},
         year = 1971,
        month = jul,
       volume = {167},
        pages = {L27},
          doi = {10.1086/180753},
       adsurl = {https://ui.adsabs.harvard.edu/abs/1971ApJ...167L..27W}
}

@ARTICLE{1971ApJ...163..235B,
       author = {{Bahcall}, John N. and {Oke}, J.~B.},
        title = "{Some Inferences from Spectrophotometry of Quasi-Stellar Sources}",
      journal = {\apj},
         year = 1971,
        month = jan,
       volume = {163},
        pages = {235},
          doi = {10.1086/150762},
       adsurl = {https://ui.adsabs.harvard.edu/abs/1971ApJ...163..235B}
}

@ARTICLE{2019ApJ...880...16K,
       author = {{Kewley}, Lisa J. and {Nicholls}, David C. and {Sutherland}, Ralph and {Rigby}, Jane R. and {Acharya}, Ayan and {Dopita}, Michael A. and {Bayliss}, Matthew B.},
        title = "{Theoretical ISM Pressure and Electron Density Diagnostics for Local and High-redshift Galaxies}",
      journal = {\apj},
         year = 2019,
        month = jul,
       volume = {880},
       number = {1},
          eid = {16},
        pages = {16},
          doi = {10.3847/1538-4357/ab16ed},
archivePrefix = {arXiv},
       eprint = {1908.05504},
 primaryClass = {astro-ph.GA},
       adsurl = {https://ui.adsabs.harvard.edu/abs/2019ApJ...880...16K}
}

@ARTICLE{2017MNRAS.471..562C,
       author = {{Congiu}, E. and {Contini}, M. and {Ciroi}, S. and {Cracco}, V. and {Berton}, M. and {Di Mille}, F. and {Frezzato}, M. and {La Mura}, G. and {Rafanelli}, P.},
        title = "{High-resolution spectroscopy of the extended narrow-line region of IC 5063 and NGC 7212}",
      journal = {\mnras},
         year = 2017,
        month = oct,
       volume = {471},
       number = {1},
        pages = {562-588},
          doi = {10.1093/mnras/stx1628},
archivePrefix = {arXiv},
       eprint = {1706.08970},
 primaryClass = {astro-ph.GA},
       adsurl = {https://ui.adsabs.harvard.edu/abs/2017MNRAS.471..562C}
}

@ARTICLE{2024MNRAS.527.6110O,
       author = {{Ormerod}, K. and {Conselice}, C.~J. and {Adams}, N.~J. and {Harvey}, T. and {Austin}, D. and {Trussler}, J. and {Ferreira}, L. and {Caruana}, J. and {Lucatelli}, G. and {Li}, Q. and {Roper}, W.~J.},
        title = "{EPOCHS VI: the size and shape evolution of galaxies since z   8 with JWST Observations}",
      journal = {\mnras},
         year = 2024,
        month = jan,
       volume = {527},
       number = {3},
        pages = {6110-6125},
          doi = {10.1093/mnras/stad3597},
archivePrefix = {arXiv},
       eprint = {2309.04377},
 primaryClass = {astro-ph.GA},
       adsurl = {https://ui.adsabs.harvard.edu/abs/2024MNRAS.527.6110O}
}

@ARTICLE{2023ApJ...956..139I,
       author = {{Isobe}, Yuki and {Ouchi}, Masami and {Nakajima}, Kimihiko and {Harikane}, Yuichi and {Ono}, Yoshiaki and {Xu}, Yi and {Zhang}, Yechi and {Umeda}, Hiroya},
        title = "{Redshift Evolution of Electron Density in the Interstellar Medium at z   0-9 Uncovered with JWST/NIRSpec Spectra and Line-spread Function Determinations}",
      journal = {\apj},
         year = 2023,
        month = oct,
       volume = {956},
       number = {2},
          eid = {139},
        pages = {139},
          doi = {10.3847/1538-4357/acf376},
archivePrefix = {arXiv},
       eprint = {2301.06811},
 primaryClass = {astro-ph.GA},
       adsurl = {https://ui.adsabs.harvard.edu/abs/2023ApJ...956..139I}
}

@ARTICLE{2025MNRAS.542.3181D,
       author = {{Dors}, O.~L. and {Oliveira}, C.~B. and {Cardaci}, M.~V. and {H{\"a}gele}, G.~F. and {Armah}, Mark and {Riffel}, R.~A. and {Ramos Vieira}, L. and {Almeida}, G.~C. and {Morais}, I.~N. and {Santos}, P.~C.},
        title = "{Metallicity of active galactic nuclei from ultraviolet and optical emission lines ─ II. Revisiting the C43 metallicity calibration and its implications}",
      journal = {\mnras},
         year = 2025,
        month = oct,
       volume = {542},
       number = {4},
        pages = {3181-3197},
          doi = {10.1093/mnras/staf1407},
archivePrefix = {arXiv},
       eprint = {2508.05397},
 primaryClass = {astro-ph.GA},
       adsurl = {https://ui.adsabs.harvard.edu/abs/2025MNRAS.542.3181D}
}

@ARTICLE{2012MNRAS.427.1266V,
       author = {{Vaona}, L. and {Ciroi}, S. and {Di Mille}, F. and {Cracco}, V. and {La Mura}, G. and {Rafanelli}, P.},
        title = "{Spectral properties of the narrow-line region in Seyfert galaxies selected from the SDSS-DR7}",
      journal = {\mnras},
         year = 2012,
        month = dec,
       volume = {427},
       number = {2},
        pages = {1266-1283},
          doi = {10.1111/j.1365-2966.2012.22060.x},
archivePrefix = {arXiv},
       eprint = {1210.5201},
 primaryClass = {astro-ph.CO},
       adsurl = {https://ui.adsabs.harvard.edu/abs/2012MNRAS.427.1266V}
}

@ARTICLE{2021MNRAS.500.2666C,
       author = {{Cerqueira-Campos}, F.~C. and {Rodr{\'\i}guez-Ardila}, A. and {Riffel}, R. and {Marinello}, M. and {Prieto}, A. and {Dahmer-Hahn}, L.~G.},
        title = "{Coronal-line forest active galactic nuclei - I. Physical properties of the emission-line regions}",
      journal = {\mnras},
         year = 2021,
        month = jan,
       volume = {500},
       number = {2},
        pages = {2666-2684},
          doi = {10.1093/mnras/staa3320},
archivePrefix = {arXiv},
       eprint = {2010.12595},
 primaryClass = {astro-ph.GA},
       adsurl = {https://ui.adsabs.harvard.edu/abs/2021MNRAS.500.2666C}
}

@ARTICLE{1991PASP..103..888C,
       author = {{Cruz-Gonzalez}, Irene and {Guichard}, Jose and {Serrano}, Alfonso and {Carrasco}, Luis},
        title = "{A Spectrophotometric Study of Seyfert Galaxies: Physical Conditions of the Emitting Gas}",
      journal = {\pasp},
         year = 1991,
        month = aug,
       volume = {103},
        pages = {888},
          doi = {10.1086/132900},
       adsurl = {https://ui.adsabs.harvard.edu/abs/1991PASP..103..888C}
}

@ARTICLE{2006ApJS..167..177D,
       author = {{Dopita}, Michael A. and {Fischera}, J{\"o}rg and {Sutherland}, Ralph S. and {Kewley}, Lisa J. and {Leitherer}, Claus and {Tuffs}, Richard J. and {Popescu}, Cristina C. and {van Breugel}, Wil and {Groves}, Brent A.},
        title = "{Modeling the Pan-Spectral Energy Distribution of Starburst Galaxies. III. Emission Line Diagnostics of Ensembles of Evolving H II Regions}",
      journal = {\apjs},
         year = 2006,
        month = dec,
       volume = {167},
       number = {2},
        pages = {177-200},
          doi = {10.1086/508261},
archivePrefix = {arXiv},
       eprint = {astro-ph/0608062},
 primaryClass = {astro-ph},
       adsurl = {https://ui.adsabs.harvard.edu/abs/2006ApJS..167..177D}
}

@ARTICLE{2011MNRAS.415.3616D,
       author = {{Dors}, Jr., O.~L. and {Krabbe}, Angela and {H{\"a}gele}, Guillermo F. and {P{\'e}rez-Montero}, Enrique},
        title = "{Analysing derived metallicities and ionization parameters from model-based determinations in ionized gaseous nebulae}",
      journal = {\mnras},
         year = 2011,
        month = aug,
       volume = {415},
       number = {4},
        pages = {3616-3626},
          doi = {10.1111/j.1365-2966.2011.18978.x},
archivePrefix = {arXiv},
       eprint = {1104.5460},
 primaryClass = {astro-ph.CO},
       adsurl = {https://ui.adsabs.harvard.edu/abs/2011MNRAS.415.3616D}
}

@ARTICLE{1990ApJS...74..731W,
       author = {{Wilson}, A.~S. and {Nath}, B.},
        title = "{A Survey of the Profiles of Narrow Emission Lines in Seyfert Galaxies}",
      journal = {\apjs},
         year = 1990,
        month = nov,
       volume = {74},
        pages = {731},
          doi = {10.1086/191515},
       adsurl = {https://ui.adsabs.harvard.edu/abs/1990ApJS...74..731W}
}

@ARTICLE{2019ApJ...886L..28M,
       author = {{Mowla}, Lamiya A. and {Nelson}, Erica J. and {van Dokkum}, Pieter and {Tadaki}, Ken-ichi},
        title = "{Anomalously Narrow Line Widths of Compact Massive Star-forming Galaxies at z {\ensuremath{\sim}} 2.3: A Possible Inclination Bias in the Size-Mass Plane}",
      journal = {\apjl},
         year = 2019,
        month = dec,
       volume = {886},
       number = {2},
          eid = {L28},
        pages = {L28},
          doi = {10.3847/2041-8213/ab54d1},
archivePrefix = {arXiv},
       eprint = {1910.10722},
 primaryClass = {astro-ph.GA},
       adsurl = {https://ui.adsabs.harvard.edu/abs/2019ApJ...886L..28M}
}

@ARTICLE{2012MNRAS.422.3475H,
       author = {{H{\"a}gele}, Guillermo F. and {Firpo}, Ver{\'o}nica and {Bosch}, Guillermo and {D{\'\i}az}, {\'A}ngeles I. and {Morrell}, Nidia},
        title = "{High-resolution spectroscopy of the blue compact dwarf galaxy Haro 15 - II. Chemodynamics}",
      journal = {\mnras},
         year = 2012,
        month = jun,
       volume = {422},
       number = {4},
        pages = {3475-3494},
          doi = {10.1111/j.1365-2966.2012.20858.x},
archivePrefix = {arXiv},
       eprint = {1203.0531},
 primaryClass = {astro-ph.CO},
       adsurl = {https://ui.adsabs.harvard.edu/abs/2012MNRAS.422.3475H}
}

@ARTICLE{2011MNRAS.414.3288F,
       author = {{Firpo}, Ver{\'o}nica and {Bosch}, Guillermo and {H{\"a}gele}, Guillermo F. and {D{\'\i}az}, {\'A}ngeles I. and {Morrell}, Nidia},
        title = "{High-resolution spectroscopy of the blue compact dwarf galaxy Haro 15 - I. Internal kinematics}",
      journal = {\mnras},
         year = 2011,
        month = jul,
       volume = {414},
       number = {4},
        pages = {3288-3297},
          doi = {10.1111/j.1365-2966.2011.18630.x},
archivePrefix = {arXiv},
       eprint = {1103.0526},
 primaryClass = {astro-ph.CO},
       adsurl = {https://ui.adsabs.harvard.edu/abs/2011MNRAS.414.3288F}
}

@ARTICLE{2014MNRAS.442.3565C,
       author = {{Ch{\'a}vez}, R. and {Terlevich}, R. and {Terlevich}, E. and {Bresolin}, F. and {Melnick}, J. and {Plionis}, M. and {Basilakos}, S.},
        title = "{The L-{\ensuremath{\sigma}} relation for massive bursts of star formation}",
      journal = {\mnras},
         year = 2014,
        month = aug,
       volume = {442},
       number = {4},
        pages = {3565-3597},
          doi = {10.1093/mnras/stu987},
archivePrefix = {arXiv},
       eprint = {1405.4010},
 primaryClass = {astro-ph.GA},
       adsurl = {https://ui.adsabs.harvard.edu/abs/2014MNRAS.442.3565C}
}

@ARTICLE{2005MNRAS.358..521M,
       author = {{Moll{\'a}}, M. and {D{\'\i}az}, A.~I.},
        title = "{A grid of chemical evolution models as a tool to interpret spiral and irregular galaxies data}",
      journal = {\mnras},
         year = 2005,
        month = apr,
       volume = {358},
       number = {2},
        pages = {521-543},
          doi = {10.1111/j.1365-2966.2005.08782.x},
archivePrefix = {arXiv},
       eprint = {astro-ph/0501370},
 primaryClass = {astro-ph},
       adsurl = {https://ui.adsabs.harvard.edu/abs/2005MNRAS.358..521M}
}

@ARTICLE{2019A&A...630A.125V,
       author = {{Vincenzo}, F. and {Miglio}, A. and {Kobayashi}, C. and {Mackereth}, J.~T. and {Montalban}, J.},
        title = "{He abundances in disc galaxies. I. Predictions from cosmological chemodynamical simulations}",
      journal = {\aap},
         year = 2019,
        month = oct,
       volume = {630},
          eid = {A125},
        pages = {A125},
          doi = {10.1051/0004-6361/201935886},
archivePrefix = {arXiv},
       eprint = {1905.08309},
 primaryClass = {astro-ph.GA},
       adsurl = {https://ui.adsabs.harvard.edu/abs/2019A&A...630A.125V}
}

@ARTICLE{2022MNRAS.514.5506D,
       author = {{Dors}, O.~L. and {Valerdi}, M. and {Freitas-Lemes}, P. and {Krabbe}, A.~C. and {Riffel}, R.~A. and {Am{\^o}res}, E.~B. and {Riffel}, R. and {Armah}, M. and {Monteiro}, A.~F. and {Oliveira}, C.~B.},
        title = "{Chemical abundances in Seyfert galaxies - IX. Helium abundance estimates}",
      journal = {\mnras},
         year = 2022,
        month = aug,
       volume = {514},
       number = {4},
        pages = {5506-5527},
          doi = {10.1093/mnras/stac1722},
archivePrefix = {arXiv},
       eprint = {2206.09836},
 primaryClass = {astro-ph.GA},
       adsurl = {https://ui.adsabs.harvard.edu/abs/2022MNRAS.514.5506D}
}

@ARTICLE{2008MNRAS.383..209H,
       author = {{H{\"a}gele}, Guillermo F. and {D{\'\i}az}, {\'A}ngeles I. and {Terlevich}, Elena and {Terlevich}, Roberto and {P{\'e}rez-Montero}, Enrique and {Cardaci}, M{\'o}nica V.},
        title = "{Precision abundance analysis of bright HII galaxies}",
      journal = {\mnras},
         year = 2008,
        month = jan,
       volume = {383},
       number = {1},
        pages = {209-229},
          doi = {10.1111/j.1365-2966.2007.12527.x},
archivePrefix = {arXiv},
       eprint = {0710.1828},
 primaryClass = {astro-ph},
       adsurl = {https://ui.adsabs.harvard.edu/abs/2008MNRAS.383..209H}
}

@ARTICLE{1979MNRAS.189...95P,
       author = {{Pagel}, B.~E.~J. and {Edmunds}, M.~G. and {Blackwell}, D.~E. and {Chun}, M.~S. and {Smith}, G.},
        title = "{On the composition of H II regions in southern galaxies - I. NGC 300 and 1365.}",
      journal = {\mnras},
         year = 1979,
        month = oct,
       volume = {189},
        pages = {95-113},
          doi = {10.1093/mnras/189.1.95},
       adsurl = {https://ui.adsabs.harvard.edu/abs/1979MNRAS.189...95P}
}

@ARTICLE{1991ApJ...380..140M,
       author = {{McGaugh}, Stacy S.},
        title = "{H II Region Abundances: Model Oxygen Line Ratios}",
      journal = {\apj},
         year = 1991,
        month = oct,
       volume = {380},
        pages = {140},
          doi = {10.1086/170569},
       adsurl = {https://ui.adsabs.harvard.edu/abs/1991ApJ...380..140M}
}

@ARTICLE{1984MNRAS.211..507E,
       author = {{Edmunds}, M.~G. and {Pagel}, B.~E.~J.},
        title = "{On the composition of H II regions in southern galaxies. III. NGC 2997 and 7793.}",
      journal = {\mnras},
         year = 1984,
        month = dec,
       volume = {211},
        pages = {507-519},
          doi = {10.1093/mnras/211.3.507},
       adsurl = {https://ui.adsabs.harvard.edu/abs/1984MNRAS.211..507E}
}

@ARTICLE{2021MNRAS.506L..11R,
       author = {{Riffel}, Rogemar A. and {Dors}, Oli L. and {Krabbe}, Angela C. and {Esteban}, C{\'e}sar},
        title = "{Electron temperature fluctuations in Seyfert galaxies}",
      journal = {\mnras},
         year = 2021,
        month = sep,
       volume = {506},
       number = {1},
        pages = {L11-L15},
          doi = {10.1093/mnrasl/slab064},
archivePrefix = {arXiv},
       eprint = {2106.03623},
 primaryClass = {astro-ph.GA},
       adsurl = {https://ui.adsabs.harvard.edu/abs/2021MNRAS.506L..11R}
}

@ARTICLE{2024A&A...684A..53B,
       author = {{Binette}, Luc and {Zovaro}, Henry R.~M. and {Villar Mart{\'\i}n}, Montserrat and {Dors}, Oli L. and {Krongold}, Yair and {Morisset}, Christophe and {Revalski}, Mitchell and {Alarie}, Alexandre and {Riffel}, Rogemar A. and {Dopita}, Michael A.},
        title = "{Constraints on the densities and temperature of the Seyfert 2 narrow line region}",
      journal = {\aap},
         year = 2024,
        month = apr,
       volume = {684},
          eid = {A53},
        pages = {A53},
          doi = {10.1051/0004-6361/202245754},
archivePrefix = {arXiv},
       eprint = {2401.06972},
 primaryClass = {astro-ph.GA},
       adsurl = {https://ui.adsabs.harvard.edu/abs/2024A&A...684A..53B}
}

@ARTICLE{2025MNRAS.540.1608D,
       author = {{Dors}, O.~L. and {Oliveira}, C.~B. and {Cardaci}, M.~V. and {H{\"a}gele}, G.~F. and {Morais}, I.~N. and {Ji}, X. and {Riffel}, R.~A. and {Riffel}, R. and {Mezcua}, M. and {Almeida}, G.~C. and {Santos}, P.~C. and {de Mellos}, M.~S.~Z.},
        title = "{Metallicity of active galactic nuclei from ultraviolet and optical emission lines {\textendash} I. Carbon abundance dependence}",
      journal = {\mnras},
         year = 2025,
        month = jun,
       volume = {540},
       number = {2},
        pages = {1608-1625},
          doi = {10.1093/mnras/staf727},
archivePrefix = {arXiv},
       eprint = {2505.00095},
 primaryClass = {astro-ph.GA},
       adsurl = {https://ui.adsabs.harvard.edu/abs/2025MNRAS.540.1608D}
}

@ARTICLE{2011A&A...530A..33V,
       author = {{Vladilo}, G. and {Abate}, C. and {Yin}, J. and {Cescutti}, G. and {Matteucci}, F.},
        title = "{Silicon depletion in damped Ly {\ensuremath{\alpha}} systems. The S/Zn method}",
      journal = {\aap},
         year = 2011,
        month = jun,
       volume = {530},
          eid = {A33},
        pages = {A33},
          doi = {10.1051/0004-6361/201016330},
       adsurl = {https://ui.adsabs.harvard.edu/abs/2011A&A...530A..33V}
}

@ARTICLE{2002A&A...391...21P,
       author = {{Pettini}, M. and {Ellison}, S.~L. and {Bergeron}, J. and {Petitjean}, P.},
        title = "{The abundances of nitrogen and oxygen in damped Lyman alpha systems}",
      journal = {\aap},
         year = 2002,
        month = aug,
       volume = {391},
        pages = {21-34},
          doi = {10.1051/0004-6361:20020809},
archivePrefix = {arXiv},
       eprint = {astro-ph/0205472},
 primaryClass = {astro-ph},
       adsurl = {https://ui.adsabs.harvard.edu/abs/2002A&A...391...21P}
}

@ARTICLE{2016A&A...596A..97D,
       author = {{De Cia}, A. and {Ledoux}, C. and {Mattsson}, L. and {Petitjean}, P. and {Srianand}, R. and {Gavignaud}, I. and {Jenkins}, E.~B.},
        title = "{Dust-depletion sequences in damped Lyman-{\ensuremath{\alpha}} absorbers. A unified picture from low-metallicity systems to the Galaxy}",
      journal = {\aap},
         year = 2016,
        month = dec,
       volume = {596},
          eid = {A97},
        pages = {A97},
          doi = {10.1051/0004-6361/201527895},
archivePrefix = {arXiv},
       eprint = {1608.08621},
 primaryClass = {astro-ph.GA},
       adsurl = {https://ui.adsabs.harvard.edu/abs/2016A&A...596A..97D}
}

@ARTICLE{1994ApJ...435..171K,
       author = {{Kraemer}, Steven B. and {Wu}, Chi-Chao and {Crenshaw}, D. Michael and {Harrington}, J. Patrick},
        title = "{IUE Spectra and Photoionization Models of the Seyfert 2 Galaxies NGC 7674 and I ZW 92}",
      journal = {\apj},
         year = 1994,
        month = nov,
       volume = {435},
        pages = {171},
          doi = {10.1086/174803},
       adsurl = {https://ui.adsabs.harvard.edu/abs/1994ApJ...435..171K}
}

@ARTICLE{2006A&A...447..863N,
       author = {{Nagao}, T. and {Maiolino}, R. and {Marconi}, A.},
        title = "{Gas metallicity in the narrow-line regions of high-redshift active galactic nuclei}",
      journal = {\aap},
         year = 2006,
        month = mar,
       volume = {447},
       number = {3},
        pages = {863-876},
          doi = {10.1051/0004-6361:20054127},
archivePrefix = {arXiv},
       eprint = {astro-ph/0508652},
 primaryClass = {astro-ph},
       adsurl = {https://ui.adsabs.harvard.edu/abs/2006A&A...447..863N}
}

@ARTICLE{2025ApJ...993..204H,
       author = {{Harikane}, Yuichi and {Sanders}, Ryan L. and {Ellis}, Richard and {Jones}, Tucker and {Ouchi}, Masami and {Laporte}, Nicolas and {Roberts-Borsani}, Guido and {Katz}, Harley and {Nakajima}, Kimihiko and {Ono}, Yoshiaki and {Gupta}, Mansi},
        title = "{JWST and ALMA Joint Analysis with [O II] {\ensuremath{\lambda}}{\ensuremath{\lambda}}3726, 3729, [O III] {\ensuremath{\lambda}}4363, [O III] 88 {\ensuremath{\mu}}m, and [O III] 52 {\ensuremath{\mu}}m: Multizone Evolution of Electron Densities at z {\ensuremath{\sim}} 0─14 and its Impact on Metallicity Measurements}",
      journal = {\apj},
         year = 2025,
        month = nov,
       volume = {993},
       number = {2},
          eid = {204},
        pages = {204},
          doi = {10.3847/1538-4357/ae0e53},
archivePrefix = {arXiv},
       eprint = {2505.09186},
 primaryClass = {astro-ph.GA},
       adsurl = {https://ui.adsabs.harvard.edu/abs/2025ApJ...993..204H}
}

@ARTICLE{2020ApJ...893...96B,
       author = {{Berg}, Danielle A. and {Pogge}, Richard W. and {Skillman}, Evan D. and {Croxall}, Kevin V. and {Moustakas}, John and {Rogers}, Noah S.~J. and {Sun}, Jiayi},
        title = "{CHAOS IV: Gas-phase Abundance Trends from the First Four CHAOS Galaxies}",
      journal = {\apj},
         year = 2020,
        month = apr,
       volume = {893},
       number = {2},
          eid = {96},
        pages = {96},
          doi = {10.3847/1538-4357/ab7eab},
archivePrefix = {arXiv},
       eprint = {2001.05002},
 primaryClass = {astro-ph.GA},
       adsurl = {https://ui.adsabs.harvard.edu/abs/2020ApJ...893...96B}
}

@ARTICLE{2020MNRAS.492..468D,
       author = {{Dors}, O.~L. and {Freitas-Lemes}, P. and {Am{\^o}res}, E.~B. and {P{\'e}rez-Montero}, E. and {Cardaci}, M.~V. and {H{\"a}gele}, G.~F. and {Armah}, M. and {Krabbe}, A.~C. and {Fa{\'u}ndez-Abans}, M.},
        title = "{Chemical abundances of Seyfert 2 AGNs - I. Comparing oxygen abundances from distinct methods using SDSS}",
      journal = {\mnras},
         year = 2020,
        month = feb,
       volume = {492},
       number = {1},
        pages = {468-479},
          doi = {10.1093/mnras/stz3492},
archivePrefix = {arXiv},
       eprint = {1912.04236},
 primaryClass = {astro-ph.GA},
       adsurl = {https://ui.adsabs.harvard.edu/abs/2020MNRAS.492..468D}
}

@ARTICLE{2018MNRAS.479.2294D,
       author = {{Dors}, O.~L. and {Agarwal}, B. and {H{\"a}gele}, G.~F. and {Cardaci}, M.~V. and {Rydberg}, Claes-Erik and {Riffel}, R.~A. and {Oliveira}, A.~S. and {Krabbe}, A.~C.},
        title = "{Nature and chemical abundances of a sample of Lyman-{\ensuremath{\alpha}} emitter objects at high redshift}",
      journal = {\mnras},
         year = 2018,
        month = sep,
       volume = {479},
       number = {2},
        pages = {2294-2307},
          doi = {10.1093/mnras/sty1658},
archivePrefix = {arXiv},
       eprint = {1806.07732},
 primaryClass = {astro-ph.GA},
       adsurl = {https://ui.adsabs.harvard.edu/abs/2018MNRAS.479.2294D}
}

@ARTICLE{2022MNRAS.513.5134N,
       author = {{Nakajima}, K. and {Maiolino}, R.},
        title = "{Diagnostics for PopIII galaxies and direct collapse black holes in the early universe}",
      journal = {\mnras},
         year = 2022,
        month = jul,
       volume = {513},
       number = {4},
        pages = {5134-5147},
          doi = {10.1093/mnras/stac1242},
archivePrefix = {arXiv},
       eprint = {2204.11870},
 primaryClass = {astro-ph.GA},
       adsurl = {https://ui.adsabs.harvard.edu/abs/2022MNRAS.513.5134N}
}

@ARTICLE{2016MNRAS.456.3354F,
       author = {{Feltre}, A. and {Charlot}, S. and {Gutkin}, J.},
        title = "{Nuclear activity versus star formation: emission-line diagnostics at ultraviolet and optical wavelengths}",
      journal = {\mnras},
         year = 2016,
        month = mar,
       volume = {456},
       number = {3},
        pages = {3354-3374},
          doi = {10.1093/mnras/stv2794},
archivePrefix = {arXiv},
       eprint = {1511.08217},
 primaryClass = {astro-ph.GA},
       adsurl = {https://ui.adsabs.harvard.edu/abs/2016MNRAS.456.3354F}
}

@ARTICLE{2025arXiv250717057B,
       author = {{Berg}, Danielle A. and {Sanders}, Ryan L. and {Shapley}, Alice E. and {Topping}, Michael W. and {Reddy}, Naveen A. and {Skillman}, Evan D. and {Aver}, Erik and {Cullen}, Fergus and {Donnan}, Callum T. and {Dunlop}, James S. and {Jones}, Tucker and {Khostovan}, Ali Ahmad and {McLeod}, Derek J. and {Narayanan}, Desika and {Oesch}, Pascal A. and {Pahl}, Anthony J. and {Pettini}, Max and {F{\"o}rster Schreiber}, N.~M. and {Stark}, Daniel P.},
        title = "{The AURORA Survey: Robust Helium Abundances at High Redshift Reveal A Subpopulation of Helium-Enhanced Galaxies in the Early Universe}",
      journal = {arXiv e-prints},
         year = 2025,
        month = jul,
          eid = {arXiv:2507.17057},
        pages = {arXiv:2507.17057},
          doi = {10.48550/arXiv.2507.17057},
archivePrefix = {arXiv},
       eprint = {2507.17057},
 primaryClass = {astro-ph.GA},
       adsurl = {https://ui.adsabs.harvard.edu/abs/2025arXiv250717057B}
}

@ARTICLE{1989ApJ...345..245C,
       author = {{Cardelli}, Jason A. and {Clayton}, Geoffrey C. and {Mathis}, John S.},
        title = "{The Relationship between Infrared, Optical, and Ultraviolet Extinction}",
      journal = {\apj},
         year = "1989",
        month = "Oct",
       volume = {345},
        pages = {245},
          doi = {10.1086/167900},
       adsurl = {https://ui.adsabs.harvard.edu/abs/1989ApJ...345..245C}
}

@book{2006agna.book.....O,
       author = {{Osterbrock}, Donald E. and {Ferland}, Gary J.},
        title     = {{Astrophysics of Gaseous Nebulae and Active Galactic Nuclei}},
  edition   = {2nd},
  publisher = {University Science Books},
  address   = {Sausalito, California, USA},
         year = 2006,
       adsurl = {https://ui.adsabs.harvard.edu/abs/2006agna.book.....O}
}

@ARTICLE{2013ApJ...774L..10K,
       author = {{Kewley}, Lisa J. and {Maier}, Christian and {Yabe}, Kiyoto and {Ohta}, Kouji and {Akiyama}, Masayuki and {Dopita}, Michael A. and {Yuan}, Tiantian},
        title = "{The Cosmic BPT Diagram: Confronting Theory with Observations}",
      journal = {\apjl},
         year = 2013,
        month = sep,
       volume = {774},
       number = {1},
          eid = {L10},
        pages = {L10},
          doi = {10.1088/2041-8205/774/1/L10},
archivePrefix = {arXiv},
       eprint = {1307.0514},
 primaryClass = {astro-ph.CO},
       adsurl = {https://ui.adsabs.harvard.edu/abs/2013ApJ...774L..10K}
}

@ARTICLE{2023MNRAS.526.3610H,
       author = {{Hirschmann}, Michaela and {Charlot}, Stephane and {Feltre}, Anna and {Curtis-Lake}, Emma and {Somerville}, Rachel S. and {Chevallard}, Jacopo and {Choi}, Ena and {Nelson}, Dylan and {Morisset}, Christophe and {Plat}, Adele and {Vidal-Garcia}, Alba},
        title = "{Emission-line properties of IllustrisTNG galaxies: from local diagnostic diagrams to high-redshift predictions for JWST}",
      journal = {\mnras},
         year = 2023,
        month = dec,
       volume = {526},
       number = {3},
        pages = {3610-3636},
          doi = {10.1093/mnras/stad2955},
archivePrefix = {arXiv},
       eprint = {2212.02522},
 primaryClass = {astro-ph.GA},
       adsurl = {https://ui.adsabs.harvard.edu/abs/2023MNRAS.526.3610H}
}

@ARTICLE{2010arXiv1004.5251L,
       author = {{Lopez-Sanchez}, Angel R. and {Esteban}, Cesar},
        title = "{Massive star formation in Wolf-Rayet galaxies: IV b. Using empirical calibrations to compute the oxygen abundance}",
      journal = {arXiv e-prints},
         year = 2010,
        month = apr,
          eid = {arXiv:1004.5251},
        pages = {arXiv:1004.5251},
          doi = {10.48550/arXiv.1004.5251},
archivePrefix = {arXiv},
       eprint = {1004.5251},
 primaryClass = {astro-ph.CO},
       adsurl = {https://ui.adsabs.harvard.edu/abs/2010arXiv1004.5251L}
}

@ARTICLE{2012ApJ...755...89R,
       author = {{Rafelski}, Marc and {Wolfe}, Arthur M. and {Prochaska}, J. Xavier and {Neeleman}, Marcel and {Mendez}, Alexander J.},
        title = "{Metallicity Evolution of Damped Ly{\ensuremath{\alpha}} Systems Out to z \raisebox{-0.5ex}\textasciitilde 5}",
      journal = {\apj},
         year = 2012,
        month = aug,
       volume = {755},
       number = {2},
          eid = {89},
        pages = {89},
          doi = {10.1088/0004-637X/755/2/89},
archivePrefix = {arXiv},
       eprint = {1205.5047},
 primaryClass = {astro-ph.CO},
       adsurl = {https://ui.adsabs.harvard.edu/abs/2012ApJ...755...89R}
}

@ARTICLE{2025ApJ...991..228H,
       author = {{Huyan}, Jianghao and {Kulkarni}, Varsha P. and {Poudel}, Suraj and {Tejos}, Nicolas and {P{\'e}roux}, Celine and {Lopez}, Sebastian},
        title = "{The Diversity of Metal Enrichment and Abundance Patterns at High Redshift: A Magellan Survey of Gas-rich Galaxies Traced by Damped Ly{\ensuremath{\alpha}} Absorbers at z {\ensuremath{\sim}} 5}",
      journal = {\apj},
         year = 2025,
        month = oct,
       volume = {991},
       number = {2},
          eid = {228},
        pages = {228},
          doi = {10.3847/1538-4357/adf63e},
archivePrefix = {arXiv},
       eprint = {2508.02940},
 primaryClass = {astro-ph.GA},
       adsurl = {https://ui.adsabs.harvard.edu/abs/2025ApJ...991..228H}
}

@ARTICLE{2003ApJ...595L...9P,
       author = {{Prochaska}, Jason X. and {Gawiser}, Eric and {Wolfe}, Arthur M. and {Castro}, Sandra and {Djorgovski}, S.~G.},
        title = "{The Age-Metallicity Relation of the Universe in Neutral Gas: The First 100 Damped Ly{\ensuremath{\alpha}} Systems}",
      journal = {\apjl},
         year = 2003,
        month = sep,
       volume = {595},
       number = {1},
        pages = {L9-L12},
          doi = {10.1086/378945},
archivePrefix = {arXiv},
       eprint = {astro-ph/0305314},
 primaryClass = {astro-ph},
       adsurl = {https://ui.adsabs.harvard.edu/abs/2003ApJ...595L...9P}
}

@ARTICLE{2016ApJ...830..158M,
       author = {{Morrison}, Sean and {Kulkarni}, Varsha P. and {Som}, Debopam and {DeMarcy}, Bryan and {Quiret}, Samuel and {P{\'e}roux}, Celine},
        title = "{Element Abundances in a Gas-rich Galaxy at z = 5: Clues to the Early Chemical Enrichment of Galaxies}",
      journal = {\apj},
         year = 2016,
        month = oct,
       volume = {830},
       number = {2},
          eid = {158},
        pages = {158},
          doi = {10.3847/0004-637X/830/2/158},
archivePrefix = {arXiv},
       eprint = {1603.03492},
 primaryClass = {astro-ph.GA},
       adsurl = {https://ui.adsabs.harvard.edu/abs/2016ApJ...830..158M}
}

@ARTICLE{2014MNRAS.443.1291D,
       author = {{Dors}, Oli L. and {Cardaci}, M{\'o}nica V. and {H{\"a}gele}, Guillermo F. and {Krabbe}, {\^A}ngela C.},
        title = "{Metallicity evolution of AGNs from UV emission lines based on a new index}",
      journal = {\mnras},
         year = 2014,
        month = sep,
       volume = {443},
       number = {2},
        pages = {1291-1300},
          doi = {10.1093/mnras/stu1218},
archivePrefix = {arXiv},
       eprint = {1406.4832},
 primaryClass = {astro-ph.GA},
       adsurl = {https://ui.adsabs.harvard.edu/abs/2014MNRAS.443.1291D}
}

@ARTICLE{2024ApJ...962...95M,
       author = {{Mingozzi}, Matilde and {James}, Bethan L. and {Berg}, Danielle A. and {Arellano-C{\'o}rdova}, Karla Z. and {Plat}, Adele and {Scarlata}, Claudia and {Aloisi}, Alessandra and {Amor{\'\i}n}, Ricardo O. and {Brinchmann}, Jarle and {Charlot}, St{\'e}phane and {Chisholm}, John and {Feltre}, Anna and {Gazagnes}, Simon and {Hayes}, Matthew and {Heckman}, Timothy and {Hernandez}, Svea and {Kewley}, Lisa J. and {Kumari}, Nimisha and {Leitherer}, Claus and {Martin}, Crystal L. and {Maseda}, Michael and {Nanayakkara}, Themiya and {Ravindranath}, Swara and {Rigby}, Jane R. and {Senchyna}, Peter and {Skillman}, Evan D. and {Sugahara}, Yuma and {Wilkins}, Stephen M. and {Wofford}, Aida and {Xu}, Xinfeng},
        title = "{CLASSY. VIII. Exploring the Source of Ionization with UV Interstellar Medium Diagnostics in Local High-z Analogs}",
      journal = {\apj},
         year = 2024,
        month = feb,
       volume = {962},
       number = {1},
          eid = {95},
        pages = {95},
          doi = {10.3847/1538-4357/ad1033},
archivePrefix = {arXiv},
       eprint = {2306.15062},
 primaryClass = {astro-ph.GA},
       adsurl = {https://ui.adsabs.harvard.edu/abs/2024ApJ...962...95M}
}

@ARTICLE{2024MNRAS.527.7217V,
       author = {{Vidal-Garc{\'\i}a}, A. and {Plat}, A. and {Curtis-Lake}, E. and {Feltre}, A. and {Hirschmann}, M. and {Chevallard}, J. and {Charlot}, S.},
        title = "{BEAGLE-AGN I: simultaneous constraints on the properties of gas in star-forming and AGN narrow-line regions in galaxies}",
      journal = {\mnras},
         year = 2024,
        month = jan,
       volume = {527},
       number = {3},
        pages = {7217-7241},
          doi = {10.1093/mnras/stad3252},
archivePrefix = {arXiv},
       eprint = {2211.13648},
 primaryClass = {astro-ph.GA},
       adsurl = {https://ui.adsabs.harvard.edu/abs/2024MNRAS.527.7217V}
}

@ARTICLE{2023ApJ...955..141C,
       author = {{Carr}, David J. and {Salzer}, John J. and {Gronwall}, Caryl and {Williams}, Anna L.},
        title = "{Metal Abundances of Intermediate-redshift Active Galactic Nuclei: Evidence for a Population of Lower-metallicity Seyfert 2 Galaxies at z = 0.3-0.4}",
      journal = {\apj},
         year = 2023,
        month = oct,
       volume = {955},
       number = {2},
          eid = {141},
        pages = {141},
          doi = {10.3847/1538-4357/aced91},
archivePrefix = {arXiv},
       eprint = {2308.06824},
 primaryClass = {astro-ph.GA},
       adsurl = {https://ui.adsabs.harvard.edu/abs/2023ApJ...955..141C}
}

@ARTICLE{1998ApJ...507L.113Y,
       author = {{Yoshii}, Yuzuru and {Tsujimoto}, Takuji and {Kawara}, Kimiaki},
        title = "{Age Dating of a High-Redshift QSO B1422+231 at Z = 3.62and Its Cosmological Implications}",
      journal = {\apjl},
         year = 1998,
        month = nov,
       volume = {507},
       number = {2},
        pages = {L113-L116},
          doi = {10.1086/311690},
archivePrefix = {arXiv},
       eprint = {astro-ph/9809047},
 primaryClass = {astro-ph},
       adsurl = {https://ui.adsabs.harvard.edu/abs/1998ApJ...507L.113Y}
}

@ARTICLE{2024ApJ...975..214J,
       author = {{Jiang}, Danyang and {Onoue}, Masafusa and {Jiang}, Linhua and {Lai}, Samuel and {Ba{\~n}ados}, Eduardo and {Becker}, George D. and {Bischetti}, Manuela and {Bosman}, Sarah E.~I. and {Davies}, Rebecca L. and {D'Odorico}, Valentina and {Farina}, Emanuele Paolo and {Haehnelt}, Martin G. and {Mazzucchelli}, Chiara and {Schindler}, Jan-Torge and {Walter}, Fabian and {Zhu}, Yongda},
        title = "{No Redshift Evolution in the Fe II/Mg II Flux Ratios of Quasars across Cosmic Time}",
      journal = {\apj},
         year = 2024,
        month = nov,
       volume = {975},
       number = {2},
          eid = {214},
        pages = {214},
          doi = {10.3847/1538-4357/ad7d09},
archivePrefix = {arXiv},
       eprint = {2409.06174},
 primaryClass = {astro-ph.GA},
       adsurl = {https://ui.adsabs.harvard.edu/abs/2024ApJ...975..214J}
}

@ARTICLE{2022MNRAS.513.1801L,
       author = {{Lai}, Samuel and {Bian}, Fuyan and {Onken}, Christopher A. and {Wolf}, Christian and {Mazzucchelli}, Chiara and {Ba{\~n}ados}, Eduardo and {Bischetti}, Manuela and {Bosman}, Sarah E.~I. and {Becker}, George and {Cupani}, Guido and {D'Odorico}, Valentina and {Eilers}, Anna-Christina and {Fan}, Xiaohui and {Farina}, Emanuele Paolo and {Onoue}, Masafusa and {Schindler}, Jan-Torge and {Walter}, Fabian and {Wang}, Feige and {Yang}, Jinyi and {Zhu}, Yongda},
        title = "{Chemical abundance of z {\ensuremath{\sim}} 6 quasar broad-line regions in the XQR-30 sample}",
      journal = {\mnras},
         year = 2022,
        month = jun,
       volume = {513},
       number = {2},
        pages = {1801-1819},
          doi = {10.1093/mnras/stac1001},
archivePrefix = {arXiv},
       eprint = {2204.03335},
 primaryClass = {astro-ph.GA},
       adsurl = {https://ui.adsabs.harvard.edu/abs/2022MNRAS.513.1801L}
}

@ARTICLE{2018A&A...616L...4M,
       author = {{Matsuoka}, K. and {Nagao}, T. and {Marconi}, A. and {Maiolino}, R. and {Mannucci}, F. and {Cresci}, G. and {Terao}, K. and {Ikeda}, H.},
        title = "{The mass-metallicity relation of high-z type-2 active galactic nuclei}",
      journal = {\aap},
         year = 2018,
        month = aug,
       volume = {616},
          eid = {L4},
        pages = {L4},
          doi = {10.1051/0004-6361/201833418},
archivePrefix = {arXiv},
       eprint = {1807.09276},
 primaryClass = {astro-ph.GA},
       adsurl = {https://ui.adsabs.harvard.edu/abs/2018A&A...616L...4M}
}

@ARTICLE{2019A&A...626A...9M,
       author = {{Mignoli}, M. and {Feltre}, A. and {Bongiorno}, A. and {Calura}, F. and {Gilli}, R. and {Vignali}, C. and {Zamorani}, G. and {Lilly}, S.~J. and {Le F{\`e}vre}, O. and {Bardelli}, S. and {Bolzonella}, M. and {Bordoloi}, R. and {Le Brun}, V. and {Caputi}, K.~I. and {Cimatti}, A. and {Diener}, C. and {Garilli}, B. and {Koekemoer}, A.~M. and {Maier}, C. and {Mainieri}, V. and {Peng}, Y. and {P{\'e}rez Montero}, E. and {Silverman}, J.~D. and {Zucca}, E.},
        title = "{Obscured AGN at 1.5 < z < 3.0 from the zCOSMOS-deep Survey . I. Properties of the emitting gas in the narrow-line region}",
      journal = {\aap},
         year = 2019,
        month = jun,
       volume = {626},
          eid = {A9},
        pages = {A9},
          doi = {10.1051/0004-6361/201935062},
archivePrefix = {arXiv},
       eprint = {1903.11085},
 primaryClass = {astro-ph.GA},
       adsurl = {https://ui.adsabs.harvard.edu/abs/2019A&A...626A...9M}
}

@ARTICLE{2018ApJ...856...46R,
       author = {{Revalski}, M. and {Crenshaw}, D.~M. and {Kraemer}, S.~B. and {Fischer}, T.~C. and {Schmitt}, H.~R. and {Machuca}, C.},
        title = "{Quantifying Feedback from Narrow Line Region Outflows in Nearby Active Galaxies. I. Spatially Resolved Mass Outflow Rates for the Seyfert 2 Galaxy Markarian 573}",
      journal = {\apj},
         year = 2018,
        month = mar,
       volume = {856},
       number = {1},
          eid = {46},
        pages = {46},
          doi = {10.3847/1538-4357/aab107},
archivePrefix = {arXiv},
       eprint = {1802.07734},
 primaryClass = {astro-ph.GA},
       adsurl = {https://ui.adsabs.harvard.edu/abs/2018ApJ...856...46R}
}

@ARTICLE{2024PASA...41...99O,
       author = {{Oliveira}, Celso B. and {Dors}, Oli and {Zinchenko}, Igor and {Cardaci}, Monica and {H{\"a}gele}, Guillermo and {Morais}, Istenio and {Santos}, Pedro and {Almeida}, Gleicy},
        title = "{Semi-empirical calibration of the oxygen abundance for LINER galaxies based on SDSS-IV MaNGA - The case for strong and weak AGN}",
      journal = {\pasa},
         year = 2024,
        month = dec,
       volume = {41},
          eid = {e099},
        pages = {e099},
          doi = {10.1017/pasa.2024.110},
archivePrefix = {arXiv},
       eprint = {2411.02043},
 primaryClass = {astro-ph.GA},
       adsurl = {https://ui.adsabs.harvard.edu/abs/2024PASA...41...99O}
}

@ARTICLE{2020MNRAS.496.2191F,
       author = {{Flury}, Sophia R. and {Moran}, Edward C.},
        title = "{Chemical abundances in active galaxies}",
      journal = {\mnras},
         year = 2020,
        month = aug,
       volume = {496},
       number = {2},
        pages = {2191-2203},
          doi = {10.1093/mnras/staa1563},
archivePrefix = {arXiv},
       eprint = {2006.01113},
 primaryClass = {astro-ph.GA},
       adsurl = {https://ui.adsabs.harvard.edu/abs/2020MNRAS.496.2191F}
}

@ARTICLE{2014MNRAS.439..771B,
       author = {{Batra}, Neelam Dhanda and {Baldwin}, Jack A.},
        title = "{The metallicities of the broad emission line regions in the nitrogen-loudest quasars}",
      journal = {\mnras},
         year = 2014,
        month = mar,
       volume = {439},
       number = {1},
        pages = {771-787},
          doi = {10.1093/mnras/stu007},
       adsurl = {https://ui.adsabs.harvard.edu/abs/2014MNRAS.439..771B}
}

@ARTICLE{2018MNRAS.480..345X,
       author = {{Xu}, Fei and {Bian}, Fuyan and {Shen}, Yue and {Zuo}, Wenwen and {Fan}, Xiaohui and {Zhu}, Zonghong},
        title = "{The evolution of chemical abundance in quasar broad line region}",
      journal = {\mnras},
         year = 2018,
        month = oct,
       volume = {480},
       number = {1},
        pages = {345-357},
          doi = {10.1093/mnras/sty1763},
archivePrefix = {arXiv},
       eprint = {1807.01978},
 primaryClass = {astro-ph.GA},
       adsurl = {https://ui.adsabs.harvard.edu/abs/2018MNRAS.480..345X}
}

@ARTICLE{2021ApJ...910..115S,
       author = {{{\'S}niegowska}, Marzena and {Marziani}, Paola and {Czerny}, Bo{\.z}ena and {Panda}, Swayamtrupta and {Mart{\'\i}nez-Aldama}, Mary Loli and {del Olmo}, Ascensi{\'o}n and {D'Onofrio}, Mauro},
        title = "{High Metal Content of Highly Accreting Quasars}",
      journal = {\apj},
         year = 2021,
        month = apr,
       volume = {910},
       number = {2},
          eid = {115},
        pages = {115},
          doi = {10.3847/1538-4357/abe1c8},
archivePrefix = {arXiv},
       eprint = {2009.14177},
 primaryClass = {astro-ph.HE},
       adsurl = {https://ui.adsabs.harvard.edu/abs/2021ApJ...910..115S}
}

@ARTICLE{2022ApJ...925..121W,
       author = {{Wang}, Shu and {Jiang}, Linhua and {Shen}, Yue and {Ho}, Luis C. and {Vestergaard}, Marianne and {Ba{\~n}ados}, Eduardo and {Willott}, Chris J. and {Wu}, Jin and {Zou}, Siwei and {Yang}, Jinyi and {Wang}, Feige and {Fan}, Xiaohui and {Wu}, Xue-Bing},
        title = "{Metallicity in Quasar Broad-line Regions at Redshift 6}",
      journal = {\apj},
         year = 2022,
        month = feb,
       volume = {925},
       number = {2},
          eid = {121},
        pages = {121},
          doi = {10.3847/1538-4357/ac3a69},
archivePrefix = {arXiv},
       eprint = {2112.07799},
 primaryClass = {astro-ph.GA},
       adsurl = {https://ui.adsabs.harvard.edu/abs/2022ApJ...925..121W}
}

@ARTICLE{2019ApJ...874...22S,
       author = {{Shin}, Jaejin and {Nagao}, Tohru and {Woo}, Jong-Hak and {Le}, Huynh Anh N.},
        title = "{The Fe II/Mg II Flux Ratio of Low-luminosity Quasars at z {\ensuremath{\sim}} 3}",
      journal = {\apj},
         year = 2019,
        month = mar,
       volume = {874},
       number = {1},
          eid = {22},
        pages = {22},
          doi = {10.3847/1538-4357/ab05da},
archivePrefix = {arXiv},
       eprint = {1902.04579},
 primaryClass = {astro-ph.GA},
       adsurl = {https://ui.adsabs.harvard.edu/abs/2019ApJ...874...22S}
}

@ARTICLE{2017ApJ...834..203S,
       author = {{Sameshima}, H. and {Yoshii}, Y. and {Kawara}, K.},
        title = "{Chemical Evolution of the Universe at 0.7 < z < 1.6 Derived from Abundance Diagnostics of the Broad-line Region of Quasars}",
      journal = {\apj},
         year = 2017,
        month = jan,
       volume = {834},
       number = {2},
          eid = {203},
        pages = {203},
          doi = {10.3847/1538-4357/834/2/203},
archivePrefix = {arXiv},
       eprint = {1611.06027},
 primaryClass = {astro-ph.GA},
       adsurl = {https://ui.adsabs.harvard.edu/abs/2017ApJ...834..203S}
}

@ARTICLE{2003ApJ...596L.155M,
       author = {{Maiolino}, R. and {Juarez}, Y. and {Mujica}, R. and {Nagar}, N.~M. and {Oliva}, E.},
        title = "{Early Star Formation Traced by the Highest Redshift Quasars}",
      journal = {\apjl},
         year = 2003,
        month = oct,
       volume = {596},
       number = {2},
        pages = {L155-L158},
          doi = {10.1086/379600},
archivePrefix = {arXiv},
       eprint = {astro-ph/0307264},
 primaryClass = {astro-ph},
       adsurl = {https://ui.adsabs.harvard.edu/abs/2003ApJ...596L.155M}
}

@ARTICLE{2009A&A...494L..25J,
       author = {{Juarez}, Y. and {Maiolino}, R. and {Mujica}, R. and {Pedani}, M. and {Marinoni}, S. and {Nagao}, T. and {Marconi}, A. and {Oliva}, E.},
        title = "{The metallicity of the most distant quasars}",
      journal = {\aap},
         year = 2009,
        month = feb,
       volume = {494},
       number = {2},
        pages = {L25-L28},
          doi = {10.1051/0004-6361:200811415},
archivePrefix = {arXiv},
       eprint = {0901.0974},
 primaryClass = {astro-ph.CO},
       adsurl = {https://ui.adsabs.harvard.edu/abs/2009A&A...494L..25J}
}

@ARTICLE{2002ApJ...564..592H,
       author = {{Hamann}, Fred and {Korista}, K.~T. and {Ferland}, G.~J. and {Warner}, Craig and {Baldwin}, Jack},
        title = "{Metallicities and Abundance Ratios from Quasar Broad Emission Lines}",
      journal = {\apj},
         year = 2002,
        month = jan,
       volume = {564},
       number = {2},
        pages = {592-603},
          doi = {10.1086/324289},
archivePrefix = {arXiv},
       eprint = {astro-ph/0109006},
 primaryClass = {astro-ph},
       adsurl = {https://ui.adsabs.harvard.edu/abs/2002ApJ...564..592H}
}

@ARTICLE{2011ApJ...739...56D,
       author = {{De Rosa}, G. and {Decarli}, R. and {Walter}, F. and {Fan}, X. and {Jiang}, L. and {Kurk}, J. and {Pasquali}, A. and {Rix}, H.~W.},
        title = "{Evidence for Non-evolving Fe II/Mg II Ratios in Rapidly Accreting z \raisebox{-0.5ex}\textasciitilde 6 QSOs}",
      journal = {\apj},
         year = 2011,
        month = oct,
       volume = {739},
       number = {2},
          eid = {56},
        pages = {56},
          doi = {10.1088/0004-637X/739/2/56},
archivePrefix = {arXiv},
       eprint = {1106.5501},
 primaryClass = {astro-ph.CO},
       adsurl = {https://ui.adsabs.harvard.edu/abs/2011ApJ...739...56D}
}

@ARTICLE{2006A&A...447..157N,
       author = {{Nagao}, T. and {Marconi}, A. and {Maiolino}, R.},
        title = "{The evolution of the broad-line region among SDSS quasars}",
      journal = {\aap},
         year = 2006,
        month = feb,
       volume = {447},
       number = {1},
        pages = {157-172},
          doi = {10.1051/0004-6361:20054024},
archivePrefix = {arXiv},
       eprint = {astro-ph/0510385},
 primaryClass = {astro-ph},
       adsurl = {https://ui.adsabs.harvard.edu/abs/2006A&A...447..157N}
}

@ARTICLE{2003ApJ...589..722D,
       author = {{Dietrich}, M. and {Hamann}, F. and {Shields}, J.~C. and {Constantin}, A. and {Heidt}, J. and {J{\"a}ger}, K. and {Vestergaard}, M. and {Wagner}, S.~J.},
        title = "{Quasar Elemental Abundances at High Redshifts}",
      journal = {\apj},
         year = 2003,
        month = jun,
       volume = {589},
       number = {2},
        pages = {722-732},
          doi = {10.1086/374662},
archivePrefix = {arXiv},
       eprint = {astro-ph/0302494},
 primaryClass = {astro-ph},
       adsurl = {https://ui.adsabs.harvard.edu/abs/2003ApJ...589..722D}
}

@ARTICLE{1993ApJ...418...11H,
       author = {{Hamann}, Fred and {Ferland}, Gary},
        title = "{The Chemical Evolution of QSOs and the Implications for Cosmology and Galaxy Formation}",
      journal = {\apj},
         year = 1993,
        month = nov,
       volume = {418},
        pages = {11},
          doi = {10.1086/173366},
       adsurl = {https://ui.adsabs.harvard.edu/abs/1993ApJ...418...11H}
}

@ARTICLE{2020ApJ...898..105O,
       author = {{Onoue}, Masafusa and {Ba{\~n}ados}, Eduardo and {Mazzucchelli}, Chiara and {Venemans}, Bram P. and {Schindler}, Jan-Torge and {Walter}, Fabian and {Hennawi}, Joseph F. and {Andika}, Irham Taufik and {Davies}, Frederick B. and {Decarli}, Roberto and {Farina}, Emanuele P. and {Jahnke}, Knud and {Nagao}, Tohru and {Tominaga}, Nozomu and {Wang}, Feige},
        title = "{No Redshift Evolution in the Broad-line-region Metallicity up to z = 7.54: Deep Near-infrared Spectroscopy of ULAS J1342+0928}",
      journal = {\apj},
         year = 2020,
        month = aug,
       volume = {898},
       number = {2},
          eid = {105},
        pages = {105},
          doi = {10.3847/1538-4357/aba193},
archivePrefix = {arXiv},
       eprint = {2006.16268},
 primaryClass = {astro-ph.GA},
       adsurl = {https://ui.adsabs.harvard.edu/abs/2020ApJ...898..105O}
}

@ARTICLE{2019MNRAS.486.5853D,
       author = {{Dors}, O.~L. and {Monteiro}, A.~F. and {Cardaci}, M.~V. and {H{\"a}gele}, G.~F. and {Krabbe}, A.~C.},
        title = "{Semi-empirical metallicity calibrations based on ultraviolet emission lines of type-2 AGNs}",
      journal = {\mnras},
         year = 2019,
        month = jul,
       volume = {486},
       number = {4},
        pages = {5853-5866},
          doi = {10.1093/mnras/stz1242},
archivePrefix = {arXiv},
       eprint = {1905.00691},
 primaryClass = {astro-ph.GA},
       adsurl = {https://ui.adsabs.harvard.edu/abs/2019MNRAS.486.5853D}
}

@ARTICLE{2006MNRAS.370...43M,
       author = {{Meiring}, Joseph D. and {Kulkarni}, Varsha P. and {Khare}, Pushpa and {Bechtold}, Jill and {York}, Donald G. and {Cui}, Jun and {Lauroesch}, James T. and {Crotts}, Arlin P.~S. and {Nakamura}, Osamu},
        title = "{Elemental abundance measurements in low-redshift damped Lyman {\ensuremath{\alpha}} absorbers}",
      journal = {\mnras},
         year = 2006,
        month = jul,
       volume = {370},
       number = {1},
        pages = {43-62},
          doi = {10.1111/j.1365-2966.2006.10500.x},
archivePrefix = {arXiv},
       eprint = {astro-ph/0604617},
 primaryClass = {astro-ph},
       adsurl = {https://ui.adsabs.harvard.edu/abs/2006MNRAS.370...43M}
}

@ARTICLE{2013ApJ...772...93K,
       author = {{Kulkarni}, Girish and {Rollinde}, Emmanuel and {Hennawi}, Joseph F. and {Vangioni}, Elisabeth},
        title = "{Chemical Enrichment of Damped Ly{\ensuremath{\alpha}} Systems as a Direct Constraint on Population III Star Formation}",
      journal = {\apj},
         year = 2013,
        month = aug,
       volume = {772},
       number = {2},
          eid = {93},
        pages = {93},
          doi = {10.1088/0004-637X/772/2/93},
archivePrefix = {arXiv},
       eprint = {1301.4201},
 primaryClass = {astro-ph.CO},
       adsurl = {https://ui.adsabs.harvard.edu/abs/2013ApJ...772...93K}
}

@ARTICLE{2001ApJ...556L..63A,
       author = {{Allende Prieto}, Carlos and {Lambert}, David L. and {Asplund}, Martin},
        title = "{The Forbidden Abundance of Oxygen in the Sun}",
      journal = {\apjl},
         year = 2001,
        month = jul,
       volume = {556},
       number = {1},
        pages = {L63-L66},
          doi = {10.1086/322874},
archivePrefix = {arXiv},
       eprint = {astro-ph/0106360},
 primaryClass = {astro-ph},
       adsurl = {https://ui.adsabs.harvard.edu/abs/2001ApJ...556L..63A}
}

@ARTICLE{2022ApJ...926...80G,
       author = {{Garg}, Prerak and {Narayanan}, Desika and {Byler}, Nell and {Sanders}, Ryan L. and {Shapley}, Alice E. and {Strom}, Allison L. and {Dav{\'e}}, Romeel and {Hirschmann}, Michaela and {Lovell}, Christopher C. and {Otter}, Justin and {Popping}, Gerg{\"o} and {Privon}, George C.},
        title = "{The BPT Diagram in Cosmological Galaxy Formation Simulations: Understanding the Physics Driving Offsets at High Redshift}",
      journal = {\apj},
         year = 2022,
        month = feb,
       volume = {926},
       number = {1},
          eid = {80},
        pages = {80},
          doi = {10.3847/1538-4357/ac43b8},
archivePrefix = {arXiv},
       eprint = {2201.03564},
 primaryClass = {astro-ph.GA},
       adsurl = {https://ui.adsabs.harvard.edu/abs/2022ApJ...926...80G}
}

@ARTICLE{2020MNRAS.496.3209D,
       author = {{Dors}, O.~L. and {Maiolino}, R. and {Cardaci}, M.~V. and {H{\"a}gele}, G.~F. and {Krabbe}, A.~C. and {P{\'e}rez-Montero}, E. and {Armah}, M.},
        title = "{Chemical abundances of Seyfert 2 AGNs - III. Reducing the oxygen abundance discrepancy}",
      journal = {\mnras},
         year = 2020,
        month = aug,
       volume = {496},
       number = {3},
        pages = {3209-3221},
          doi = {10.1093/mnras/staa1781},
archivePrefix = {arXiv},
       eprint = {2006.09152},
 primaryClass = {astro-ph.GA},
       adsurl = {https://ui.adsabs.harvard.edu/abs/2020MNRAS.496.3209D}
}

@ARTICLE{2001ApJ...556..121K,
       author = {{Kewley}, L.~J. and {Dopita}, M.~A. and {Sutherland}, R.~S. and {Heisler}, C.~A. and {Trevena}, J.},
        title = "{Theoretical Modeling of Starburst Galaxies}",
      journal = {\apj},
         year = 2001,
        month = jul,
       volume = {556},
       number = {1},
        pages = {121-140},
          doi = {10.1086/321545},
archivePrefix = {arXiv},
       eprint = {astro-ph/0106324},
 primaryClass = {astro-ph},
       adsurl = {https://ui.adsabs.harvard.edu/abs/2001ApJ...556..121K}
}

@ARTICLE{2003MNRAS.346.1055K,
       author = {{Kauffmann}, Guinevere and {Heckman}, Timothy M. and {Tremonti}, Christy and {Brinchmann}, Jarle and {Charlot}, St{\'e}phane and {White}, Simon D.~M. and {Ridgway}, Susan E. and {Brinkmann}, Jon and {Fukugita}, Masataka and {Hall}, Patrick B. and {Ivezi{\'c}}, {\v{Z}}eljko and {Richards}, Gordon T. and {Schneider}, Donald P.},
        title = "{The host galaxies of active galactic nuclei}",
      journal = {\mnras},
         year = 2003,
        month = dec,
       volume = {346},
       number = {4},
        pages = {1055-1077},
          doi = {10.1111/j.1365-2966.2003.07154.x},
archivePrefix = {arXiv},
       eprint = {astro-ph/0304239},
 primaryClass = {astro-ph},
       adsurl = {https://ui.adsabs.harvard.edu/abs/2003MNRAS.346.1055K}
}

@ARTICLE{1981PASP...93....5B,
       author = {{Baldwin}, J.~A. and {Phillips}, M.~M. and {Terlevich}, R.},
        title = "{Classification parameters for the emission-line spectra of extragalactic objects.}",
      journal = {\pasp},
         year = 1981,
        month = feb,
       volume = {93},
        pages = {5-19},
          doi = {10.1086/130766},
       adsurl = {https://ui.adsabs.harvard.edu/abs/1981PASP...93....5B}
}

@ARTICLE{1974ApJ...191..309S,
       author = {{Shields}, G.~A.},
        title = "{X-ray ionization and the helium abundance in 3C 120.}",
      journal = {\apj},
         year = 1974,
        month = jul,
       volume = {191},
        pages = {309-316},
          doi = {10.1086/152969},
       adsurl = {https://ui.adsabs.harvard.edu/abs/1974ApJ...191..309S}
}





\bsp	
\label{lastpage}
\end{document}